\documentclass[aps,prb,twocolumn,floatfix]{revtex4}

\usepackage{graphicx}
\usepackage{figsize}   % used by graphicx

\begin{document}
\title{Magneto crystalline anisotropies in (Ga,Mn)As: A systematic theoretical study and comparison with experiment}
\author{J. Zemen$^{1}$, J. Ku\v{c}era$^{1}$, K. Olejn\'{\i}k$^{1}$, T. Jungwirth$^{1,2}$}
\affiliation{$^{1}$Institute of Physics ASCR, v. v. i., Cukrovarnick\'a 10,
162 00 Praha 6, Czech Republic}
\affiliation{$^{2}$School of Physics and Astronomy, University of Nottingham, Nottingham NG7 2RD, UK}
\date{\today}

\begin{abstract}
We present a theoretical survey of magnetocrystalline anisotropies in (Ga,Mn)As epilayers and compare the calculations to available experimental data. Our model is based on an envelope function description of the valence band holes and a spin representation for their kinetic-exchange interaction with localised electrons on Mn$^{2+}$ ions, treated in the mean-field approximation.
For epilayers with growth induced lattice-matching strains we study in-plane to out-of-plane easy-axis reorientations as a function of Mn local-moment concentration, hole concentration, and temperature. Next we focus on the competition of in-plane cubic and uniaxial anisotropies. We add an in-plane shear strain to the effective Hamiltonian in order to capture measured data in bare, unpatterned epilayers, and we  provide microscopic justification for this  approach. The model is then extended by an in-plane uniaxial strain and used to directly describe experiments with strains controlled by postgrowth lithography or attaching a piezo stressor. The calculated easy-axis directions and anisotropy fields are in semiquantitative agreement with experiment in a wide parameter range.
\end{abstract}
\maketitle

\section{Introduction}
%abstract 
Dilute moment ferromagnetic semiconductors, such as (Ga,Mn)As, are particularly favourable systems for the research in basic spintronics phenomena and towards potential applications in memory and information processing technologies.
For typical doping levels~1-10\% of Mn the magnetic dipole interactions and corresponding shape anisotropies are 10-100 times weaker in (Ga,Mn)As than in conventional dense-moment ferromagnets. Consequently, magnetocrystalline anisotropy plays a decisive role in the process of magnetisation reversal.
Despite the low saturation magnetisation the magnetic anisotropy fields reach $\sim$~10-100mT due to the large spin-orbit coupling.

The dependence of magnetic properties of (Ga,Mn)As epilayers on doping, external electric fields, temperature, and on strain has been explained by means of an effective model of Mn local moments anti-ferromagnetically coupled to valence band hole spins. The virtual crystal $\textbf{k} \cdot \textbf{p}$ approximation for hole states and mean-field treatment of their exchange interaction with Mn $d$-shell moments allow for efficient numerical simulations. \cite{Dietl:2001_b,Abolfath:2001_a,Jungwirth:2006_a,Wenisch:2007_a} 
The approach has proved useful in researching many thermodynamic and magneto-transport properties of (Ga,Mn)As samples with metallic conductivities,\cite{Jungwirth:2006_a} such as the measured transition temperatures,\cite{Dietl:2000_a,Jungwirth:2002_b,Jungwirth:2005_a,Jungwirth:2005_b} the anomalous Hall effect,\cite{Jungwirth:2003_b,Jungwirth:2002_a,Dietl:2003_c,Sinova:2004_c} anisotropic magneto resistance,\cite{Jungwirth:2003_b,Jungwirth:2002_c,Dietl:2003_c,Sinova:2004_c,Ruster:2005_a} spin-stiffness,\cite{Konig:2001_a} ferromagnetic domain wall widths,\cite{Dietl:2001_c,Sugawara:2008_a} Gilbert damping coefficient,\cite{Sinova:2003_a,Sinova:2004_b} and  magneto-optical coefficients.\cite{Dietl:2001_b,Sinova:2002_a,Sinova:2003_a,Lang:2005_a,Sinova:2004_c} In this study we systematically explore the reliability of the effective model in predicting the magnetocrystalline anisotropies of (Ga,Mn)As epilayer and micro-devices. In our comparisons to experiment we include an extensive collection of available published and unpublished measured data.%\cite{} 

Sec.~\ref{se_theory} reviews key elements of the physical model of (Ga,Mn)As and of the corresponding effective Hamiltonian used in our study. Special attention is given to mechanisms breaking the cubic symmetry of an ideal zinc-blende (Ga,Mn)As crystal. The lattice mismatch between the epilayer and the substrate, producing a growth-direction strain, is responsible for the broken symmetry between in-plane and out-of-plane cubic axes. Microscopic mechanism which breaks the remaining in-plane square symmetry in unpatterned epilayers is not fully understood. However, it can be modelled by introducing an additional uniaxial in-plane strain in the Hamiltonian. In Sec.~\ref{su_beyondcub} we discuss the correspondence of this effective approach and a generic  $\textbf{k} \cdot \textbf{p}$ Hamiltonian with the lowered symmetry of the $p$-orbital states which form the top of the spin-orbit coupled valence band. Sec.~\ref{su_shapeaniso} provides brief estimates of the shape anisotropy in thin-film (Ga,Mn)As epilayers and micro(nano)-bar devices.

Sections~\ref{se_unpatterned} and \ref{se_lithopiezo} give the survey and analysis of theoretical and experimental data over a wide range of strains, Mn moment concentrations, hole densities, and temperatures.
Sec.~\ref{su_pmaima} focuses on the easy-axis switching between the in-plane and out-of-plane directions.
Sec.~\ref{su_ima} studies the competition of cubic and uniaxial in-plane anisotropies.
Sec.~\ref{su_afields} provides comparison based on anisotropy fields extracted by fitting the calculated and experimental data to the phenomenological formula for the magnetic anisotropy energy.
Sec.~\ref{se_lithopiezo} studies in-plane easy axis reorientations in systems with additional in-plane uniaxial strain introduced experimentally by post-growth treatment of epilayers.
Finally, in Sec.~\ref{se_conclusion} we draw conclusions and discuss the limitations of our theoretical understanding of magnetic anisotropies in (Ga,Mn)As.

\section{Magnetic Anisotropy Modelling}
\label{se_theory}
We use the effective Hamiltonian approach to calculate the magneto-crystalline anisotropy energy of a system of itinerant carriers exchange coupled to Mn local moments. The $\textbf{k} \cdot \textbf{p}$ approximation is well suited for the description of hole states near the top of the valence band in a (III,Mn)V semiconductor. The strong spin-orbit interaction makes the band structure sensitive to the direction of the magnetisation. The Hamiltonian reads:
\begin{equation}
{\cal H}={\cal H}_{KL}+J_{pd}\sum_{I} {\bf S}_I\cdot \hat{\bf s}({\bf r}) \delta({\bf r}-{\bf R_I})+{\cal H}_{str}.
\label{Heff}
\end{equation}
${\cal  H}_{KL}$ is the six-band Kohn-Luttinger Hamiltonian\cite{Luttinger:1955_a} including the spin-orbit coupling (see Appendix~\ref{app_KLHam}). We use GaAs values for the Luttinger parameters.\cite{Vurgaftman:2001_a}. ${\cal H}_{str}$ is the strain Hamiltonian discussed in the following section. The second term in Eq.~(\ref{Heff}) is the short-range antiferromagnetic kinetic-exchange interaction between localised spin ${\bf S}_I$ ($S=5/2$) on the Mn$^{2+}$ ions and the itinerant hole spin $\hat{\bf s}$, parametrised by a constant\cite{Vurgaftman:2001_a} $J_{pd}=55$~meVm$^{-3}$. In the mean-field approximation it becomes $J_{pd} N_{Mn} \langle S \rangle \hat{\bf M} \cdot \hat{\bf s}$. The explicit form of the 6$\times$6 spin matrices $\hat{\bf s}$ is given in Ref.~[\onlinecite{Abolfath:2001_a}]. $\hat{\bf M}$ is the magnetisation unit vector and $ N_{Mn}=4x/a_0^3$ is the concentration of Mn atoms in Ga$_{1-x}$Mn$_x$As ($a_0$ is the lattice constant). Note that the Fermi temperature in the studied systems is much higher than the Curie temperature so the smearing of Fermi-Dirac distribution function is negligible. Therefore, finite temperature enters our model only in the form of decreasing the magnitude of magnetisation $ |{\bf M}| = S B_S (J_{pd} \langle \hat{\bf s} \rangle / k_B T)$, where $B_S$ is the Brillouin function, $\langle\hat{\bf s}\rangle$ is the hole spin-density calculated from the mean-field form of Eq.~(\ref{Heff}).

We emphasise that the above model description is based on the canonical Schrieffer-Wolf transformation of the many-body Anderson Hamiltonian. For (Ga,Mn)As the transformation replaces the microscopic hybridisation of Mn $d$-orbitals with As and Ga $sp$-orbitals by the effective spin-spin kinetic-exchange interaction of  $L=0,S=5/2$ local Mn-moments with host valence band states.\cite{Jungwirth:2006_a} Therefore, the local moments in the effective model carry zero spin-orbit interaction and the magneto-crystalline anisotropy is entirely due to the spin-orbit coupled valence-band holes. The $\hat{\bf M}$-dependent total energy density, which determines the magneto-crystalline anisotropy, is calculated by summing one-particle energies for all occupied hole states in the valence band,
\begin{equation}
E_{tot}({\bf M}) = \sum_{n=1}^{m} \int E_n({\bf k},{\bf M})f(E_n({\bf k},{\bf M}))d^3k, 
\label{integral}
\end{equation}
where $1 \leq m \leq 6$ is the number of occupied bands $f(E_n({\bf k}))$ is the Fermi distribution function at zero temperature.

\subsection{Beyond the cubic symmetry of the GaAs host}
\label{su_beyondcub}
The $\textbf{k} \cdot \textbf{p}$ method provides straightforward means of incorporating elastic strains,\cite{Dietl:2001_b,Bir:1974_a,Chow:1999_a} which we now discuss in more detail. Small deformation of the crystal lattice can be described by a transformation of coordinates:
\begin{equation}
r'_{\alpha} = r_{\alpha} + \sum_{\beta}e_{\alpha\beta}r_{\beta},
\label{r_exp}
\end{equation}
where $e_{\alpha\beta}$ is the strain tensor. Expressing ${\cal H}_{KL}$ in ${\bf r}^{\prime}$ coordinates leads to extra terms dependent on the strain that can be treated perturbatively. The resulting strain Hamiltonian has the same structure as the Kohn-Luttinger Hamiltonian with $k_ik_j$ replaced by $e_{ij}$. (For detailed description of ${\cal H}_{str}$ see Eq.~(\ref{strain_lutpar}) in the Appendix.) 

Lattice matching strain induced by the epitaxial growth breaks the symmetry between in-plane and out-of-plane cubic axes. Corresponding non-zero components of the strain tensor read $e_{xx}= e_{yy} \equiv e_0 = -\frac{c_{11}}{2c_{11}}e_{zz} =(a_s-a_0)/a_0$ where $a_s$ and $a_0$ are the lattice constant of the substrate and the relaxed epilayer, respectively, and $c_{12}$, $c_{11}$ are the elastic moduli.\cite{Vurgaftman:2001_a} Typical magnitudes are $e_0 \sim 10^{-4}-10^{-2}$.

As we discuss in Sec.~\ref{se_lithopiezo}, relaxing the growth strain in microbars in transverse direction produces a uniaxial symmetry breaking in the plane, described by a combination of $e_{xx}\neq e_{yy}$ and $e_{xy}$ strains, depending on the crystal orientation of the microbar.\cite{Wenisch:2007_a,Humpfner:2006_a,Wunderlich:2007_c,Pappert:2007_a} The magnitudes range between zero and the growth strain. Additional in-plane uniaxial anisotropy effects can be also induced by piezo stressors.\cite{Goennenwein:2008_a,Bihler:2008_a,Rushforth:2008_a,Overby:2008_a} The typical magnitude achieved by commercial stressors\cite{Shayegan:2003_a} at low temperature is of the order of $10^{-4}$.

An unpatterned bulk (Ga,Mn)As epilayer can also show broken in-plane symmetry, most frequently between the [110] and [1$\overline{1}$0] directions (see e.g. Refs.~[\onlinecite{Sawicki:2004_b,Sawicki:2004_a,Wang:2005_e,Welp:2004_a,Welp:2003_a,Hamaya:2005_a,Stanciu:2005_a,Thevenard:2007_a,Pappert:2006_b,Gould:2008_a}]). For convenience and for direct comparison with effects mentioned in the previous paragraph we model this ``intrinsic'' in-plane uniaxial anisotropy by $e_{xy}^{int}$. We fix its sign and magnitude for a given wafer by fitting to the corresponding measured anisotropy coefficients. To narrow down the number of fitted values for $e_{xy}^{int}$ in the extensive set of experimental data which we analyse, we assume that $e_{xy}^{int}$ describes effectively a symmetry breaking mechanism induced during growth and its value does not change upon the post-growth treatments, including annealing, hydrogenation, lithography or piezo-stressing.

We point out that an in-plane strain has not been detected experimentally in the bare unpatterned (Ga,Mn)As epilayers. It is indeed unlikely to occur as the substrate imposes the cubic symmetry. The possibility of transfer of the shear strain from the substrate to the epilayer was ruled out by the following test experiment. A 50~nm (Ga,Mn)As film was grown on GaAs substrate. An identical film was grown on the opposite side of the neighbouring part of the same substrate. Both samples developed uniaxial magnetic anisotropy along a diagonal but the easy axes were orthogonal to each other. If there were a uniaxial strain in the substrate responsible for the uniaxial anisotropy in the epilayer, the easy axes in the two samples would be collinear. Nevertheless, we argue below that the effective modelling via $e_{xy}^{int}$ provides a meaningful description of the ``intrinsic'' uniaxial anisotropy.

We compare the effective Hamiltonian corresponding to the $e_{xy}^{int}$ strain with a $\textbf{k} \cdot \textbf{p}$ Hamiltonian in which, without introducing the macroscopic lattice distortion, the $[\overline{1}10]/[110]$ symmetry is broken. In the derivation of the 6-band Kohn-Luttinger Hamiltonian originating from the As $p$-orbitals (denoted by $|X\rangle$, $|Y\rangle$, and $|Z\rangle$), the $\textbf{k} \cdot \textbf{p}$ term is treated perturbatively to second order:
\begin{equation}
\langle i|{\cal H}_{kp}|j \rangle=\frac{\hbar^2}{m_0^2}\sum_{l\notin \left\lbrace X,Y,Z\right\rbrace }\frac{\langle i|k \cdot p |l \rangle \langle l|k \cdot p|j \rangle}{E_i-E_l},
\label{perturbkp}
\end{equation}
where the diagonal terms of the unperturbed 6-band Hamiltonian corresponding to atomic orbital levels are set to zero. The symmetries of the tetrahedron (zinc-blend) point group $T_d$ narrow down the number of non-vanishing independent matrix elements, represented by Kohn-Luttinger parameters. The summation over neighbouring energy levels runs only through the $\Gamma_1$ and $\Gamma_4$ states of the conduction band as other levels are excluded due to the parity of the wave functions or by the large separation in energy. After including the spin-orbit interaction and transforming to a basis of total momentum eigen-states we obtain  the Hamiltonian ${\cal H}_{KL}$ (see Eqs.~(\ref{123456}) and (\ref{hl}) in the Appendix) with three independent Luttinger parameters $\gamma_1$,  $\gamma_2$, and $\gamma_3$, plus a spin-orbit splitting parameters $\Delta_{so}$.\cite{Chuang_1995_a,Chow:1999_a}

\begin{figure}
\includegraphics[scale=0.8]{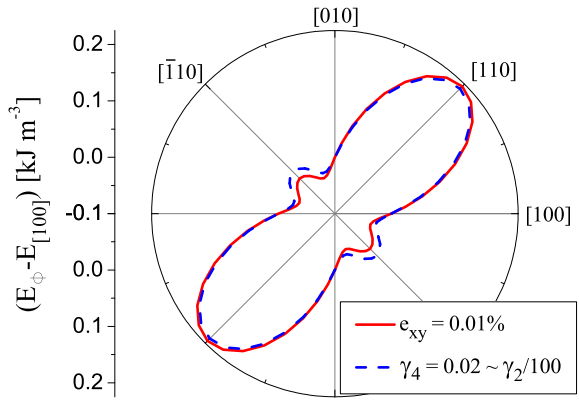}
\caption{(Color online) Modification of the originally cubic in-plane magnetic anisotropy by adding a uniaxial anisotropy due to the shear strain $e_{xy}$ or due to the local potential $V=xy\xi$. $e_0=-0.3\%$, $p=3\times 10^{20}$~cm$^{-3}$, $x=3\%$, $\gamma_4$ is the additional Luttinger parameter resulting from the in-plane symmetry lowering and $\gamma_2$ is one of the Luttinger parameters for GaAs (see text and Eq.~(\ref{lutpar_def}) in the Appendix).}
\label{micro}
\end{figure}

If the tetrahedral symmetry of the  GaAs lattice is broken the number of independent parameters increases. Let us consider a perturbation to the crystal potential that removes two of the $C_2$ elements of group $T_d$ (rotations by $180^{\circ}$ about the $[100]$ and $[010]$ axes). The corresponding potential takes a form $V = xy\xi$, which mixes the $\Gamma_1$ and $\Gamma_4 (z)$ states of the conduction band considered in the summation in Eq.~(\ref{perturbkp}) and leaves $\Gamma_4 (x)$ and $\Gamma_4 (y)$ states unchanged. ($\xi$ is a fast decreasing radial function.) Such inter-mixing of surrounding states represents the local symmetry lowering of the environment of the valence band $p$-orbitals. The summation over the perturbed states, $\alpha \Gamma_1 + \beta \Gamma_4(z)$, $-\beta \Gamma_1 + \alpha \Gamma_4(z)$, $\Gamma_4(x)$, $\Gamma_4(y)$ in Eq.~\ref{perturbkp} gives rise to extra terms in the Hamiltonian $\tilde{\cal H}_{kp}$. (The original form ${\cal H}_{kp}$ is given in Eq.~(\ref{HKL_no_so}) in the spin degenerate basis listed by Eq.~(\ref{xyz}) in the Appendix.) Assuming a weak local potential $V$, $\alpha>>\beta$, we can neglect terms of quadratic and higher order dependence on $V$ and obtain:
\begin{widetext}
\begin{equation}
\tilde{\cal H}_{kp} = \left(\begin{array}{ccc} 
Ak_x^2+B(k_z^2+k_y^2)+2Dk_xk_y & Ck_xk_y+D(k_x^2+k_y^2) & Ck_xk_z \\ 
Ck_yk_x+D(k_x^2+k_y^2) & Ak_y^2+B(k_z^2+k_x^2)+2Dk_xk_y & Ck_yk_z \\ 
Ck_zk_x & Ck_zk_y & Ak_z^2+B(k_x^2+k_y^2) \\
\end{array}\right),
\label{HKL_new}
\end{equation}
\end{widetext}
where
\begin{equation} 
D \sim \langle X |p_y| \Gamma_4 (z) \rangle \langle \Gamma_1 |p_x| X \rangle.
\end{equation}
See Eq.~(\ref{D}) in the Appendix giving the full expression for the paramater $D$. 
Elements containing the parameter $D$ change the dependence of the original Kohn-Luttinger Hamiltonian on the $k$-vector. After considering the spin-orbit coupling 
%(see Eq.~(\ref{sym_lutpar}) in the Appendix for details) 
we find that the original Kohn-Luttinger Hamiltonian with ${\cal H}_{str}$ corresponding to $e_{xy}^{int}$ has the same form as the corrected Kohn-Luttinger Hamiltonian $\tilde{\cal H}_{KL}$ with the microscopic symmetry breaking potential $V$ included if we neglect the contribution of this potential to the diagonal elements and replace the term $D(k_x^2+k_y^2)$ by a constant term proportional to $e_{xy}$.

Fig.~\ref{micro} illustrates that the in-plane anisotropy energy profile due to the local potential $V$ can indeed be accurately obtained by the mapping on the effective shear strain Hamiltonian. For the particular set of material parameters and $e_{xy}^{int}=0.01$\% considered in Fig.~\ref{micro}, the new Luttinger parameter $\gamma_4 \approx \gamma_2/100$, where $\gamma_4 = -2Dm_0/3\hbar^2$ (see Eq.~(\ref{lutpar_def}) in the Appendix for the definition of $\gamma_2$ and the other Luttinger parameters). As we discuss in the following section, effective modelling using the strain Hamiltonian with the constant $e_{xy}^{int}$  term is sufficient to capture semiquantitatively many of the observed experimental trends. Here we have demonstarted, that the model effectively describes a microscopic symmetry breaking mechanism yielding quantitatively the same in-plane anisotropy energy profiles without the presumption of a macroscopic lattice distortion.

\subsection{Shape anisotropy evaluation}
\label{su_shapeaniso}
We conclude this theoretical modelling section by briefly discussing the role of shape anisotropy in (Ga,Mn)As thin films and microstructures. Magnetic shape anisotropy is due to the long range dipolar interaction. Surface divergence of magnetisation $\textbf{M}$ gives rise to demagnetising field $\textbf{H}^D(\textbf{M},\textbf{r})$. In homogeneously magnetised bodies of general shape the demagnetising field is a function of magnetisation magnitude and direction with respect to the sample. In ellipsoidal bodies the function becomes linear in $\textbf{M}$ and $\textbf{H}^D(\textbf{M})$ is uniform in the body:
\begin{equation}
H^D_i(\textbf{M})=-\sum_j N_{ij} M_j.
\label{hd_lin}
\end{equation}
Tensor $N_{ij}$ is the so called demagnetising factor. In rectangular prisms the linear formula (\ref{hd_lin}) is a good approximation and the non-uniform demagnetising factor can be replaced by its spatial average. For the magnetostatic energy density of a homogeneously magnetised rectangular prism we get:
\begin{equation}
E^D(\textbf{M})=-\frac{1}{2} \mu \sum_{ij}N_{ij}(a,b,c) M_i M_j,
\label{ed_const}
\end{equation}
where we assume a prism extending over the volume $-a<x<a$, $-b<y<b$ and $-c<z<c$ in a Cartesian coordinate system. Ref.~[\onlinecite{Aharoni:1997_a}] shows the expression for $N_{ij}(a,b,c)$ in such prism.

\begin{figure}[h]
\includegraphics[scale=0.34]{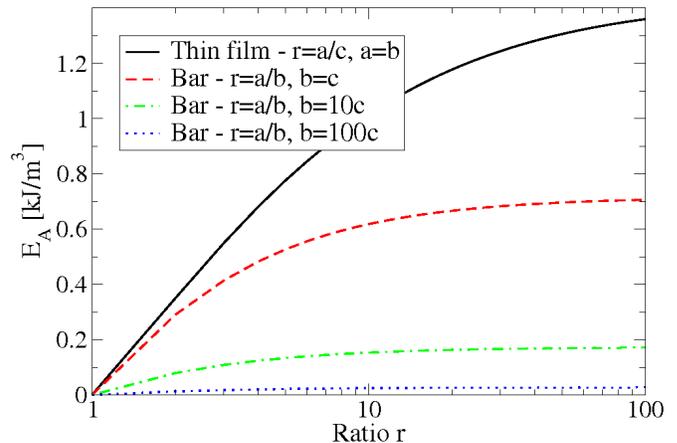}
\caption{Shape anisotropy $E_A = E^D(M_1)-E^D(M_2)$ of a film of a thickness $c$ and a long bar of length $a$ and width $b$ as a function of the dimension-less ratio $r$ as defined in the caption. The curves were obtained using the demagnetising factor approximation of Ref.~[\onlinecite{Aharoni:1997_a}] for $|\textbf{M}|=0.06$T which corresponds to Mn doping of $x=5\%$ at $T=0$K.}
\label{demag}
\end{figure}

Fig.~\ref{demag} shows the calculated shape anisotropy energy $E_A = E^D(M_1)-E^D(M_2)$ for a (i) thin film with $a=b>c$ and with magnetisation out-of-plane or in-plane ($M_1=(0,0,M)$, $M_2=(M,0,0)$), and (ii) for a bar with $a>b \sim c$ and with magnetisation in-plane ($M_1=(0,M,0)$, $M_2=(M,0,0)$). In the former case the shape anisotropy favours in-plane easy-axis direction while in the latter case the easy-axis tends to align along the bar.

As a result of the relatively low saturation magnetisation of the dilute magnetic semiconductor, the in-plane vs. out-of-plane shape anisotropy $E_A$ is only about $1.4$ kJ/m$^3$ ($0.06$~T) for Mn doping $x=5\%$ and $c<a/100$. This is in agreement with the limit of infinite 2D sheet, where the formula for shape anisotropy energy per unit volume simplifies to $E_A=\frac{\mu_0}{2}\textbf{M}^2 \cos^2{\theta}$. $\theta$ is the angle that the saturation magnetisation $\textbf{M}$ subtends to the plane normal. The in-plane anisotropy of a bar is even weaker and decreases with relative widening of the bar.

In general, the shape anisotropies in the (Ga,Mn)As dilute-moment  ferromagnet are weak compared to the spin-orbit coupling induced magneto-crystalline anisotropies and can be often neglected.

%  Most sensitive is PEA vs IEA at x=2%, but there still: Shape aniso ~ 0.265 kJ/m3 < Crystalline aniso ~ 0.79 kJ/m3

\section{Magnetic easy axes in unpatterned samples}
\label{se_unpatterned}

A large amount of experimental data on magnetic anisotropy in (Ga,Mn)As has accumulated over the past years. Comparison of these results with predictions of the effective Hamiltonian model is not straightforward due to the presence of unintentional compensating defects in (Ga,Mn)As epilayers. Most importantly, a fraction of Mn is incorporated in interstitial positions. These impurities tend to form pairs with Mn$_{\rm Ga}$ acceptors in as-grown systems with approximately zero net moment of the pair, resulting in an effective local-moment doping $x_{eff}=x_s - x_i$.\cite{Jungwirth:2005_b} Here $x_s$ and $x_i$ are partial concentrations of substitutional and interstitial Mn, respectively. In as-grown materials, the partial concentration $x_i$ increases with the total Mn concentration, $x_{tot}=x_s + x_i$. For $x_{tot}>1.5\%$, $dx_i/dx \approx 0.2$.\cite{Jungwirth:2005_b} We emphasise that in theory the Mn local moment doping labelled as "$x$" corresponds to the density of uncompensated local moments, i.e., to $x_{eff}$ in the notation used above. Mn doping "$x$" quoted in experimental works refers typically to the total nominal Mn doping, i.e., to $x_{tot}$. When comparing theory and experiment this distinction has to be considered.

%For brevity of classification of theoretical results, we use $x$ as the concentration of uncompensated Mn local moments, but it should be kept in mind, that this value corresponds to $x_{eff}$. (In discussion of experiments in all following sections the Mn concentration is denoted $x$ and corresponds to the nominal concentration $x_{tot}$ or the measurement method is specified.) 
Although interstitial Mn can be removed by low-temperature annealing, $x_{eff}$ will remain smaller than the total nominal Mn doping. The interstitial Mn impurities are double donors. Assuming no other sources of charge compensation the hole density is given by $p=(x_s - 2x_i)4/a_0^3$.\cite{Jungwirth:2005_b}

The concentration of ferromagnetically ordered Mn local moments and holes is not accurately controlled during growth or determined post growth.\cite{Jungwirth:2005_a} We acknowledge this uncertainty when comparing available magnetometry results with theory. Throughout the paper we test the relevance of our model over a wide parameter range, focusing on general trends rather than on matching results directly based on the material parameters assumed in the experimental papers. 

\subsection{In-plane vs. out-of-plane magnetic easy axis}
\label{su_pmaima}

In this section we study the switching between in-plane and perpendicular-to-plane directions of the magnetic easy axis. (Anisotropies within the growth plain of a sample are studied in Sec. \ref{su_ima}.)
Early experiments were suggesting that the in-plane vs. perpendicular-to-plane easy axis direction is determined exclusively by the sign of the growth induced strain in the sample. The in-plane easy axis (IEA) develops for compressive growth strain $e_0 = (a_{s}-a_0)/a_0 <0$. Tensile growth strain, $e_0>0$, results in the perpendicular-to-plane easy axis (PEA). This simple picture was subsequently corrected by experimental results reported for example in Refs. [\onlinecite{Ohno:2000_a,Sawicki:2004_b,Sawicki:2006_a,Thevenard:2005_a,Thevenard:2006_a}]. Sign changes in the magnetic anisotropy for the same sign of the growth strain were observed with varying Mn concentration, hole density, and temperature. 

\begin{figure}
\includegraphics[scale=0.81]{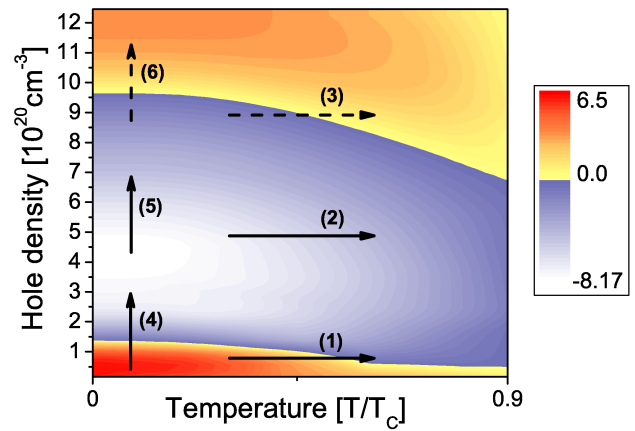}
\caption{(Color online) Anisotropy energy $\Delta E = E(M_{||}) - E(M_{\perp})$ [kJm$^{-3}$] calculated for $x=8\%$, $e_0=-0.2\%$, $e_{xy}=0$. Positive(negative) $\Delta E$ corresponds to IEA(PEA). Arrows mark anisotropy transitions driven by change of temperature or hole density.}
\label{gx8}
\end{figure}

\begin{figure}
\includegraphics[scale=0.81]{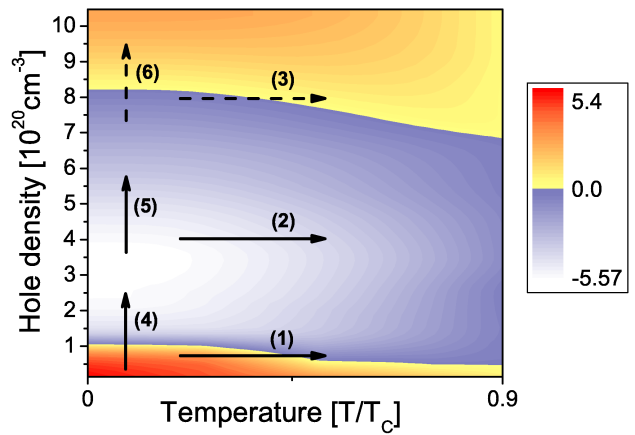}
\caption{(Color online) Anisotropy energy $\Delta E = E(M_{||}) - E(M_{\perp})$ [kJm$^{-3}$] calculated for $x=6\%$, $e_0=-0.2\%$, $e_{xy}=0$. Positive(negative) $\Delta E$ corresponds to IEA(PEA). Arrows mark anisotropy transitions driven by change of temperature or hole density.}
\label{gx6}
\end{figure}

\begin{figure}
\includegraphics[scale=0.81]{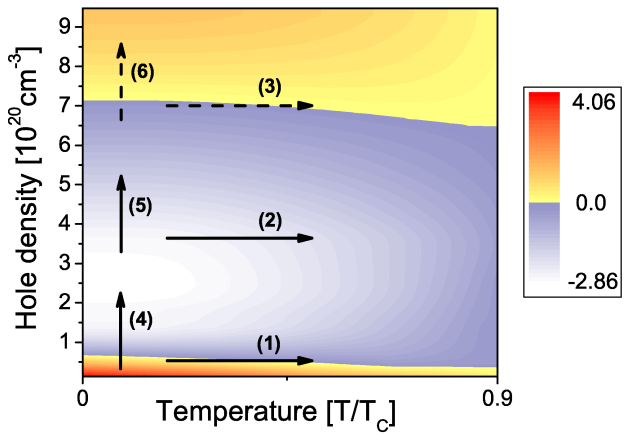}
\caption{(Color online) Anisotropy energy $\Delta E = E(M_{||}) - E(M_{\perp})$ [kJm$^{-3}$] calculated for $x=4\%$, $e_0=-0.2\%$, $e_{xy}=0$. Positive(negative) $\Delta E$ corresponds to IEA(PEA). Arrows mark anisotropy transitions driven by change of temperature or hole density.}
\label{gx4}
\end{figure}

\begin{figure}
\includegraphics[scale=0.81]{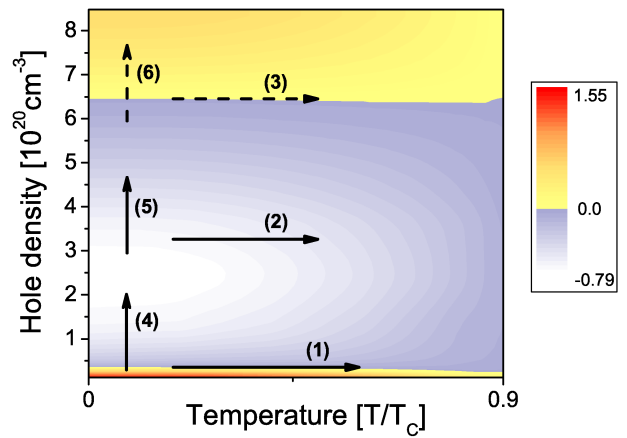}
\caption{(Color online) Anisotropy energy $\Delta E = E(M_{||}) - E(M_{\perp})$ [kJm$^{-3}$] calculated for $x=2\%$, $e_0=-0.2\%$, $e_{xy}=0$. Positive(negative) $\Delta E$ corresponds to IEA(PEA). Arrows mark anisotropy transitions driven by change of temperature or hole density.}
\label{gx2}
\end{figure}

An overview of theoretical easy axis reorientations driven by changes of the material parameters is given in  Figs.~\ref{gx8} - \ref{gx2}. In the plots we show the difference $\Delta E$ between total hole energy density for the magnetisation lying in-plane ($E_{tot}(M_{||})$) and out of plane ($E_{tot}(M_{\perp})$) as a function of the hole density and temperature. ($E_{tot}(M_{||})$ is always the smaller of $E_{tot}$ for magnetisation along the $[100]$ and the $[110]$ axis.)
% it would be more apropriate to compare projections of easy axis but the error is small due to strong growth strain... sqrt(2)*K_[001] ~ K_c1 aby se pretahovali...
We include calculations for four Mn local moment concentrations to facilitate the comparison with experimental data of different nominal Mn concentrations and different degree of annealing, which also increases the number of uncompensated local moments as discussed above. We note that the calculated magneto-crystalline anisotropies are almost precisely linear in the growth strain and therefore the boundaries between IEA and PEA in the Figs.~\ref{gx8} - \ref{gx2} depend only very weakly on the magnitude of the growth strain, certainly up to the typical experimental values $|e_0|<1\%$. Magneto-crystalline anisotropy diagrams presented in this section for a compressive strain $e_0=-0.2\%$ are therefore generic for all typical strains, with the IEA and PEA switching places for tensile strain.

Solid arrows in Figs.~\ref{gx8} - \ref{gx2} mark easy-axis behaviour as a function of temperature and doping that has been observed experimentally. The dashed arrows correspond to theoretical anisotropy variations that have not been observed experimentally. At low hole densities, increasing temperature (marked by arrow (1)) induces a reorientation of the easy axis from a perpendicular-to-plane to an in-plane direction. With decreasing $x$ this transition shifts to lower hole densities; at $x=2$\% the theoretical densities allowing for such a transition reach unrealistically low values for a ferromagnetic (Ga,Mn)As material with metallic conduction. Warming up the partially compensated samples (marked by arrow (2)) has no reorientation effect and the easy axis stays in-plane. There are no exceptions to this behaviour at different Mn concentrations. Finally, increasing temperature of a very weakly compensated (fully annealed) sample can cause switching of the theoretical easy direction from in-plane to perpendicular-to-plane (marked by arrow(3)), with the exception of the low Mn concentrations.

The techniques used to increase the hole density in the experimental works discussed in this section are the postgrowth sample annealing and annealing followed by hydrogen passivation/depassivation.\cite{Thevenard:2005_a} The latter method yields solely a change of hole density, whereas the former is associated also with an increase of the effective Mn concentration and a decrease of the growth strain. The growth strain is caused to a large extent by Mn atoms in interstitial positions,\cite{Masek:2005_a} which are removed by the annealing. The simultaneous increase of hole density and effective Mn concentration due to annealing implies a transfer between the phase diagrams of Figs.~\ref{gx8} - \ref{gx2} accompanying the transitions marked by arrows (4) - (6). We argue that the remarkable similarity of the four diagrams assures a meaningfull qualitative comparison with the effect of annealing even within a given diagram.

We now discuss individual measurements and compare with  theoretical diagrams in Figs.~\ref{gx8} - \ref{gx2}. Ref.~[\onlinecite{Thevenard:2005_a}] reports experiments in a 50~nm thick (Ga,Mn)As epilayer nominally doped to $x=6 - 7\%$ and grown on a GaAs substrate under compressive strain. The sample is first annealed to lower the number of interstitial Mn, then hydrogenated to passivate virtually all itinerant holes and finally depassivated in subsequent steps by annealing. The hole density was not measured but for the given Mn doping we expect the density in the range of $p\sim 10^{20} - 10^{21}$cm$^{-3}$ after depassivation. The low temperature ($T=4$~K) reorientation from PEA to IEA induced by successive depassivations and detected indirectly by anomalous Hall effect measurement in Ref.~[\onlinecite{Thevenard:2005_a}] matches the transition marked by arrow (4) in Figs.~\ref{gx8} - \ref{gx4}.

Magnetic hysteresis loops measured by the Hall resistivity in Ref.~[\onlinecite{Thevenard:2006_a}] reveal easy axis reorientations induced by annealing or increasing temperature in material with nominal Mn doping $x=7\%$. This (Ga,Mn)As epilayer was grown on a (In,Ga)As buffer which leads to a tensile strain. (Recall that the anisotropy energy $\Delta E$ is an odd function of the growth strain so the IEA and PEA regions have to be interchanged in Figs.~\ref{gx8}-\ref{gx2} when considering tensile strain.) Again, the hole density is not known and can be estimated to $p\sim 10^{20} - 10^{21}$cm$^{-3}$. After annealing, the material exhibits perpendicular-to-plane easy axis at 4~K and no reorientation occurs during heating up to 115~K ($T_C\approx 120-130$~K in this material). Such behaviour corresponds to arrow (2) of  Fig. \ref{gx6} or Fig. \ref{gx8}. The as-grown sample has IEA at 4~K and PEA at 22~K. This easy axis reorientation corresponds to arrow (1), again considering a tensile strain. The as-grown and annealed samples both share PEA at elevated temperature. Such a stability of the easy axis while changing the hole density corresponds to arrow (5). Theoretical anisotropy variations described by arrows (3) and (6) are not observed in Ref.~[\onlinecite{Thevenard:2006_a}]

Ref.~[\onlinecite{Sawicki:2004_b}] presents measurements in compressively strained (Ga,Mn)As epilayers grown on a GaAs substrate. The reported nominal Mn concentrations are $x=5.3\%$ and $x=3\%$ with compressive growth strain $e_0=-0.27\%$ and $e_0=-0.16\%$, respectively,  as inferred from x-ray diffraction measurement of the lattice parameter. The higher doped material was partially annealed for several different annealing times. The hole density was not measured but likely increases substantially with annealing. The as-grown $x=5.3\%$ sample at 5~K exhibits PEA, which changes to IEA upon warming up to 22~K. This anisotropy variation is not observed for samples subject to long annealing times. Such a result is consistent with Ref.~[\onlinecite{Thevenard:2006_a}] and corresponds to the theoretical predictions marked by arrows (1) and (2) of Fig. \ref{gx4} for increasing temperature of the as-grown and annealed sample, respectively. Again, the effect of annealing is in good agreement with anisotropy behaviour predicted for low (high) temperature represented by arrow (4) (arrow (5)), however, there is no experimental counterpart of transitions marked by arrows (3) and (6). The sample doped to $x=3\%$ was not annealed and no transition from PEA to IEA is observed upon warming. The behaviour corresponds to arrow (2) in Fig. \ref{gx2} or \ref{gx4}.

Ref.~[\onlinecite{Sawicki:2003_a}] already reports a successful comparison of measured magnetic anisotropy and theoretical predictions.\cite{Dietl:2001_b} Among other samples, it presents a compressively strained (Ga,Mn)As epilayer with nominal Mn concentration $x=2.3\%$ (inferred from x-ray diffraction measurement). A superconducting quantum interference device (SQUID) measurement of this as-grown sample shows PEA at 5~K and IEA at 25~K, corresponding to anisotropy variation marked by arrow (1) in Fig. \ref{gx2} (occurring only for a very narrow hole density interval).

Ref.~[\onlinecite{Liu:2003_a}] presents (Ga,Mn)As epilayers with compressive and tensile strain grown on GaAs and (In,Ga)As buffers, respectively, with nominal Mn concentration $x=3\%$ inferred from reflection high energy electron diffraction (RHEED) oscillations measured during the molecular-beam epitaxy (MBE) growth. Two of the samples are annealed and magnetic anisotropy is investigated at 5~K. The tensile strained sample has its easy axis aligned perpendicular to the growth plane and the compressively strained sample has an in-plane easy axis. This observation is in good agreement with our theoretical modelling. %The other samples are compressively strained and only in-plane magnetic anisotropy is studied.

Finally, Ref.~[\onlinecite{Ohno:2000_a}] shows a transition from PEA to IMA upon increasing temperature or change of hole concentration (induced by gating in this case). The sample is a (In,Mn)As epilayer grown on an InAs, and its magnetic anisotropy is described consistently by our model when the appropriate band parameters are used.

\subsection{In-plane anisotropy: Competition of cubic and uniaxial components}
\label{su_ima}

%The uniaxial in-plane anisotropy was first observed and described in Ref.~[\onlinecite{Sawicki:2004_b}]. Warming up the as-grown sample with nominal Mn concentration  $x=5\%$ led to a rotation of the two easy axes from a direction close to the main crystal axes towards the $[110]$ diagonal where they merged at a temperature close to $T_C$.

\begin{figure}
\includegraphics[scale=0.8]{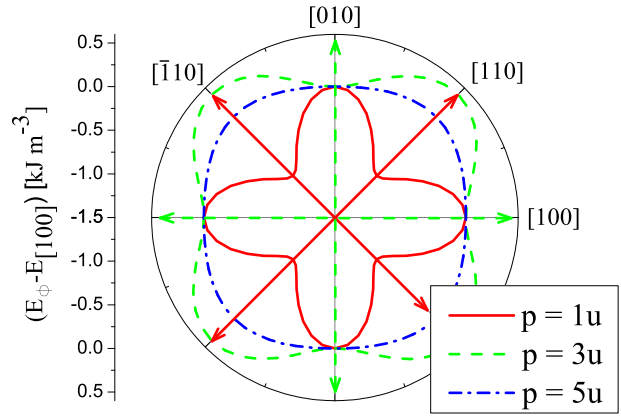}
\caption{(Color online) Magnetic anisotropy energy $\Delta E = E_{\phi}-E_{[100]}$ as a function of the in-plane magnetisation orientation ${\bf M}=|{\bf M}|[\cos\phi,\sin\phi,0]$ and its dependence on material parameters. Magnetic easy axes (marked by arrows) change their direction upon change of hole density $p$ given in units u~$\equiv 10^{20}$~cm$^{-3}$ at Mn local moment concentration $x=5\%$, shear strain $e_{xy}=0$, and zero temperature.}
\label{polar_a}
\end{figure}

\begin{figure}
\includegraphics[scale=0.8]{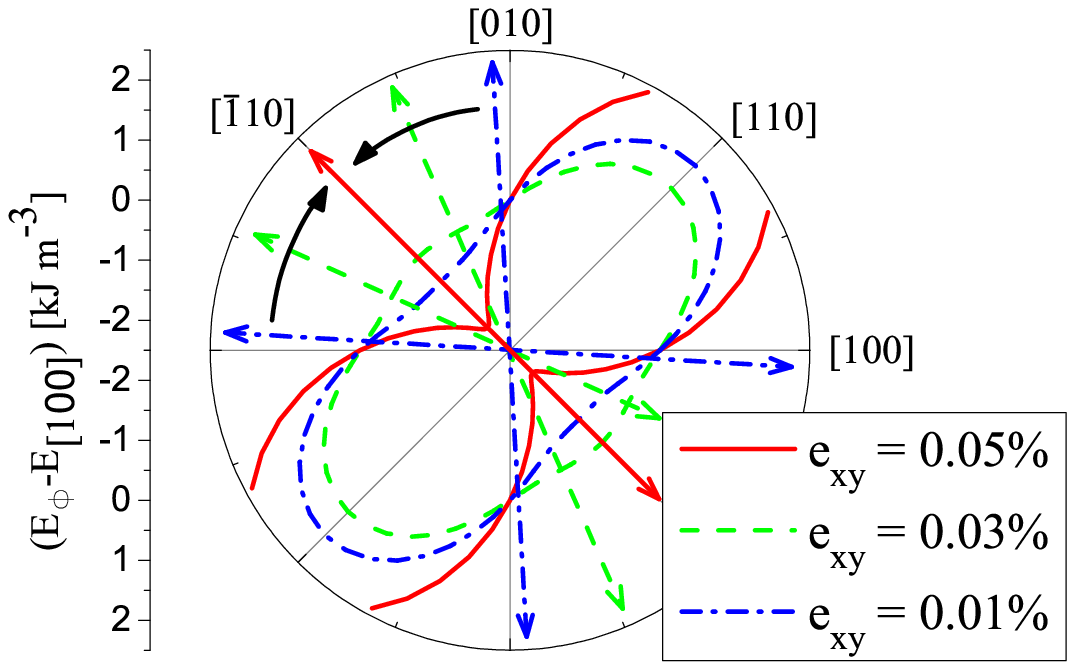}
\caption{(Color online) Magnetic anisotropy energy $\Delta E = E_{\phi}-E_{[100]}$ as a function of the in-plane magnetisation orientation ${\bf M}=|{\bf M}|[\cos\phi,\sin\phi,0]$ and its dependence on material parameters. Magnetic easy axes (marked by arrows) change their direction upon change of magnitude of shear strain $e_{xy}>0$ at Mn local moment concentration $x=5\%$, hole density $p=3\times 10^{20}$~cm$^{-3}$, and zero temperature.}
\label{polar_b}
\end{figure}

\begin{figure}
\includegraphics[scale=0.8]{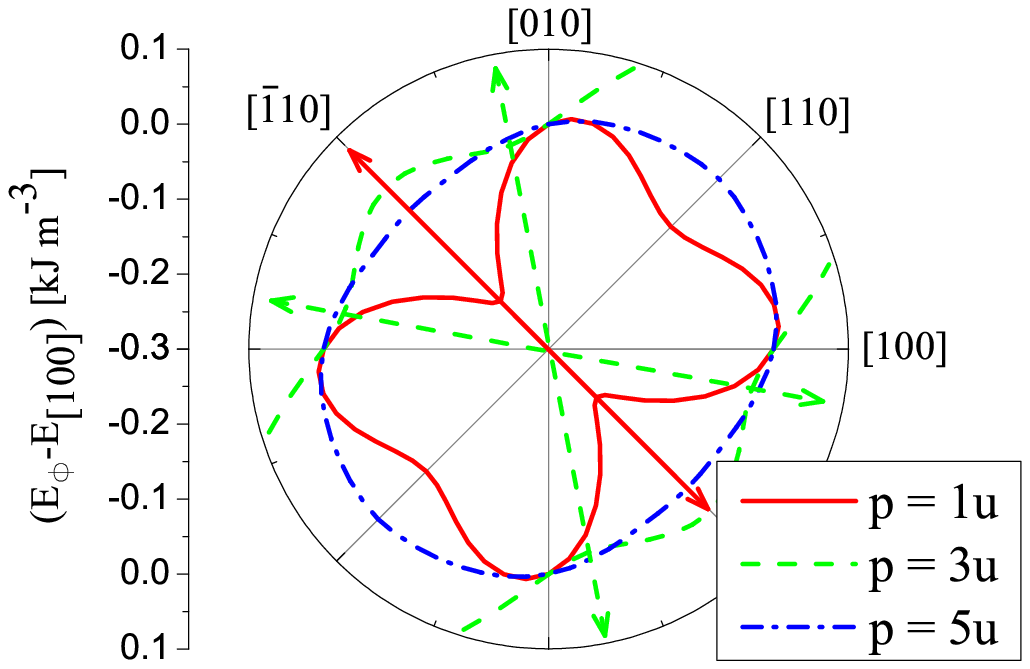}
\caption{(Color online) Magnetic anisotropy energy $\Delta E = E_{\phi}-E_{[100]}$ as a function of the in-plane magnetisation orientation ${\bf M}=|{\bf M}|[\cos\phi,\sin\phi,0]$ and its dependence on material parameters. Magnetic easy axes (marked by arrows) change their direction upon change of hole density $p$ given in units u~$\equiv 10^{20}$~cm$^{-3}$, at Mn local moment concentration $x=3\%$, shear strain $e_{xy}=0.01\%$, and zero temperature.}
\label{polar_c}
\end{figure}

\begin{figure}
\includegraphics[scale=0.8]{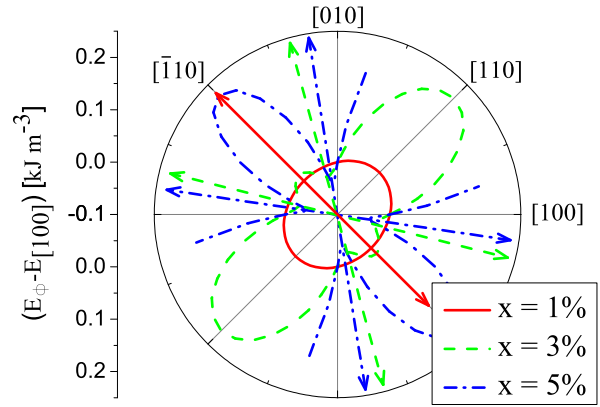}
\caption{(Color online) Magnetic anisotropy energy $\Delta E = E_{\phi}-E_{[100]}$ as a function of the in-plane magnetisation orientation ${\bf M}=|{\bf M}|[\cos\phi,\sin\phi,0]$ and its dependence on material parameters. Magnetic easy axes (marked by arrows) change their direction upon change of Mn local moment concentration $x$ at hole density $p=3\times 10^{20}$~cm$^{-3}$, shear strain $e_{xy}=0.01\%$, and zero temperature.}
\label{polar_d}
\end{figure}

As we discussed in the previous section, the magnetic easy axis(axes) is in the plane of (Ga,Mn)As/GaAs films  over a wide range of dopings. Experimental works in bare (Ga,Mn)As epilayers discussed in this section show that the in-plane magnetic anisotropy has cubic and uniaxial components. Typically, the strongest uniaxial term is along the in-plane diagonal ($[110]$/$[1\overline{1}0]$) direction. (A weak uniaxial component along the main crystal axes ($[100]$/$[010]$) has also been detected.\cite{Pappert:2006_b,Gould:2008_a}) The theoretical model used so far to describe the easy axis reorientation between the in-plane and out-of-plane alignment, assuming the growth strain, can account only for the cubic in-plane anisotropy component. In this case we find two easy axes perpendicular to each other either along the main crystal axes or along the diagonals depending on the Mn concentration and hole density, as shown in Fig. \ref{polar_a}. In order to account for the uniaxial component of the in-plane $[110]$/$[1\overline{1}0]$ anisotropy in bare (Ga,Mn)As epilayers the elastic shear strain $e_{xy}$ is incorporated into our model as discussed in Sec.~\ref{se_theory}. (For brevity we omit the index "int" in the following text and reintroduce the index only when additional real in-plane strains are present due to micro-patterning or attached piezo-stressors.) The superposition of the two components results in a rich phenomenology of magnetic easy axis alignments as reviewed in Fig.~\ref{polar_b} - \ref{polar_d}.

Fig.~\ref{polar_b} shows an example with easy axes aligned close to the main crystal axes $[100]$ and $[010]$ at Mn local moment concentration $x=5\%$, hole density $p=3\times10^{20}$cm$^{-3}$, and a weak shear strain $e_{xy}=0.01\%$. For a stronger shear strain $e_{xy}=0.03\%$ the cubic anisotropy is no longer dominant and the easy axes ``rotate'' symmetrically towards the diagonal $[1\overline{1}0]$ direction until they merge for $e_{xy} \gtrsim 0.05\%$.
As explained in detail in Sec.~\ref{se_theory}, the magnitude and sign of the intrinsic shear strain $e_{xy}$ enter as free parameters when modelling in-plane anisotropies of bare epilayers.

The relative strength of uniaxial and cubic anisotropy terms depends also on the hole density and Mn concentration as shown by Fig.~\ref{polar_c} and \ref{polar_d}, respectively. Both anisotropies are non-monotonous functions of $x$ and $p$, compared to the linear dependence of uniaxial anisotropy on the shear strain. We do not show explicitly the effect of increasing temperature which in the mean-field theory is equivalent to decreasing the effective Mn concentration while keeping the hole density constant (as explained in Sec.~\ref{se_theory}).

%\subsection{in-plane components}
%\label{ima}

We begin the comparison of theory and experiment by analysing experimental studies of in-plane magnetic anisotropy in bare samples without lithographically or piezo-electrically induced in-plane uniaxial strain. Experimental results are summarised in Tab.~\ref{exp_pap}. Samples are identified by nominal Mn concentration and hole density or annealing as given by the authors. Typically, the hole density is in the range 10$^{20}$ - 10$^{21}$cm$^{-3}$. All samples are thin (Ga,Mn)As epilayers deposited by MBE on a GaAs substrate. According to our calculations, the compressive growth strain has a negligible effect on the interplay of cubic and uniaxial in-plane anisotropies.

Tab.~\ref{exp_pap} shows the largest measured projection of the easy axis (axes) on the main crystal directions ($[100]$, $[010]$, $[110]$, $[1\overline{1}0]$) in the corresponding sample. (Note that unlike our theoretical calculations of the full in-plane anisotropy profile, most experiments listed in Tab.~\ref{exp_pap} report only projections of the magnetisation to the main crystal directions.  
%This reduction of the complete in-plane angular profile of magnetic anisotropy, as presented by Fig.~\ref{polar} in case of the calculated data, is due to the type of magnetometry used in most of the analysed samples. (The SQUID magnetometer detects only projections of the magnetisation. 
Studies using anisotropic magneto-resistance (AMR) to map the easy axis direction precisely are discussed in Sec.~\ref{su_afields} and \ref{se_lithopiezo}.)
%At the same time, the simplified description of in-plane anisotropies is more suitable for comparison with theoretical predictions.
Tab.~\ref{exp_pap} includes a column labelled as EA$_0$ giving the largest easy axis projection at low temperatures (typically 4~K) and a column labelled as EA$_{T_C}$ corresponding to measurements at temperatures close to $T_C$. This simplified overview of the temperature-dependence of the in-plane anisotropies reflects the nature of available experimental data. The ferromagnetic resonance (FMR) spectra are typically provided only at one high and one low temperature. Moreover, available SQUID data reveal at most one transition between main crystal directions corresponding to the largest projection of the magnetisation in the whole temperature interval. Sample No.~25 in Tab.~\ref{exp_pap} which shows two transitions is the only exception to this trend.

From Tab.~\ref{exp_pap} we infer the following general trend in the experimentally observed in-plane anisotropies: 
At low temperatures the in-plane anisotropy is dominated by its cubic component. In most cases, this leads to two equivalent easy axes aligned close to $[100]$ and $[010]$ directions. Only in a few samples the cubic anisotropy yields easy-axis directions along the [110]/[1-10] diagonals at low temperature. The two diagonals are not equivalent, however, due to the additional uniaxial anisotropy component.\cite{Stanciu:2005_a, Hamaya:2006_a, olejnik:unpub_a, wang:unpub_a} At high temperatures the uniaxial anisotropy dominates giving rise to only one diagonal easy axis. %We observe no prevalence of $[110]$ or $[\overline{1}10]$ direction.
Finally we note that  Refs.~[\onlinecite{Welp:2004_a, Thevenard:2007_a}] do not identify the correspondence between the in-plane diagonal easy-axis and one of the two non-equivalent crystallographic axes [110] and [1-10]  (these measurements  are marked as $\otimes$ in Tab.~\ref{exp_pap}). This ambiguity does not affect the comparison with our  modelling of  unpatterned bare films  since the shear strain $e_{xy}$ determining which of the two diagonals is magnetically easier is a free effective parameter of the theory.
Possibility of error in assigning the two non-equivalent diagonal crystallographic axes is acknowledged by the authors of Ref.~[\onlinecite{Sawicki:2004_b}], where switching roles of the diagonals makes the results consistent with later works of the group.

\begin{table} [h]
  	\begin{tabular}{|c|c|c|c|c|c|c|c|c|c|} \hline
		No. & Ref. & $x$[\%] & $p$[$^*$] & EA$_{lT}$ & EA$_{hT}$ & Fig. & A$_p$ & A$_{lT}$ & A$_{hT}$ \\ \hline \hline
		1. & [\onlinecite{Sawicki:2004_a}] & $2$ & ag & $+$ & $\nwarrow$ & \ref{tdep_strains_a} & (1) & (2) & (3) \\ \cline{1-8}
		2. & [\onlinecite{Sawicki:2004_a}] & $2$ & an & $+$ & $\nwarrow$ & \ref{tdep_strains_a} & (1) & & \\ \hline \hline
		3. & [\onlinecite{Wang:2005_e}] & $2$ & $3.5$ & $+$ & $\nwarrow$ & \ref{tdep_strains_a} & (1) &  &  \\ \hline \hline
		4. & [\onlinecite{Welp:2004_a}] & $2$ & ag & $+$ & $\otimes$ & \ref{tdep_strains_a} & (1) &  &  \\ \hline \hline
		5. & [\onlinecite{Liu:2005_d}] & $2$ & $1.1$ & $+$ & $\nearrow$ & \ref{tdep_strains_a} & (1)$^n$ & & \\ \hline \hline
		6. & [\onlinecite{Hamaya:2005_a}] & $2$ & $4$ & $+$ & $\nearrow$ & \ref{tdep_strains_a} & (1)$^n$ &  &  \\ \hline \hline
		7. & [\onlinecite{Sawicki:2004_b}] & $3$  & ag  & $+$ & $\nwarrow$ & \ref{tdep_main_a} & (1) &  &  \\ \hline \hline
		8. & [\onlinecite{Welp:2003_a}] & $3$ & ag & $+$ & $\nearrow$ & \ref{tdep_main_a} & (1)$^n$ &  &  \\ \hline \hline
		9. & [\onlinecite{Hamaya:2006_a}]  & $4$ & $3.5$ & $+$ & & \ref{tdep_main_a} &  & (2) &  \\ \cline{1-8}
		10. & [\onlinecite{Hamaya:2006_a}]  & $4$ & $5$ & $+$ & & \ref{tdep_main_a} &  &  & \\ \hline \hline
		11. & [\onlinecite{wang:unpub_a}] & $5$ & ag & $+$ & $\nwarrow$ & \ref{tdep_main_b} & (2) & (5) & (6) \\ \cline{1-8}
		12. & [\onlinecite{wang:unpub_a}] & $5$ & an & $\nearrow$ & $\nearrow$ & \ref{tdep_main_b} & (3) & & \\ \hline \hline
%		1. & [\onlinecite{Hamaya:2006_a}]  & $5.3\%$ & $4$ & $+$ & & (\ref{tdep_x3_01})(2) \\ \hline
%		1. & [\onlinecite{Hamaya:2006_a}]  & $5.3\%$ & $10$ & $\nearrow$ & & (\ref{tdep_x3_01})(2)$^p$ \\
		13. & [\onlinecite{Sawicki:2004_a}] & $5$ & ag & $+$ & $\nwarrow$ & \ref{tdep_main_b} & (2) & (4) & (6) \\ \cline{1-8}
		14. & [\onlinecite{Sawicki:2004_a}] & $5$ & an & $+$ & $\nearrow$ & \ref{tdep_main_b} & (2)$^n$ & & \\ \hline \hline
		15. & [\onlinecite{Stanciu:2005_a}] & $6$ & ag & $+$ & $\nwarrow$ & \ref{tdep_main_b} & (2) & (5) & (6)  \\ \cline{1-8}
		16. & [\onlinecite{Stanciu:2005_a}] & $6$ & an & $\nearrow$ & $\nearrow$ & \ref{tdep_main_b} & (3) & & \\ \hline \hline
		17. & [\onlinecite{Thevenard:2007_a}] & $7$ & $0.75$ & $+$ & $\otimes$ & \ref{tdep_main_c} & (3) &  &  \\ \hline
		18. & [\onlinecite{Thevenard:2007_a}] & $7$ & $2$ & $+$ & $\otimes$ & \ref{tdep_main_c} & (3) &  &  \\ \hline
		19. & [\onlinecite{Thevenard:2007_a}] & $7$ & $8.8$ & $+$ & $\otimes$ & \ref{tdep_main_c} & (4) &  &  \\ \hline
		20. & [\onlinecite{Thevenard:2007_a}] & $7$ & $12$ & $+$ & $\otimes$ & \ref{tdep_main_c} & (4) &  &  \\ \hline \hline
		21. & [\onlinecite{Hamaya:2006_a}]  & $7$ & $3.6$ & $+$ & & \ref{tdep_main_c} & & (6) &  \\ \cline{1-8}
		22. & [\onlinecite{Hamaya:2006_a}]  & $7$ & $11$ & $\nearrow$ & & \ref{tdep_main_c} &  &  & \\ \hline \hline
		23. & [\onlinecite{olejnik:unpub_a}] & $8$ & ag & $+$ & $\nwarrow$ & \ref{tdep_strains_b} & (1) & (3) & (4) \\ \cline{1-8}
		24. & [\onlinecite{olejnik:unpub_a}]  & $8$ & an & $\nwarrow$ & $\nwarrow$ & \ref{tdep_strains_b} & (2) & & \\ \hline \hline
		25. & [\onlinecite{Sawicki:2004_a}] & $8$ & an & $+$ & $\nearrow$ & \ref{tdep_main_d} & (4) & & \\ \hline
		\end{tabular}
	\caption{Experimental in-plane magneto-crystalline anisotropies at low temperature $EA_{lT}$, and high temperature $EA_{hT}$ extracted from SQUID or FMR measurements: largest easy axis projection along $[100]$ and $[010]$ axes $(+)$, along $[1\overline{1}0]$ axis $(\nwarrow)$, along $[110]$ axis $(\nearrow)$, and along one of the $[110]$/$[1\overline{1}0]$ diagonals not distinguished in the experiment $(\otimes)$. Nominal Mn concentrations $x$ reported in experimental studies are rounded down to percents. Hole density $p$ [$^*$] is given in units of $10^{20}$cm$^{-3}$. If the hole density is unknown the as-grown and annealed samples are indicated by ``ag'' and ``an'', respectively. Samples are ordered according to Mn concentration and hole density (annealed sample follows the as-grown counterpart when it exists). The last four columns label the experimental data in a way which facilitates direct comparison with  transitions highlighted by arrows in the theory Figs.~\ref{tdep_main_a} - \ref{tdep_strains_b}. Numbers in columns A$_p$, A$_{lT}$, and A$_{hT}$ point to corresponding theory transitions marked by horizontal arrows, vertical arrows at low $T$, and vertical arrows at high $T$, respectively. The index $^n$ indicates correspondence of the given arrow to modelling with negative value of $e_{xy}$.}
        %Indices $^p$ and $^x$ indicate displacement of the arrow with respect to Mn concentration and hole density, assumed in the experiment.}
	\label{exp_pap}
\end{table}

\begin{figure}[h]
\includegraphics[scale=0.48]{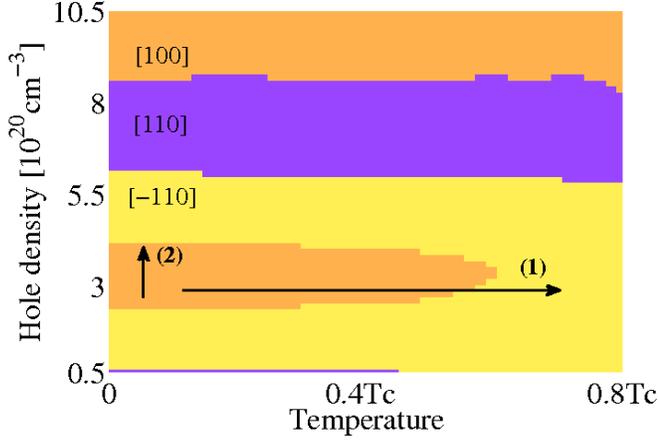}
\caption{(Color online) Theoretical hole density - temperature diagrams of crystal directions with the largest projection of the magnetic easy axis at $x=3\%$, $e_{xy}=0.01\%$, $e_0=-0.2\%$. Arrows mark anisotropy behaviour driven by change of temperature or hole density explaining experimentally observed behaviour surveyed in Tab.~\ref{exp_pap}.}
\label{tdep_main_a}
\end{figure}

\begin{figure}[h]
\includegraphics[scale=0.48]{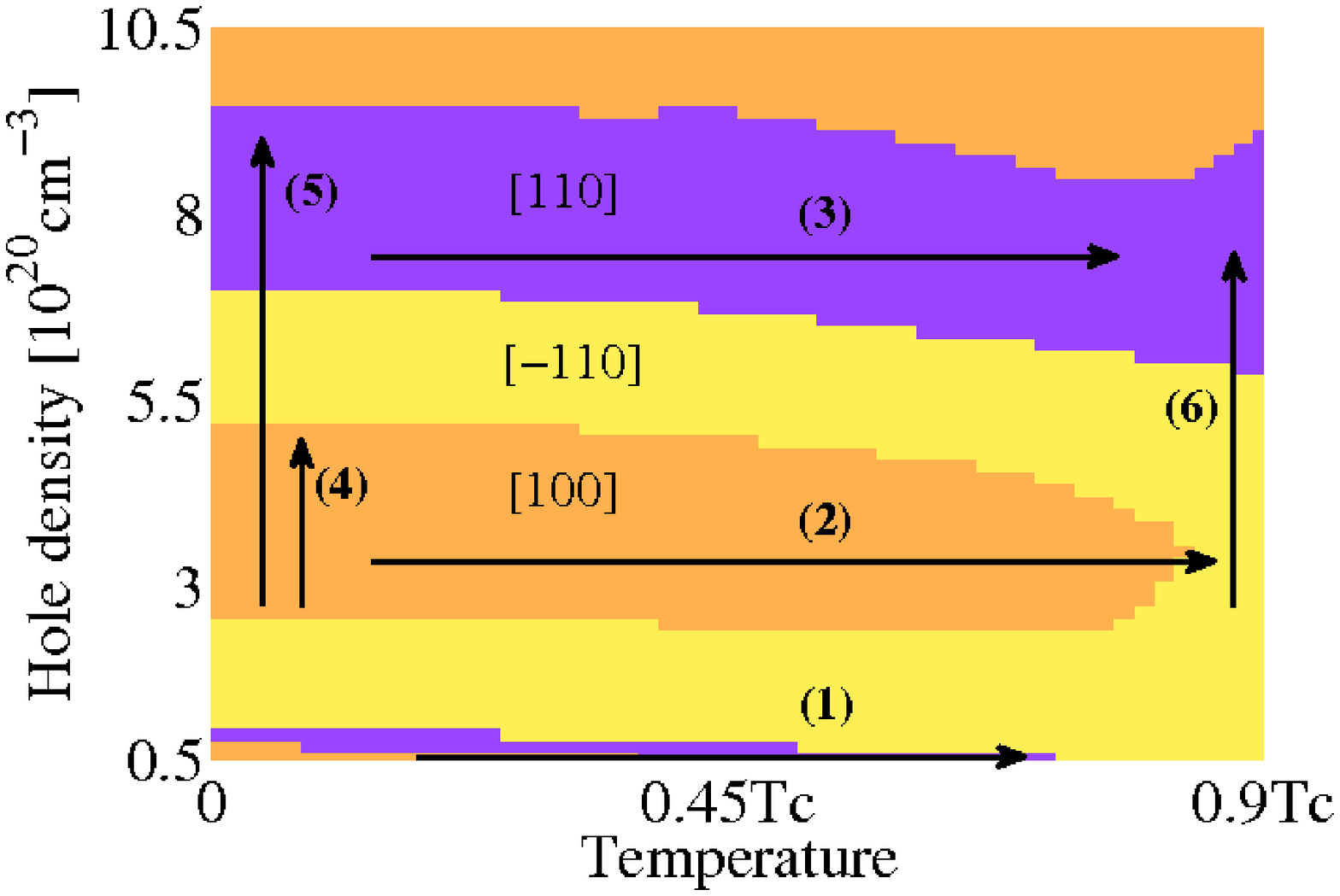}
\caption{(Color online) Theoretical hole density - temperature diagrams of crystal directions with the largest projection of the magnetic easy axis at $x=5\%$, $e_{xy}=0.01\%$, $e_0=-0.2\%$. Arrows mark anisotropy behaviour driven by change of temperature or hole density explaining experimentally observed behaviour surveyed in Tab.~\ref{exp_pap}.}
\label{tdep_main_b}
\end{figure}

\begin{figure}[h]
\includegraphics[scale=0.48]{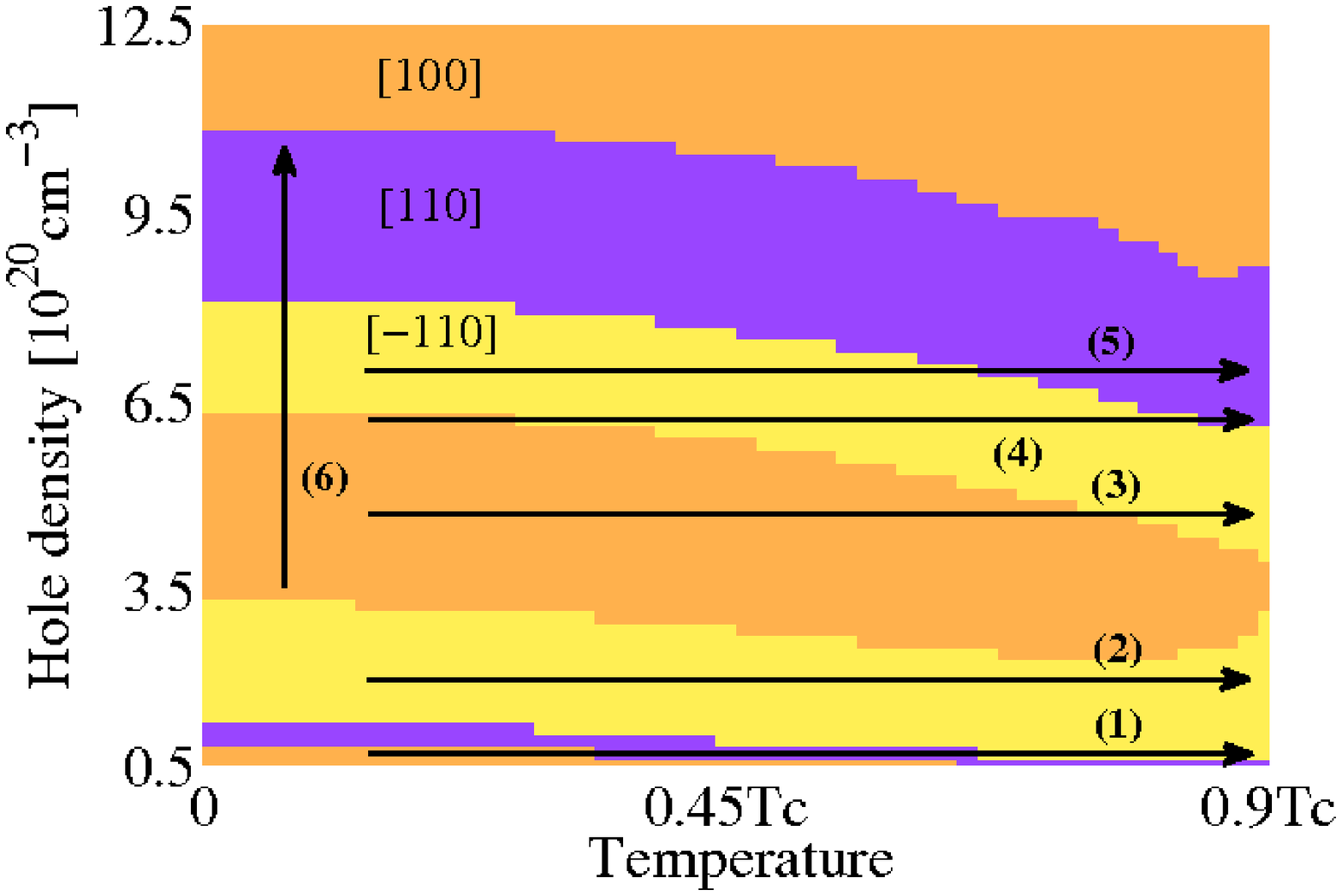}
\caption{(Color online) Theoretical hole density - temperature diagrams of crystal directions with the largest projection of the magnetic easy axis at $x=7\%$, $e_{xy}=0.01\%$, $e_0=-0.2\%$. Arrows mark anisotropy behaviour driven by change of temperature or hole density explaining experimentally observed behaviour surveyed in Tab.~\ref{exp_pap}.}
\label{tdep_main_c}
\end{figure}

\begin{figure}[h]
\includegraphics[scale=0.48]{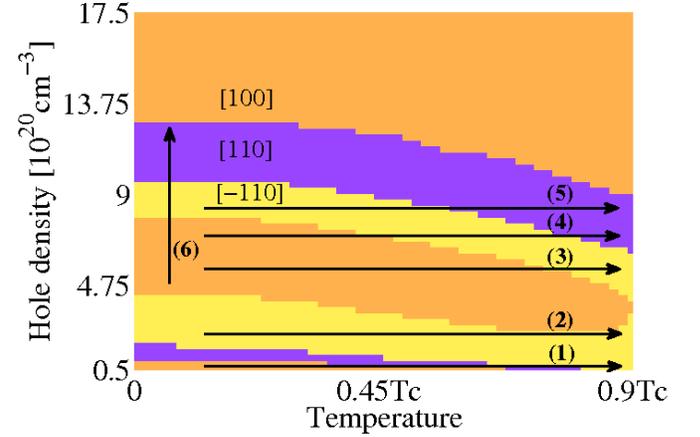}
\caption{(Color online) Theoretical hole density - temperature diagrams of crystal directions with the largest projection of the magnetic easy axis at $x=9\%$, $e_{xy}=0.01\%$, $e_0=-0.2\%$. Arrows mark anisotropy behaviour driven by change of temperature or hole density explaining experimentally observed behaviour surveyed in Tab.~\ref{exp_pap}.}
\label{tdep_main_d}
\end{figure}

\begin{figure}[h]
\includegraphics[scale=0.48]{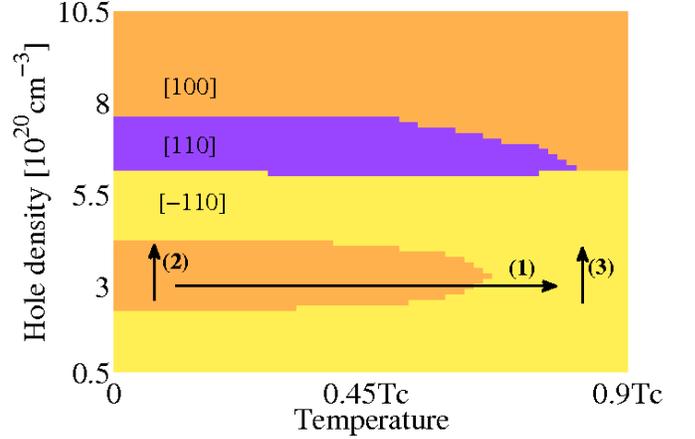}
\caption{(Color online) Theoretical hole density - temperature diagrams of crystal directions with the largest projection of the magnetic easy axis at $x=2\%$, $e_{xy}=0.005\%$, $e_0=-0.2\%$. Arrows mark anisotropy behaviour driven by change of temperature or hole density explaining experimentally observed behaviour surveyed in Tab.~\ref{exp_pap}.}
\label{tdep_strains_a}
\end{figure}

\begin{figure}[h]
\includegraphics[scale=0.48]{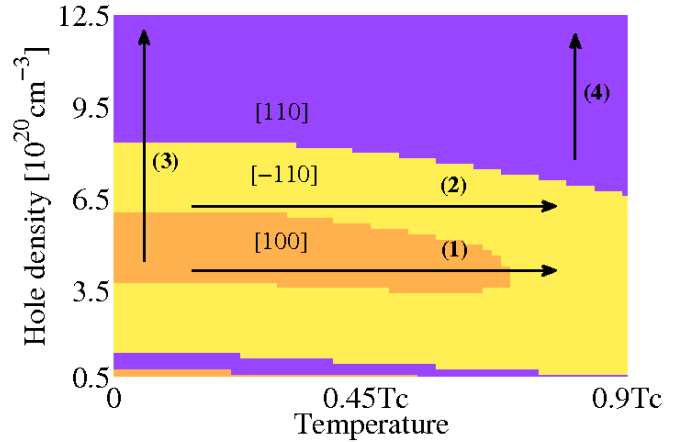}
\caption{(Color online) Theoretical hole density - temperature diagrams of crystal directions with the largest projection of the magnetic easy axis at $x=7\%$, $e_{xy}=0.03\%$, $e_0=-0.2\%$. Arrows mark anisotropy behaviour driven by change of temperature or hole density explaining experimentally observed behaviour surveyed in Tab.~\ref{exp_pap}.}
\label{tdep_strains_b}
\end{figure}

Following the strategy for presenting experimental data in Tab.~\ref{exp_pap}, we plot in Figs.~\ref{tdep_main_a} - \ref{tdep_strains_b} theoretical diagrams indicating crystallographic axes ($[100]$,$[110]$ or $[1\overline{1}0]$) with the largest projection of magnetisation as a function of the hole density and temperature. The comparison with experimental results in Tab.~\ref{exp_pap} is facilitated by numbered arrows added to the diagrams, which correspond to switchings between crystallographic directions with the largest projection of the easy-axis, driven by increasing temperature (horizontal arrows) and hole density (vertical arrows).

Figs.~\ref{tdep_main_a} - \ref{tdep_main_d} present diagrams for different Mn concentrations and for $e_{xy}=0.01\%$. Anisotropy transitions seen in the figures are consistent with majority of the reviewed experimental works, {\em i.e.}, the arrows correspond to the experimentally observed transitions and their placement in the diagrams is reasonably close to the relevant experimental parameters. Figs.~\ref{tdep_main_a} - \ref{tdep_main_d} also demonstrate how the transition from the $[100]$ to the $[1\overline{1}0]$ direction moves to higher temperatures with increasing Mn local moment concentration.

Figs.~\ref{tdep_strains_a} and~\ref{tdep_strains_b} address samples where the observed transition cannot be modelled by $e_{xy}=0.01\%$. Four of the low doped samples in Refs.~[\onlinecite{Sawicki:2004_a, Welp:2004_a, Wang:2005_e}] are modelled by a weaker strain, whereas one of the highly doped samples in Ref.~[\onlinecite{olejnik:unpub_a}] is modelled by a stronger strain. 
%We treat the shear strain as a fitting parameter: Different magnitudes are required in order to maintain a small parameter region dominated by cubic anisotropy with e.a. along the main crystal axes at low temperature for each Mn concentration as seen in experiment. Finding a universal value or at least a consistent dependence of the shear strain magnitude upon Mn doping/compensation ratio would be a support of the shear stain hypothesis.

Now we discuss in detail the theoretical diagrams in Figs.~\ref{tdep_main_a} - \ref{tdep_main_d} and compare to individual samples from Table~\ref{exp_pap}, referred to as T\ref{exp_pap}-No.
Fig.~\ref{tdep_main_a} maps in-plane magnetic anisotropy at Mn local moment concentration $x=3\%$ and shear strain $e_{xy}=0.01\%$. The easy axis reorientation of the as-grown sample T\ref{exp_pap}-7 corresponds to arrow (1) in Fig.~\ref{tdep_main_a}. Arrow (2) in Fig.~\ref{tdep_main_a} highlights the finite range of hole densities for which the largest projection of the easy-axes stays along the [100] and [010] directions at low temperature, consistent with the behaviour of the as-grown and annealed sample T\ref{exp_pap}-9 and T\ref{exp_pap}-10. (Note that hole densities in samples T\ref{exp_pap}-9 and T\ref{exp_pap}-10 were measured by the electrochemical capacitance-voltage profiling.) The transition from the largest easy axis projection along the cube edges to the $[110]$ diagonal observed in as-grown sample T\ref{exp_pap}-8 with increasing temperature has no analogy in Fig.~\ref{tdep_main_a} or Fig.~\ref{tdep_main_b}. The FMR measurement does not indicate switching of the easy axis alignment between the diagonals at any intermediate temperature. This behaviour can be explained only if the opposite sign of the shear strain is used to model the intrinsic symmetry breaking mechanism. Then the easy axis transition of T\ref{exp_pap}-8 would correspond to arrow (1) in Fig.~\ref{tdep_main_a}.

The behaviour of as-grown samples T\ref{exp_pap}-11,13,15 corresponds to arrow (2) in Fig.~\ref{tdep_main_b}. The annealed samples T\ref{exp_pap}-12,16 exhibit the rarely experimentally observed domination of uniaxial anisotropy for the whole temperature range. This behaviour is also consistently captured by the theory as highlighted by arrow (3) in Fig.~\ref{tdep_main_b}. Sample T\ref{exp_pap}-14 has a dominant cubic anisotropy preferring [100]/[010] magnetisation directions at low temperature and the easy axis aligns closer to the $[110]$ direction at high temperatures. Similarly to sample T\ref{exp_pap}-8, this transition has no analogy in Fig.~\ref{tdep_main_a} or Fig.~\ref{tdep_main_b}, however, it can be explained assuming that the [110]/[1-10] symmetry breaking mechanism has opposite sign in this material and therefore should be modelled by a negative value of the effective strain $e_{xy}$. Then the easy axis transition of T\ref{exp_pap}-14 would correspond to arrow (2) in Fig.~\ref{tdep_main_b}. Another possibility is to assume the same sign of $e_{xy}$ as for the above samples and associate the transition in sample T\ref{exp_pap}-14 with arrow (4) in Figs.~\ref{tdep_main_c} and \ref{tdep_main_d}. Note, however, that the intermediate-temperature anisotropy state with the largest magnetisation projection along the $[1\overline{1}0]$ diagonal seen when following the theory trend along arrow (4)  has not been reported in the experimental study of sample T\ref{exp_pap}-14. Arrows (4)-(6) in Fig.~\ref{tdep_main_b} correspond to measured anisotropy behaviour driven by increasing hole density in pairs of as-grown and annealed samples T\ref{exp_pap}-11,12, T\ref{exp_pap}-13,14, and T\ref{exp_pap}-15,16.

At the upper end of the investigated effective Mn concentration interval the theoretical alignment of magnetic easy axes is mapped by Figs.~\ref{tdep_main_c} and \ref{tdep_main_d}. Samples T\ref{exp_pap}-17 to T\ref{exp_pap}-20 nominally doped to $x=7\%$ were all annealed after growth, passivated by hydrogen plasma, and then gradually depassivated to achieve different hole densities (measured by high-field Hall effect). Magnetic anisotropies were determined by FMR. The assignment of the in-plane diagonal directions to the non-equivalent $[110]$ and $[1\overline{1}0]$ crystallographic axes is not specified in this experimental work; recall that this ambiguity is not crucial for the present discussion.
% the same diagonal was easier for both temperatures in low hole density samples, whereas the diagonals switched roles upon warming up at higher hole densities. Therefore arrow (3) in Fig.~\ref{tdep_main}(c) corresponds to easy axis reorientation observed for samples with low hole density T\ref{exp_pap}-17,18 rather than arrow (1) in the same figure which has more appropriate hole density but represents more complicated easy axis reorientation pattern. 
The transition observed in these samples from a cubic ($[100]$/$[010]$ easy directions) dominated anisotropy at low temperatures to a uniaxial behaviour at high temperatures is captured by arrows (3) and (4) in Figs.~\ref{tdep_main_c} and \ref{tdep_main_d}. Importantly, the depassivated higher hole density samples T\ref{exp_pap}-19 and T\ref{exp_pap}-20 show an additional switching of the easy-axis from one to the other diagonal direction at intermediate temperatures, consistent with the theoretical temperature dependence along the arrow (4). 
%Arrow (4) in Fig.~\ref{tdep_main}(c) corresponds very well to the highly depassivated (high hole density) samples T\ref{exp_pap}-19,20 as $[1\overline{1}0]$ is easier than $[110]$ direction at intermediate temperatures and vice versa close to $T_C$. 
This double transition behaviour was also detected in the annealed sample T\ref{exp_pap}-25, where the temperature dependent magnetisation projections were measured by SQUID. In this experiment it is identified that the easy-axis first rotates towards the $[1\overline{1}0]$ direction at intermediate temperatures and then switches to the $[110]$ direction at high temperatures, consistent with the behaviour marked by arrow (4) in Figs.~\ref{tdep_main_c} and \ref{tdep_main_d}.

Samples T\ref{exp_pap}-21,22 are measured only at low temperature. Easy axis reorientation from $[100]$ to $[110]$ direction is driven by increase of hole density, which corresponds to arrow (6) in Fig.~\ref{tdep_main_c} or \ref{tdep_main_d}. The hole density was measured by the electrochemical capacitance-voltage method.

In-plane anisotropies of samples with $x\approx 2\%$ are modelled in Fig.~\ref{tdep_strains_a}. To obtain the cubic anisotropy dominated region at low temperatures and a transition to the uniaxial behaviour at high temperatures, as observed in samples T\ref{exp_pap}-1 to T\ref{exp_pap}-6, we take for this low Mn doping $e_{xy}=0.005\%$. (The effective strain $e_{xy}=0.01\%$ would lead to easy axis along $[1\overline{1}0]$ over the entire temperature range and for $e_{xy}=0.001\%$ the cubic anisotropy region would extend up to very high temperatures.)
Arrow (1) in Fig.~\ref{tdep_strains_a} corresponds to easy axis switching from the $[100]$ to the $[1\overline{1}0]$ direction in samples T\ref{exp_pap}-1,2,3. Arrows (2) and (3) in Fig.~\ref{tdep_strains_a} mark the behaviour of the easy axis driven by increasing hole density when annealing the sample T\ref{exp_pap}-1 to obtain the sample T\ref{exp_pap}-2 at low and high temperature, respectively. Sample T\ref{exp_pap}-4 assumes the $[1\overline{1}0]$ diagonal always harder than the $[110]$ diagonal. A transition from cubic to uniaxial dominated anisotropy is observed upon increasing the temperature. This behaviour corresponds to arrow (1) in Figs.~\ref{tdep_strains_a}. (The hole density of sample  T\ref{exp_pap}-3, $p=3.5\times10^{20}$cm$^{-3}$, was determined by low-temperature high-field Hall effect measurements, however, it was not measured for samples T\ref{exp_pap}-1,2,4.)

Samples T\ref{exp_pap}-5 and T\ref{exp_pap}-6 have their easy axis aligned closer to the $[100]$/$[010]$ directions at low temperatures and to the $[110]$ direction at higher temperatures, similarly to sample T\ref{exp_pap}-8. The SQUID measurement of magnetisation projections for the whole range of temperature does not indicate the easy axis alignment close to the  $[1\overline{1}0]$ direction at any intermediate temperature. The hole density of the sample T\ref{exp_pap}-5, $p=1.1\times10^{20} $cm$^{-3}$, is measured by Hall effect (at room temperature) and its Mn concentration is inferred from X-ray diffraction measurement of the lattice constant. The hole density of the sample T\ref{exp_pap}-6 is $p=4\times10^{20}$ cm$^{-3}$ (measured by the electrochemical capacitance-voltage method at room temperature) and we estimate the Mn concentration from the reported critical temperature, $T_C=62$~K, after annealing. The described experimental behaviour does not correspond to predicted anisotropy transitions for relevant hole densities, Mn local moment concentrations, and positive shear strain. The behaviour can be explained, however, if the opposite sign of the shear strain is used to model the intrinsic symmetry breaking mechanism at low Mn concentration. Then the easy axis transition of T\ref{exp_pap}-5,6 would correspond to arrow (1) in Fig.~\ref{tdep_strains_a}.

Finally we comment on the less frequent behaviour observed in the annealed sample T\ref{exp_pap}-24. While its as-grown counterpart T\ref{exp_pap}-23 shows the commonly seen transition from the cubic dominated anisotropy to the uniaxial anisotropy with increasing temperature, marked by arrow (1) in Fig.~\ref{tdep_strains_b}, the annealed material has its easy axis aligned close to the $[1\overline{1}0]$ direction over the entire studied temperature range. Arrow (2) in Fig.~\ref{tdep_strains_b} provides an interpretation of this behaviour if we increase the magnitude of the effective shear strain. At $e_{xy}=0.03$\% the cubic anisotropy dominated region is already strongly diminished and for $e_{xy}=0.05$\% it vanishes completely. Arrows (3) and (4) then highlight within the same diagram the consistent description of the evolution of the experimental anisotropies, both at low and high temperatures, from the as-grown low hole density sample T\ref{exp_pap}-23 to  the annealed high hole density sample T\ref{exp_pap}-24.
%A singular type of anisotropy was observed in samples T\ref{exp_pap}-23,24. Arrow (3) in Fig.~\ref{tdep_main}(c) or arrow (1) in Fig.~\ref{tdep_strains}(b) correspond to the as-grown sample T\ref{exp_pap}-23 similarly to samples mentioned earlier, however, the annealed sample T\ref{exp_pap}-24 has its easy axis aligned with $[1\overline{1}0]$ direction for the whole temperature range. This type of anisotropy corresponds to arrow (2) in Fig.~\ref{tdep_strains}(b) for the shear strain increased to $e_{xy}=0.03\%$. The stronger strain diminishes the central cubic dominated region in the phasediagrams, which would vanish completely for $e_{xy}=0.05\%$. Arrows (3) and (4) relate to e.a. reorientation driven by increasing hole density at low and high temperature, respectively.

To summarise this section, our theoretical modelling provides a consistent overall picture of the rich phenomenology of magneto-crystalline anisotropies in unpatterned (Ga,Mn)As epilayers. Our understanding is limited, however, to only a  semiquantitative level, owing to the approximate nature of the mean-field kinetic-exchange model, ambiguities in experimental material parameters of the studied films, and unknown microscopic origin of the in-plane uniaxial symmetry breaking mechanism. We remark that the effective shear strain we include to phenomenologically account for the experimental $[110]$/$[1\overline{1}0]$ uniaxial anisotropy scales with Mn doping ($e_{xy} \simeq 0.005 x$). It brings additional confidence in this modelling approach as it is most likely the incorporation of Mn which breaks the cubic symmetry of the lattice. The magnitude of the effective strain parameter falls into the range $0.005\% < e_{xy} < 0.05\%$ and the anisotropy behaviour consistent with most experimental works is modelled with positive sign of $e_{xy}$.
%We can conclude that the calculated reorientation patterns of the magnetic easy axis are in good agreement with the general trend observed in the experimental data. The quantitative comparison on the level of material parameters is hindered by a large uncertainty in the experimentally determined hole densities and effective Mn concentration. It is reasonable to distinguish high and low hole density samples. We notice better agreement of calculated and measured easy axis reorientation patterns at low hole densities. At high hole densities the temperature driven changes of anisotropy correspond to the theoretical predictions calculated for hole densities a factor of two lower than expected in annealed highly doped samples. Only temperature driven e.a. reorientations in samples T\ref{exp_pap}-6,14 show no correspondence with our modelling.
%The analysis and qualitative comparison of theoretical and experimental data allows us to estimate the magnitude and sign of the shear strain assumed by our model. It turns out to be positive and scale linearly with Mn doping $e_{xy} \simeq 0.005 \times x [\%]$. It lies in the range $0.005\% < e_{xy} < 0.05\%$. This is well below the limit $e_{xy}<0.1\%$ imposed by the requirement of negligible effect of additional shear strain on the in-plane vs. out-of-plane anisotropy and an order of magnitude below the typical value of growth induced biaxial strain.

\begin{figure}
\includegraphics[scale=0.45]{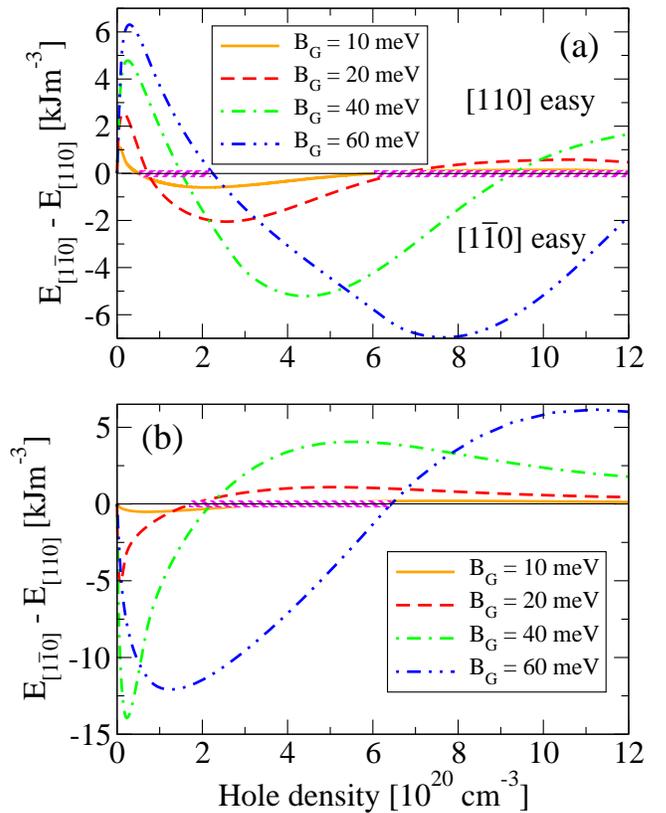}
\caption{(Color online) In-plane uniaxial anisotropy as a function hole density at zero temperature, $e_{xy}=0.05\%$, and $e_0=0$ calculated in this work (a) and in Ref.~[\onlinecite{Sawicki:2004_a}] (b). Curves are labelled by the valence-band spin-splitting parameter B$_G \equiv J_{pd}N_{Mn}S/6$ to allow for simple comparison with Ref.~[\onlinecite{Sawicki:2004_a}]. (B$_G=4.98x$ in meV and in percent, respectively.) Dashed intervals of the horizontal axis mark regions where a change of temperature (inversely proportional to B$_G$) can lead to the $[\overline{1}10] \leftrightarrow [110]$ easy axis reorientation.}
\label{err}
\end{figure}

We conclude this section by a remark on numerical simulations of the $[110]$ to $[1\overline{1}0]$ easy axis transition performed in Ref.~[\onlinecite{Sawicki:2004_a}]. The physical model employed by the authors of Ref.~[\onlinecite{Sawicki:2004_a}] is identical to ours, nevertheless, the results of the calculations do not quantitatively match ours, as illustrated in Fig.~\ref{err}. We have clarified with the authors of Ref.~[\onlinecite{Sawicki:2004_a}] the numerical origin of the discrepancy. This helpful exercise  has provided an independent confirmation of the accuracy, within the applied physical model, of the  theoretical results presented in the current paper. (To compare Fig.~\ref{err} to the original plot in Ref.~[\onlinecite{Sawicki:2004_a}] use the conversion to units of normalised anisotropy field $H_{un}/M=2(E_{[1\overline{1}0]}-E_{[110]})/(\mu_0M^2)$.)
%Ref.~[\onlinecite{Sawicki:2004_a}] presents a simulation of temperature induced switching from easy axis along $[110]$ to $[\overline{1}10]$ direction. Fig.~5 of the paper shows results calculated using a mean-field kinetic-exchange model, which differs from ours only in the additional shear strain term describing the in-plane uniaxial anisotropy. We use the same band parameters, deformation potentials, $e_{xy}=0.05\%$ and $e_0=0$. The band splitting $B_G=xJ_{pd}N_{Mn}S/6$. Fig.~\ref{err} compares the original calculation and the current one. The authors of Ref.~[\onlinecite{Sawicki:2004_a}] agree, that the current results are relevant for explanation of the particular experimental observation. Our modelling predicts the easy axis switching for two separate narrow intervals of hole density instead of a single broader interval. Both calculations can account for $[100] \longmapsto [\overline{1}10] \longmapsto [110]$ switching but the original simulation are in less agreement with samples with very low or very high hole density, e.g. Ref.~[\onlinecite{Liu:2005_d,olejnik:unpub_a}].
%we do not use 100% the same parameters but the difference has negligible effect

%%%%%%%%%%%%%%%%%%%%%%%%%%%%%%%%%%%%%%%%%%%%%%%%%%%%%%%%

\subsection{Anisotropy fields}
\label{su_afields}
Having analysed the in-plane and out-of-plane  anisotropies based on the direction of easy axes, we turn our attention to the relative strength of the anisotropy components, i.e., to the anisotropy energies. The components of magnetocrystalline anisotropy can be described in terms of a simple phenomenological model separating the free energy density $F(\hat{M})$ into components of distinct symmetry. Each component is described by a periodic function with a corresponding coefficient. We find that angular dependencies of the energies obtained from our microscopic modelling can be approximated accurately even in the first and second order of expansion into periodic functions of uniaxial and cubic symmetry, respectively.

The coefficients can be determined experimentally, e.g., by analysing the FMR spectra,\cite{Thevenard:2007_a,Liu:2005_d,Liu:2004_b,Liu:2003_a} from AMR\cite{Daeubler:2007_a,Owen:2008_a} or by fitting SQUID magnetometry data to an appropriate phenomenological formula for anisotropy energy.\cite{Wang:2005_e,olejnik:unpub_b} In this subsection we extract the relevant coefficients from the calculated anisotropies, track their dependence on material parameters and compare theory to experiment on this level.

We start with identifying the types of anisotropy terms considered in our expansion of the anisotropy energy. The cubic anisotropy due to the crystal symmetry of the zinc-blende structure is described using terms invariant under permutation of the coordinate indices $x$, $y$, and $z$. The independent first, second and third order cubic terms read: $K_{c1}\left(n_x^2 n_y^2 + n_x^2 n_z^2 + n_z^2 n_y^2 \right)$, $K_{c2}\left(n_x^2 n_y^2 n_z^2 \right)$, and $K_{c3}\left(n_x^4 n_y^4 + n_x^4 n_z^4 + n_y^4 n_z^4 \right)$, respectively, where $n_x= \cos\phi\sin\theta$, $n_y= \sin\phi\sin\theta$, and $n_z=\cos\theta$ are components of the magnetisation unit vector $\hat{M}$ (the angles $\theta$ and $\phi$ are measured from the $[001]$ and $[100]$ axis, respectively). See the Appendix~\ref{cubic_const} for details on the mutual independence of all cubic terms.

As mentioned in previous sections, the cubic anisotropy of the host crystal lattice is accompanied by different types of uniaxial anisotropy. A generic term corresponding to uniaxial anisotropy along a given unit vector $\hat{U}$ depends on the even powers of the dot product $(\hat{M} \cdot \hat{U})$. The first and second order terms read: $K_{u1}(\hat{M} \cdot \hat{U})^2$ and $K_{u2}(\hat{M} \cdot \hat{U})^4$. The particular cases of uniaxial anisotropy terms and their correspondence to lattice strains will be described later in this section.

Before we present the calculated values of the cubic anisotropy coefficient, we introduce the so called anisotropy fields which are often used in literature instead of the energy coefficients. In this section we plot the anisotropy fields in 0ersteds (Oe) to make the comparison with experiment more convenient. The relation of the anisotropy fields $H_a$ to the energy coefficients $K_a$ reads: $H_a=2K_a/M$.

\begin{figure}[h]
\begin{tabular}{c}
\includegraphics[scale=0.42]{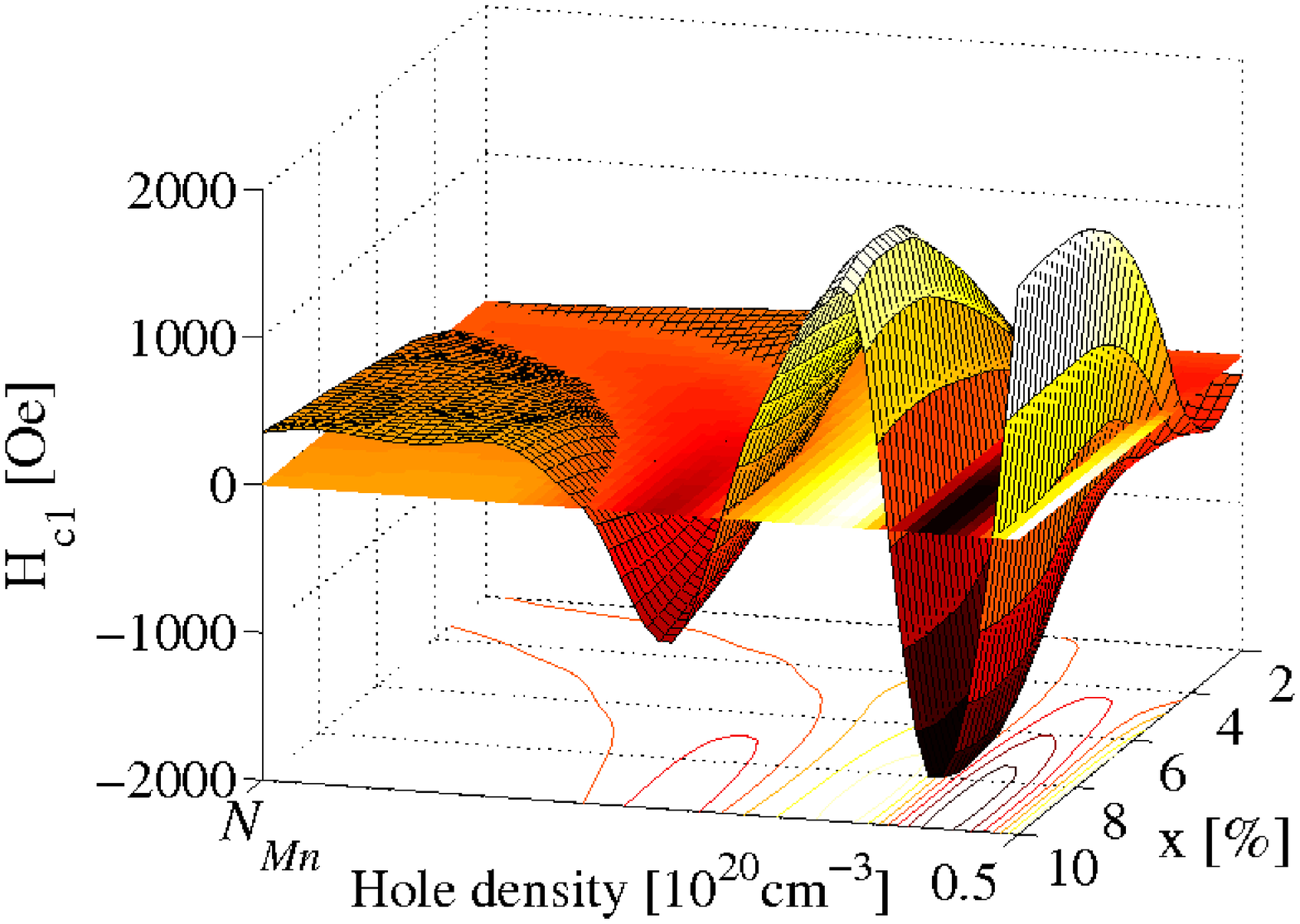} \\
%matlab/KCKU/IMA/TZERO/phase_kcku_print.m (kc...K_{c1})
%calculated on more places, used data in zemen/PROG/FERMIP/FERMIPSEQ/PRB07/PDEP/PHASEDIAGRAM/MAGANISONEW
\includegraphics[scale=0.39]{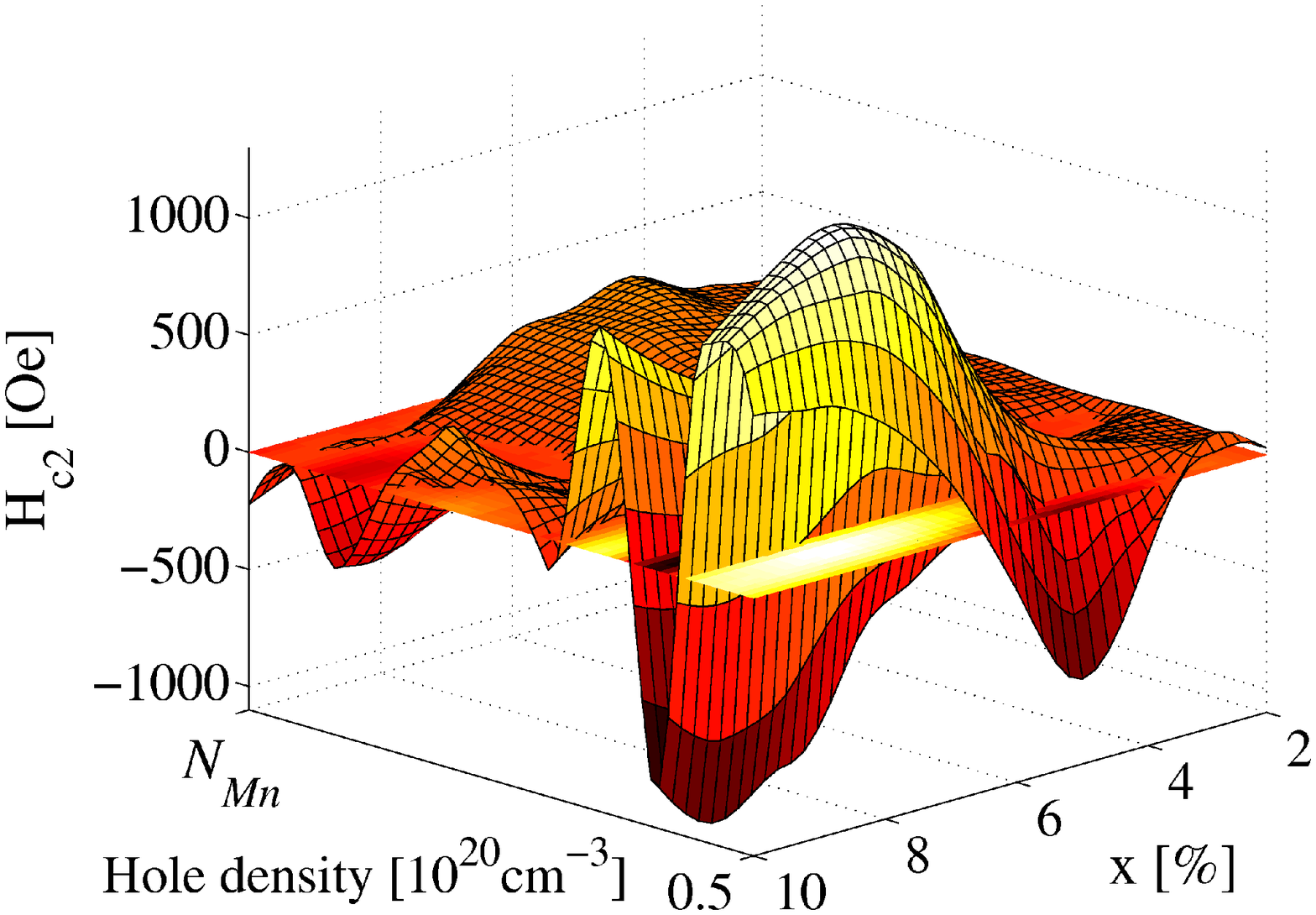} \\
%matlab/KCKU/PMA/ABOLF_PHASEDIAGRAM/PURECUB
%calculated on more places, used data in /PROG/FERMIP/FERMIPSEQ/PRB07/PDEP/PHASEDIAGRAM/MAGANISONEW/PURECUB
\end{tabular}
\caption{(color online) Lowest order cubic anisotropy field $H_{c1}$ and second order cubic anisotropy field $H_{c2}$ calculated as functions of hole density $p$ (up to zero compensation $p=N_{Mn}$) and Mn local moment concentration $x$ at zero temperature;}
\label{gen_kc}
\end{figure}

Fig.~\ref{gen_kc} shows $H_{c1}$ and $H_{c2}$ as functions of hole density $p$ and Mn local moment concentration $x$ at zero temperature. Both coefficients oscillate as function of the hole density $p$. As discussed in detail in Ref.~[\onlinecite{Abolfath:2001_a}] the anisotropies tend to weaken with increasing population of higher bands which give competing contributions. Consistent with this trend the amplitude of the oscillations increases with increasing $x$ and decreasing $p$. The upper limit of the hole density $p=N_{Mn}$ corresponds to no charge compensation (Recall, $N_{Mn} \approx 2.21x$ in $10^{20}$cm$^{-3}$ for $x$ in percent).

Our modelling predicts the extremal magnitude of the second order cubic term $H_{c2}$ a factor of two smaller than the extremal magnitude of the first order term $H_{c1}$. Upon increasing the hole density the amplitude of oscillations of $H_{c2}$ decreases faster than in case of $H_{c1}$. The third order cubic anisotropy field $H_{c3}$ is negligible compared to $H_{c1}$ and $H_{c2}$ for all studied combinations of the material parameters. To our knowledge, $H_{c2}$ and $H_{c3}$ have not been resolved experimentally. We emphasise that the second order cubic term does not contribute to the anisotropy energy for magnetisation vectors not belonging to the main crystal plains. The dependence of all three calculated cubic terms on the lattice strains of typical magnitudes (up to $1\%$) is negligible.

Now, we focus on classification of distinct uniaxial anisotropy components and their relation to lattice strains lowering the underlying cubic symmetry of the zinc-blende structure. We have already mentioned that typically the strongest symmetry breaking mechanism is the growth strain (introduced in Sec.~\ref{se_theory}). It is relevant for the in-plane versus out-of-plane alignment of the magnetic easy axis. We have also mentioned the in-plane uniaxial anisotropy between the $[110]$ and the $[1\overline{1}0]$ axes. Its origin is not known, however, we have modelled it using the shear strain which is about a factor of ten weaker than the typical growth strain.

Some (Ga,Mn)As epilayers\cite{Pappert:2006_b,Gould:2008_a} also show a very weakly broken symmetry between the main crystal axes $[100]$ and $[010]$. We will introduce here a uniaxial strain that can account for this type of anisotropy, however, our main motivation for introducing this third strain tensor is to complete an in-plane strain basis. This basis is used in Sec.~\ref{se_lithopiezo} to describe all types of lattice in-plane strains induced experimentally by growth and post-growth processing of the (Ga,Mn)As epilayers. Once the strain tensors and corresponding anisotropy contributions to the free energy are introduced, it will be shown that the chosen basis has the advantage of collinearity of the strain and of the resulting anisotropy component. Finally, in this subsection the numerical data and comparison with experiment will be presented for the bare unpatterned epilayers. The patterned structures will be discussed in Sec.~\ref{se_lithopiezo}.

%Matrix $R_{\theta}$ represents rotation by $90^{\circ}$ about the $[010]$ axis and matrix $R_{\phi}$ represents rotation by $45^{\circ}$ about the $[001]$ axis. $\varepsilon$ stands for one-dimensional expansion along arbitrary direction and $\varepsilon'=(2c_{12}/c_{11}-1)\varepsilon$ results from the assumption of zero net force resulting from the epilayer - substrate lattice missmatch.\cite{Chow:1999_a} $(2c_{12}/c_{11}-1)=-0.073$\cite{Vurgaftman:2001_a} in (Ga,Mn)As. $c_{12}$ and $c_{11}$ are the elastic moduli.\cite{Chow:1999_a}

Firstly, we recall the growth strain introduced in Eq.~(\ref{r_exp}). It is usually referred to as the biaxial pseudomorphic strain as it is due to the lattice missmatch between the substrate and the epilayer. The doped crystal is forced to certain dimensions by the substrate in the two in-plane directions whereas it can relax in the perpendicular-to-plane direction keeping the requirement of zero net force acting on the crystal: $0=c_{12}e_{xx}+c_{12}e_{yy}+c_{11}e_{zz}$. The corresponding strain tensor:
\begin{eqnarray}
\label{strain1}
{\bf e}^g & = & \left(\begin{array}{ccc}
e_0 & 0 & 0 \\
0 & e_0 & 0 \\
0 & 0 & -2\frac{c_{12}}{c_{11}}e_0 \\
\end{array} \right) \;
\end{eqnarray}
describes an expansion (contraction) along the $[100]$ and $[010]$ axes for positive (negative) $e_0$ accompanied by a contraction (expansion) along the $[001]$ axis. Parameters $c_{11}$ and $c_{12}$ are the elastic moduli. The growth strain enters our model via the strain Hamiltonian ${\cal H}_{str}$ (see Eq.~(\ref{Heff})) and induces a uniaxial anisotropy component which can be described in the lowest order by an energy term $-K_{[001]}n_z^2 = -K_{[001]}\cos^2 \theta$.

The shear strain, first introduced in Sec.~\ref{su_beyondcub}, is represented by a tensor:
\begin{eqnarray}
\label{strain2}
{\bf e}^s & = &  \left(\begin{array}{ccc} 
0 & \kappa & 0 \\
\kappa & 0 & 0 \\
0 & 0 & 0 \\
\end{array}\right). \;
\end{eqnarray}
Positive (negative) $\kappa$ corresponds to turning a square into a diamond with the longer (shorter) diagonal along the $[110]$ axis. We have used this type of strain as the ``intrinsic`` shear strain $e^{int}_{xy}$ to model the difference in energy for magnetisation aligned with the two in-plane diagonals. It results in uniaxial anisotropy along the diagonals, described in analogy to the growth strain by a term $-K_{[110]}(n_y-n_x)^2/2 = -K_{[110]} \sin^2(\phi-\pi/4) \sin^2 \theta$.

Finally, we write down the third element of the in-plane strain basis:
\begin{eqnarray}
\label{strain3}
 {\bf e}^u & = &  \left(\begin{array}{ccc} 
\lambda & 0 & 0 \\
0 & -\lambda & 0 \\
0 & 0 & 0 \\
\end{array}\right). \;
\end{eqnarray}
Positive (negative) $\lambda$ corresponds to turning a square into a rectangle where expansion (contraction) along the $[100]$ axis is accompanied by a contraction (expansion) along the $[010]$ axis of the same magnitude. Much like in case of the growth strain and the shear strain, the requirement of zero net force acting on the crystal is kept but this time it results in $e_{zz}=0$. The strain ${\bf e}^u$ induces uniaxial anisotropy along the main crystal axes, described by a term $-K_{[100]}n_y^2 = -K_{[100]} \sin^2\phi \sin^2 \theta$.

Let us remark that strain tensors in Eqs.~(\ref{strain1}-\ref{strain3}) are expressed in Cartesian coordinates fixed to the main crystallographic axes. Strains ${\bf e}^s$ and ${\bf e}^u$ for $\kappa = \lambda$ are related by a rotation about the $[001]$ axis by $\pi/4$, however, the cubic crystal is not invariant under such rotation so the two strains induce anisotropies with magnitudes $K_{[100]}$ and $K_{[110]}$ which are different in general. The growth strain ${\bf e}^g$, the shear strain ${\bf e}^s$, and the uniaxial strain ${\bf e}^u$ can be characterised by a single direction of deformation and induce uniaxial anisotropy components aligned with that particular direction. We found that higher order uniaxial terms are small unless we approach experimentally unrealistic large values of exchange splitting (large $x$) and hole compensation (low~$p$).

In total, we can write our phenomenological formula approximating accurately the calculated free energy density of an originally cubic system subject to three types of strain as a sum of distinct anisotropy components:
\begin{eqnarray}
F(\hat{M}) & = & K_{c1} \left(n_x^2 n_y^2 + n_x^2 n_z^2 + n_z^2 n_y^2 \right) + K_{c2} \left(n_x^2 n_y^2 n_z^2 \right)- \nonumber \\
& - & K_{[001]} n_z^2 - \frac{K_{[110]}}{2} (n_y-n_x)^2 - K_{[100]} n_y^2. \;
\label{free_en_us}
\end{eqnarray}
By definition of the terms, a positive coefficient $K_{[001]}$ prefers perpendicular-to-plane easy axis (PEA); positive $K_{[110]}$ and $K_{[100]}$ prefer easy axis lying in-plane (IEA) aligned closer to $[1\overline{1}0]$ and $[010]$ axis, respectively. Note that the anisotropy terms entering the phenomenological formula follow a sign convention consistent with with existing literature.\cite{Thevenard:2007_a,Liu:2005_d,Liu:2004_b,Liu:2003_a} 

We now provide the microscopic justification for the choice of the elements ${\bf e}^s$ and ${\bf e}^u$ of the in-plane strain basis and corresponding phenomenological uniaxial terms. This will be based on symmetries of the Kohn-Luttinger Hamiltonian ${\cal H}_{KL}$ and the strain Hamiltonian ${\cal H}_{str}$ as shown in Eqs.~(\ref{sym_lutpar}) and~(\ref{strain_lutpar}), respectively, which relates the band structure to a general in-plane strain with the components $e_{xx}$, $e_{yy}$, and $e_{xy}$.

First let us point out that the basis element ${\bf e}^g$ (the growth strain) is invariant under rotation about the $[001]$ axis and according to our calculation does not influence the in-plane direction of the easy axis (in the linear regime of small deformations). We continue by showing that for ${\bf e}^s$ and ${\bf e}^u$, the strains and the corresponding magnetocrystalline anisotropy components are indeed collinear and that this collinearity applies only for the special cases of uniaxial symmetries along the in-plane diagonals or main axes. Let us assume a rotation of the tensor ${\bf e}^u$ by an arbitrary angle $\omega$ about the $[001]$ axis:
\begin{eqnarray}
\label{strain_rotation}
{\bf e}^u(\omega) & = & R^T_{\omega}\left(\begin{array}{ccc} 
\lambda & 0 & 0 \\
0 & -\lambda & 0 \\
0 & 0 & 0 \\
\end{array}\right) R_{\omega} \\
 & = & \left(\begin{array}{ccc} 
\lambda \cos 2\omega & \lambda \sin 2\omega & 0 \\
\lambda \sin 2\omega & - \lambda \cos 2\omega & 0 \\
0 & 0 & 0 \\
\end{array}\right), \nonumber \;
\end{eqnarray}
where $R_{\omega}$ is the rotation matrix. (The same analysis applies to a rotation of  ${\bf e}^s$). The parameters $e_{xx}=-e_{yy}=\lambda\cos 2\omega$ and $e_{xy}=\lambda\sin 2\omega$ enter the strain Hamiltonian (see Eq.~(\ref{strain_lutpar}) in the Appendix) only via the matrix element:
\begin{eqnarray}
c^s &=& \frac{a_2}{2}\sqrt{3}(e_{yy} - e_{xx}) + ia_3 e_{xy} \nonumber \\
 &=& -\lambda \left[ a_2\sqrt{3} \cos (2\omega) - ia_3 \sin (2\omega) \right], \;
\end{eqnarray}
where $a_2 \sqrt{3}  \neq a_3$ are strain Luttinger constants. Moreover, the strain component $e_{xy}$ quantifying the shear strain enters only Im$(c^s)$, whereas the components $e_{xx}=-e_{yy}$ enter only Re$(c^s)$. According to our calculation the imaginary and real part of $c^s$ generate independent uniaxial anisotropy components along the $[110]$ and $[100]$ axis, respectively. Their combined effect can be understood based on an analogy of the in-plane rotation of the strain tensor ${\bf e}^u$ and an in-plane rotation of a $k$-vector.

As mentioned in Sec.~\ref{se_theory} the Kohn-Luttinger Hamiltonian ${\cal H}_{KL}$ and the strain Hamiltonian ${\cal H}_{str}$ have the same structure. We write here explicitly the matrix component $c$ of the Hamiltonian ${\cal H}_{KL}$ analogous to $c^s$ as a function of the in-plane angle of the $k$-vector ${\bf k}=|{\bf k}|[\cos\phi,\sin\phi,0]$. The element reads:
\begin{eqnarray}
c &=& \frac{\sqrt{3}\hbar^2}{2m}\big[\gamma_2(k_x^2 - k_y^2) - 2i(\gamma_3 k_x k_y)\big] \nonumber \\
 &=& \frac{\sqrt{3}\hbar^2}{2m}k^2\big[\gamma_2 \cos 2\phi- i\gamma_3 \sin 2\phi \big], \;
\end{eqnarray}
where again $\gamma_2 \neq \gamma_3$ are Luttinger constants describing a cubic crystal. For $\gamma_2 = \gamma_3$ the Hamiltonian ${\cal H}_{KL}$ has spherical symmetry. Similarly, if $a_2 \sqrt{3} = a_3$, the strain Hamiltonian ${\cal H}_{str}$ is spherically symmetric and the contributions of Im$(c^s)$ and Re$(c^s)$ to the anisotropy of the system combine in such a way that the resulting uniaxial term is collinear with the strain ${\bf e}^u(\omega)$ rotated with respect to the crystallographic axes by an arbitrary in-plane angle $\omega$.

Clearly, the underlying cubic symmetry of the host crystal causes a non-collinearity of the uniaxial strain along a general in-plane direction and the corresponding anisotropy component. Moreover, the misalignment is a function of Mn local moment concentration, hole density and temperature. We discuss further this misalignment in more detail in Sec.~\ref{se_lithopiezo}. Here we point out the distinct exception when $\omega$ is an integer multiple of $\pi/4$ and either the real or the imaginary part of $c^s$ vanish rendering the strain Hamiltonian effectively spherically symmetric. We choose quite naturally the simple forms of ${\bf e}^u(\omega)$ with $\omega = 0$ and $\omega = \pi/4$ as elements of the in-plane strain basis. For a different choice of the basis elements than in Eqs.~(\ref{strain2}) and~(\ref{strain3}), setting up the phenomenological formula would be more complicated.

We can now resume our discussion of the interplay of the cubic and uniaxial anisotropy components. Adding the uniaxial terms leads to rotation or imbalance of the original (cubic) easy axes as shown in Sec.~\ref{su_ima} in Fig.~\ref{polar_a}.
\begin{figure}[h]
\begin{tabular}{c}
\includegraphics[scale=0.42]{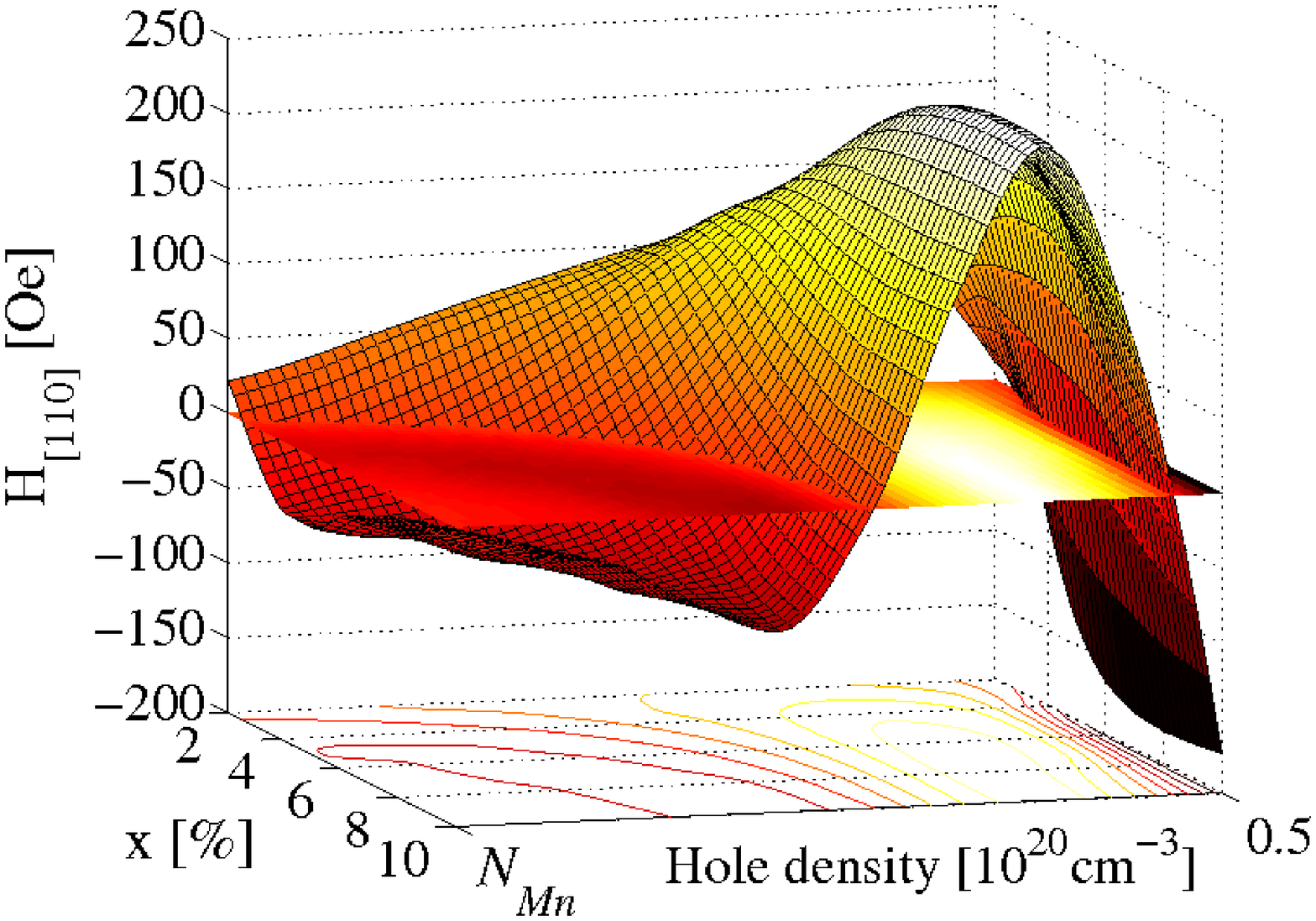}\\  % (b) $K_u$
\includegraphics[scale=0.42]{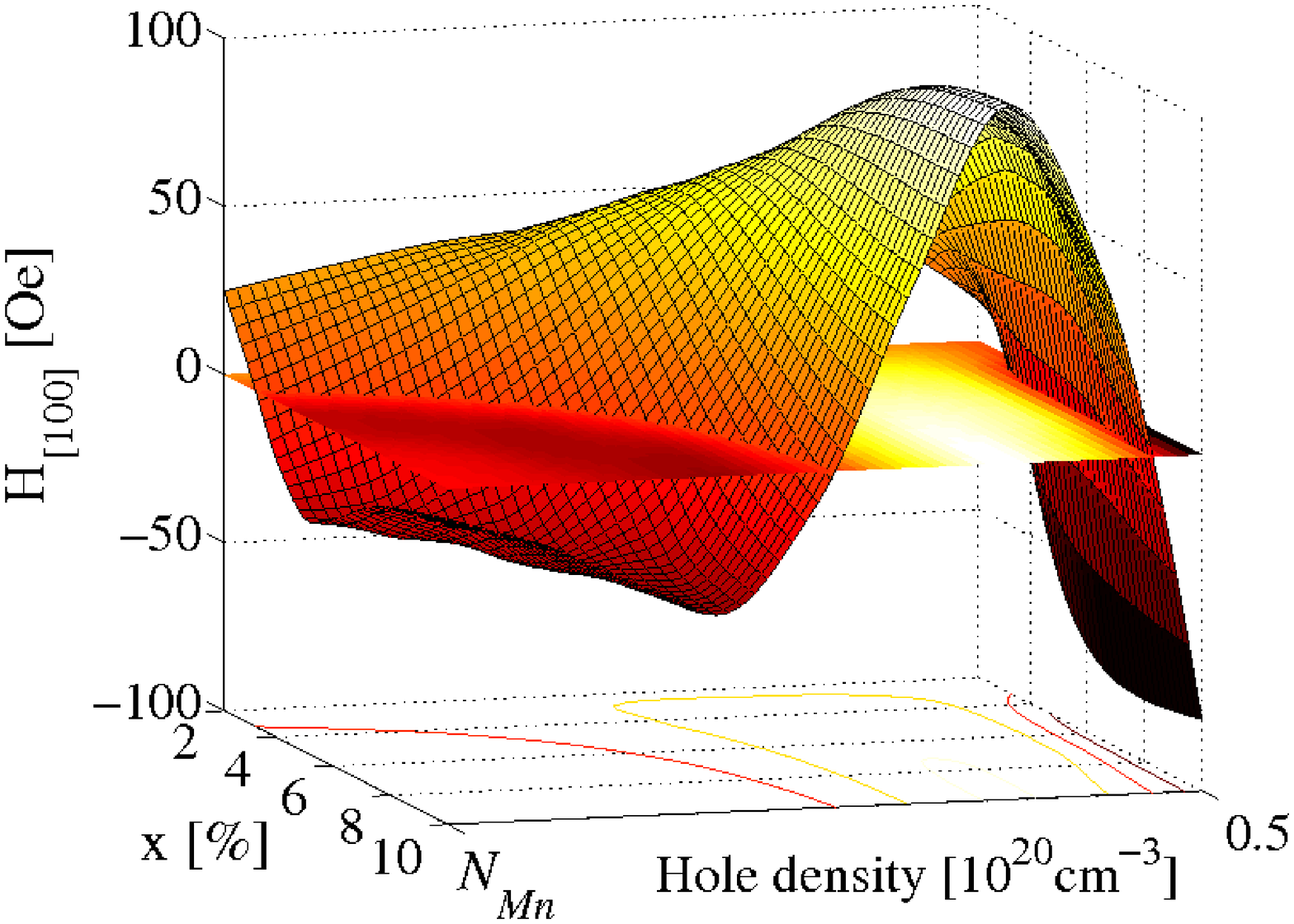}\\ 
%calculated in zemen/PROG/FERMIP/FERMIPSEQ/PRB07/PDEP/PHASEDIAGRAM/MAGANISONEW
%calculated in zemen/PROG/FERMIP/FERMIPSEQ/PRB07/PATTERN/MAGANISODIRXXYY/SHEAR_ALONG_100
%matlab/KCKU/IMA/TZERO/phase_kcku_print.m (ku...K[110])
%matlab/KCKU/PMA/ABOLF_PHASEDIAGRAM/phase_abolf_xxyy_print.m (kxu...K[100])
% data for K[001] not plotted but extracted from zemen/PROG/FERMIP/FERMIPSEQ/PRB07/IMA_PMA_TDEP/WURGAFT/MAGANISONEWXY and compared to K[110] in matlab/KCKU/PMA/ABOLF_PHASEDIAGRAM/abolf_phase.m
\end{tabular}
\caption{(color online) Calculated anisotropy fields $H_{[110]}$ and $H_{[100]}$ as functions of hole density $p$ (up to zero compensation $p=N_{Mn}$) and Mn local moment concentration $x$ at zero temperature and $e_0=-0.2\%$. For $H_{[110]}$ the in-plane strains are $\kappa = 0.01\%$ and $\lambda = 0$ ($e_{xy} = 0.01\%$, $e_{xx}=e_{yy}=e_0$), while $H_{[100]}$ is found for $\kappa = 0$ and $\lambda = 0.01\%$ ($e_{xy}=0$, $e_{xx}=e_0+0.01\%$, $e_{yy}=e_0-0.01\%$).}
\label{gen_kukxu}
\end{figure}

Fig.~\ref{gen_kukxu} shows $H_{[110]}=2K_{[110]}/M$ and $H_{[100]}=2K_{[100]}/M$ as functions of hole density $p$ and Mn local moment concentration $x$ at zero temperature. Both anisotropy fields denpend on material parameters in a qualitatively very similar manner. Moreover, we observe similar dependence on the doping parameters also in case of the field $H_{[001]}$ (not plotted). All the three fields oscillate as functions of hole density. The period of the oscillation is longer than in case of $H_{c1}$. In general, the amplitude of the oscillations decreases with decreasing Mn local moment concentration.

The uniaxial fields are linearly dependent on the strain from which they originate, unless the strains are very large ($>1\%$). For the shear strain of the value $e_{xy} = \kappa \approx 0.01\%$, which is the typical magnitude in our modelling, and zero temperature, the extremal values of $H_{[110]}$ are an order of magnitude smaller than the extremal values of $H_{c1} \sim 10^3$~Oe. For typical compressive growth strain $e_0 \approx -0.2\%$ of an as-grown $5\%$ Mn doped epilayer and zero temperature the extremal values of $H_{[001]}$ are of the same order as $H_{c1}$. When the magnitude of the uniaxial strain along $[100]$ axis is set to $(e_{xx}-e_{yy})/2=e_{xy}$, or equivalently $\kappa=\lambda$, $H_{[100]}$ is approximately a factor of two smaller than $H_{[110]}$.

%remark about IEA vs PEA. intermediate out-of-plane easy axis occurs at low H[001] and positive Kc1 or large kc2... diagrams in section ... are OK.

To quantify the observed similarity in the calculated dependencies of the uniaxial anisotropy coefficients on $x$, $p$, and strains, we can write approximate relationships:
\begin{eqnarray}
K_{[001]}(x,p,e_0) & \simeq & q_{[001]}(x,p) e_0, \nonumber \\
K_{[100]}(x,p,\lambda) & \simeq & q_{[100]}(x,p) \lambda, \nonumber \\
K_{[110]}(x,p,\kappa) & \simeq & q_{[110]}(x,p) \kappa.
\label{axp_e0}
\end{eqnarray}
Note, that each anisotropy component depends only on one type of strain, which is due to the choice of the basis in the strain space (see Eqs.~(\ref{strain1}),~(\ref{strain2}), and~(\ref{strain3})). (Such exclusive dependence of a particular uniaxial anisotropy component on the corresponding strain is, indeed, obtained also from simulations of systems subject to combinations of all three types of strain.) The linearity of anisotropy coefficients as functions of lattice strains is limited to small elastic deformations of the lattice. The approximation cannot be used for strains greater than $1\%$ as revealed also by calculations in Ref.~[\onlinecite{Dietl:2001_b}]. Experiment confirms the linear behaviour in case of the growth strain up to $e_0 \approx \pm 0.3\%$.\cite{Daeubler:2007_a} Linear dependence on in-plane uniaxial strains is corroborated by experiments discussed in Sec.~\ref{se_lithopiezo}.

\begin{figure}
\begin{tabular}{c}
\includegraphics[scale=0.32]{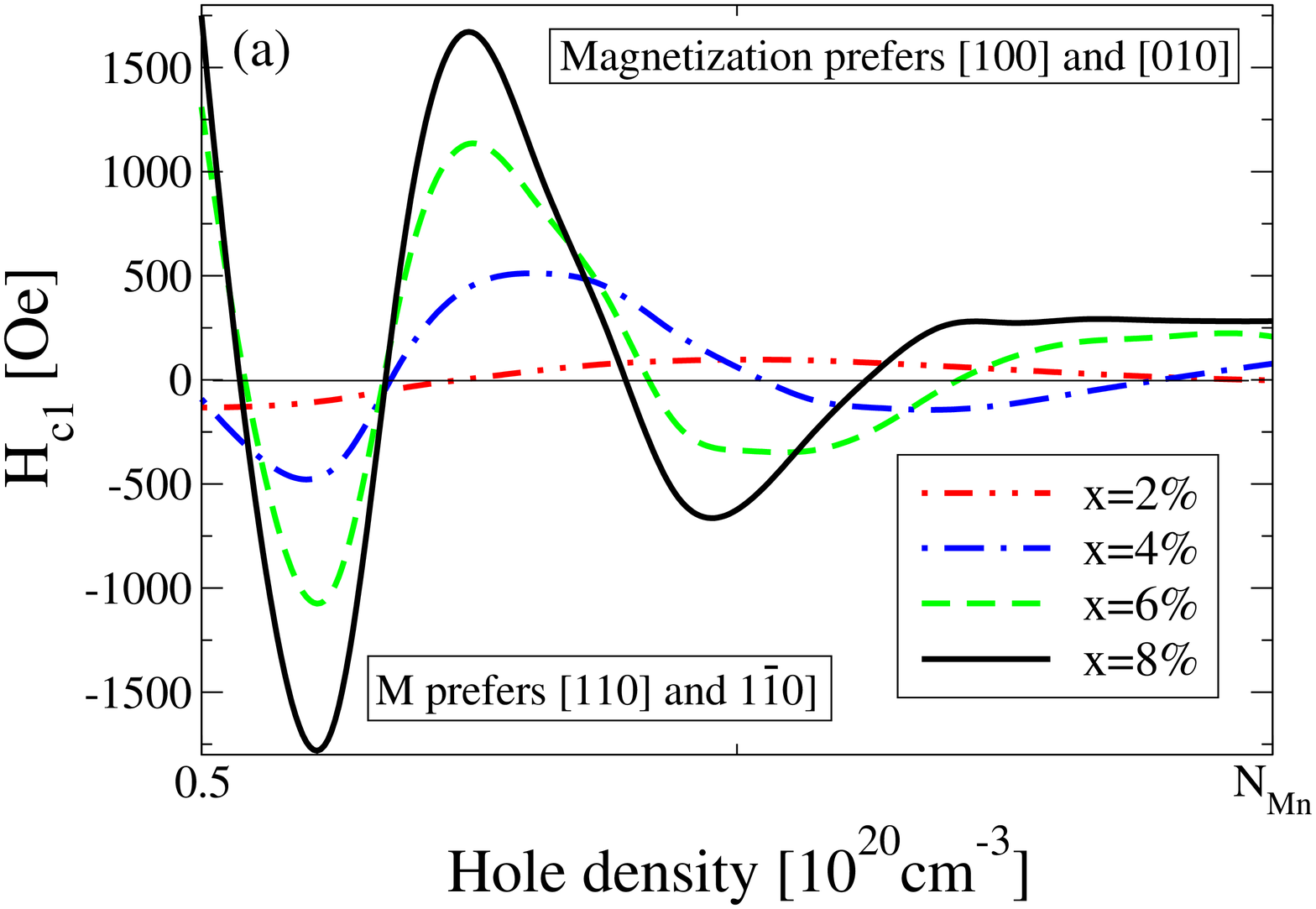} \\ % (a)
 \\
\includegraphics[scale=0.32]{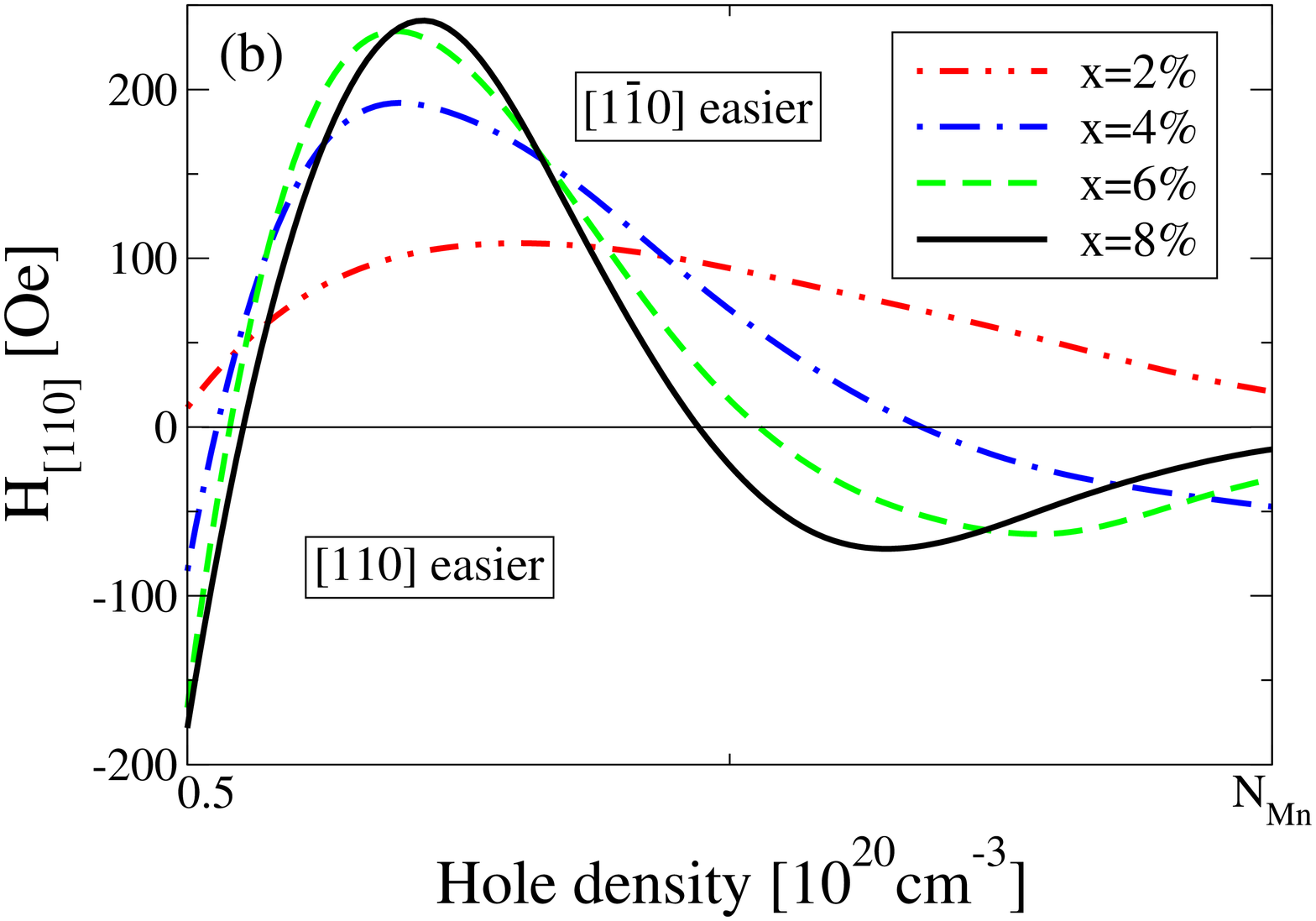} \\ % (b)
 \\
\includegraphics[scale=0.32]{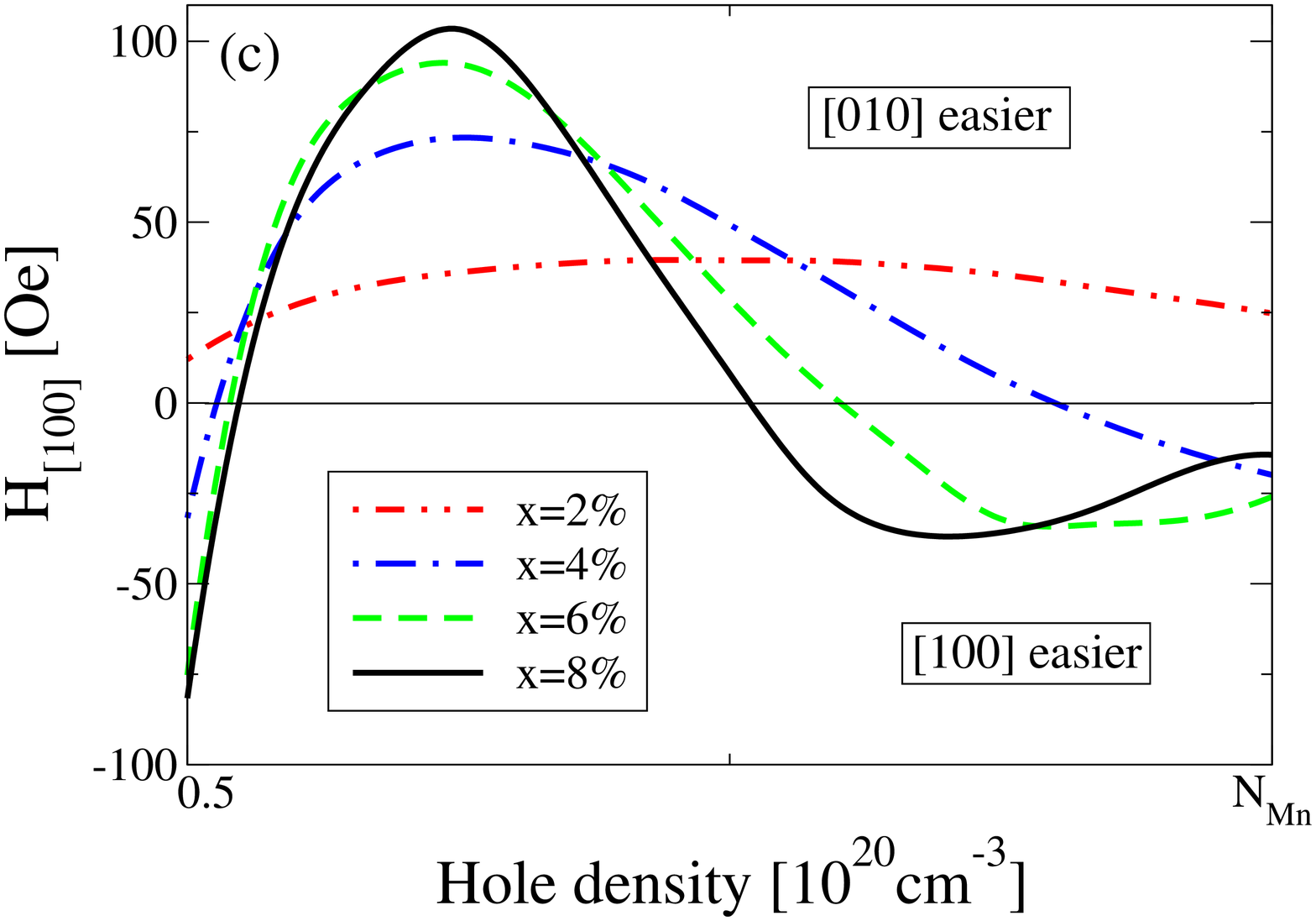} \\
%sources in zemen/WORK/FERMIP/FERMIPSEQ/PRB07/PDEP/KCKU_PHASEDIAGRAM/MAGANISONEW/CUTS_AT_GIVEN_X
\end{tabular}
\caption{Anisotropy fields $H_{c1}$, $H_{[110]}$, and $H_{[100]}$ as function of hole density (up to zero compensation $p=N_{Mn}$) at four Mn local moment concentrations $x$, zero temperature and growth strain $e_0=-0.2\%$. For $H_{[110]}$ the in-plane strains are $\kappa = 0.01\%$ and $\lambda = 0$, while $H_{[100]}$ is found for $\kappa = 0$ and $\lambda = 0.01\%$. (The field $H_{c1}$ is not a function of lattice strains but the same values as for calculation of $H_{[110]}$ were used.) ''Critical`` hole densities, where the anisotropy fields change sign, are dependent on Mn local moment concentration.}
\label{g_pdep_kcku}
\end{figure}

In addition to the linearity with respect to strain, we observe universal dependence of the three uniaxial anisotropy coefficients on hole density and Mn local moment concentration. It can be expressed using the anisotropy functions:
\begin{equation}
q_{[001]}(x,p) \simeq q_{[100]}(x,p) \simeq 0.43q_{[110]}(x,p).
\label{axp_axp}
\end{equation}
The anisotropy function $q_{[110]}(x,p)$ due to shear strain is approximately twice as large as the anisotropy functions $q_{[100]}(x,p)$ and $q_{[001]}(x,p)$. A general property of these functions is that at medium hole densities a relative compression yields a tendency of the easy axis to align with that direction. On the other hand, for very low and high hole densities, the magnetisation prefers alignment parallel to the direction of lattice expansion.

We caution that Eqs.~(\ref{axp_e0}) and (\ref{axp_axp}) are included to promote the general understanding of the anisotropic behaviour of the strained crystal but are not precise. The relative error of the approximation given by Eq.~(\ref{axp_axp}) averaged over the $x-p$ space shown in Fig.~\ref{gen_kukxu} is less than $20\%$, however, the relative error can be much larger at a given combination of $x$ and $p$ where the anisotropy coefficients fall to zero.

To finish the analysis of the theoretical results we include Fig.~\ref{g_pdep_kcku} to improve the legibility of the data. The individual curves correspond to cuts through the 3D plots in Figs.~\ref{gen_kc} and~\ref{gen_kukxu} at fixed Mn local moment concentrations. As already mentioned, the dependence of anisotropy fields on hole density is oscillatory. Note that the critical hole densities, where the sign inversion occurs, shift away from the extremal values, i.e., zero hole density and zero compensation $p=N_{Mn}$, with increasing $x$.

Neglecting the complexity of the dependence of the band structure  on M (whether changed by doping or temperature), one would expect the cubic anisotropy coefficient $K_{c1}$ to be proportional to $M^4$ and uniaxial anisotropy coefficients $K_{[001]}$, $K_{[100]}$, and $K_{[110]}$ to $M^2$. In Fig.~\ref{g_pdep_kcku} we can identify intervals of hole density where any change in Mn concentration, and therefore in M, does not induce a sign change of the anisotropy fields and the functional forms of $K_a(M)$ are roughly consistent with the above expectations. For other hole density intervals, however, the behaviour is highly non-trivial and the function $K_a(M)$ can even change sign.

We now proceed to the discussion of how the theoretically expected phenomenology detailed above is reflected in experiments in bare unpatterned (Ga,Mn)As epilayers. The experimental results\cite{Thevenard:2007_a,Liu:2005_d,Liu:2004_b,Liu:2003_a} are often analysed using the following version of the phenomenological formula:
\begin{eqnarray}
F(\hat{M}) = -2\pi M^2\sin^2 \theta - K_{2\perp}\cos^2 \theta - \frac{1}{2}K_{4\perp} \cos^4 \theta \nonumber \\
- \frac{1}{2} K_{4\parallel} \frac{3+\cos 4 \phi}{4}\sin^4 \theta - K_{2\parallel} \sin^2(\phi-\pi/4) \sin^2 \theta, \;
\label{free_en}
\end{eqnarray}
where angle $\theta$ and $\phi$ are measured, as above, from the $[001]$ and $[100]$ axis, respectively. The first term in Eq.~(\ref{free_en}) corresponds to the shape anisotropy described in Sec.~\ref{su_shapeaniso} and not included in Eq.~(\ref{free_en_us}). The uniaxial anisotropy coefficients $K_{2\perp}$ and $K_{2\parallel}$ correspond to the coefficients $K_{[001]}$ and $K_{[110]}$ in the phenomenological formula Eq.~(\ref{free_en_us}), respectively. To identify the third and fourth term in Eq.~(\ref{free_en}) we rewrite those terms as (see also Eq.~\ref{cubic1}):
\begin{eqnarray}
\label{rewrite_cub}
 & & - \frac{1}{2} K_{4\parallel} \left( \frac{3+\cos 4 \phi}{4}\sin^4 \theta + \cos^4 \theta \right) - \\
 & & - \frac{1}{2} \left( K_{4\perp} - K_{4\parallel} \right) \cos^4 \theta = \nonumber \\
 & = & - \frac{1}{2} K_{4\parallel}\left(n_x^4 + n_y^4 + n_z^4 \right) - \nonumber \\
 & & -\frac{1}{2} \left( K_{4\perp} - K_{4\parallel} \right) n_z^4 = \nonumber \\
 & \equiv & K_{c1}\left(n_x^2 n_y^2 + n_x^2 n_z^2 + n_z^2 n_y^2 \right) - \nonumber \\
 & & -\frac{1}{2} K_{[001]_2} n_z^4 + c, \nonumber \;
\end{eqnarray}
where $c$ is an angle independent constant. From here we see that the coefficient $K_{4\parallel}$ corresponds to the lowest order cubic coefficient $K_{c1}$ in Eq.~(\ref{free_en_us}) and $K_{4\perp} - K_{4\parallel} \equiv K_{[001]_2}$ corresponds to the second order uniaxial anisotropy coefficient $K_{u2}$ for $\hat{U} \parallel [001]$.
We point out that omission of the second order cubic term (and other higher order terms) can make the determination of $K_{[001]_2}$ from fitting the data to the phenomenological formula in Eq.~(\ref{free_en}) unreliable. Moreover, the accurate extraction of the coefficient $K_{[001]_2}$ can be difficult in samples with large value of the first order coefficient $K_{[001]}$.\cite{Liu:2003_a} We therefore only note that $K_{[001]_2}$ extracted from the experiment\cite{Thevenard:2007_a,Liu:2005_d,Liu:2004_b} never dominates the anisotropy, consistent with our calculations, and do not discuss the coefficient further in more detail.

The predicted strong dependence of $K_{[001]}$, $K_{[110]}$, and $K_{c1}$ on hole density, Mn local moment concentration and temperature is consistently observed in many experimental papers. %\cite{Daeubler:2007_a,Thevenard:2007_a,Liu:2003_a,Liu:2005_d,Liu:2004_b,Owen:2008_a,Ohno:2008_a, Sugawara:2008_a,Wang:2005_e} %there should be Gould and Pappert too...
We start with experiments where the out-of-plane anisotropy is studied. Measurements focusing mainly on the in-plane anisotropies are discussed at the end of this section and in Sec.~\ref{se_lithopiezo} for patterned or piezo-strained samples.

The coefficient $K_{[001]}$ is extracted in Ref.~[\onlinecite{Daeubler:2007_a}] using detailed angle-resolved magnetotransport measurements at 4~K for different growth strains in as-grown and annealed, 180~nm thick samples with identical nominal Mn concentration $x \approx 5\%$. The growth strain ranging from $e_0=-0.22\%$ (compressive) to $e_0=0.34\%$ (tensile) is achieved by MBE growth of (Ga,Mn)As on (In,Ga)As/GaAs templates. The observed linear dependence of $K_{[001]}$ on $e_0$ agrees on the large range of $e_0$ with the prediction given in Eq.~(\ref{axp_e0}). The calculated and measured gradients are of the same order of magnitude and sign, and depend on the hole density. The off-set at zero strain in the measured dependence of $K_{[001]}$ on $e_0$ in Ref.~[\onlinecite{Daeubler:2007_a}] is due to the shape anisotropy.

Ref.~[\onlinecite{Thevenard:2007_a}] presents 50~nm thick, annealed samples with nominal Mn doping $x=7\%$. All the samples are first passivated by hydrogen and then depassivated for different times to achieve different hole densities while keeping the growth strain the same. The FMR spectroscopy is carried out for in-plane and out-of-plane configurations. There is qualitative agreement of calculation and measurement on the level of the directions of the easy axes as discussed in the previous subsection. The sign change of the uniaxial anisotropy fields driven by increase of temperature is observed. The measured coefficients $K_{[001]}$ and $K_{c1}$  are of the same order of magnitude as the calculated ones and $K_{[001]} \approx K_{c1}$ is consistent with the weaker growth strain in annealed samples.

Ref.~[\onlinecite{Liu:2004_b}] presents an as-grown, 6~nm thick film nominally doped with Mn to $x=6\%$, grown on Ga$_{0.76}$Al$_{0.24}$As barrier doped with Be. Increasing the Be doping increases the hole density without changing the Mn local moment concentration. The fitting of the FMR spectra is done using the coefficients $K_{[001]}$ and $K_{c1}$ and the g-factor of the Mn. The anisotropy field corresponding to the coefficient $K_{[001]}$ reaches value as high as $\approx 6000$~Oe at 4~K. Large values of $K_{[001]}$ is consistent with expected large growth strain in a thin as-grown sample.\cite{Masek:2003_a,Masek:2005_a} However, for the measured $K_{[001]}$ our calculations would imply strain $e_0 \sim 1\%$ which is an order of magnitude larger than typical strains in as-grown $x=6\%$ (Ga,Mn)As materials. Other effects are therefore likely to contribute to $K_{[001]}$ in this sample. (Confinement effect or inhomogeneities are among the likely candidates.) The experimental $K_{[001]}$ ($K_{c1}$) increases (decreases) with increasing hole density which is in agreement with our modelling of highly compensated samples.

Observation of qualitatively consistent behaviour of the anisotropies with the theory but unexpectedly large magnitudes of the anisotropy fields applies also to thick samples studied by FMR in Refs.~[\onlinecite{Liu:2005_d,Liu:2003_a}]. Temperature dependence of the anisotropy fields is studied by FMR in Ref.~[\onlinecite{Liu:2005_d}] for a low doped ($x \approx 2\%$), as-grown, 200~nm thick (Ga,Mn)As film. Only the combined contribution of shape anisotropy and $K_{[001]}$ was resolved. The easy axis stays in-plane for all studied temperatures which is consistent with predicted crystalline anisotropy as well as the shape anisotropy dominating at weak growth strains. The uniaxial in-plane anisotropy is of the predicted magnitude but its sign corresponds to modelling by the less frequent negative intrinsic shear strain.

Ref.~[\onlinecite{Liu:2003_a}] discussed in Sec.~\ref{su_pmaima} on the level of easy axis orientation shows, among other samples, 300~nm thick annealed epilayers with nominal Mn concentration $x=3\%$ deposited on GaAs and (Ga,In)As substrate under compressive and tensile growth strain, respectively. The strain is measured by x-ray diffraction, however, the predicted linear dependence of $K_{[001]}$ on the growth strain (Eq.~(\ref{axp_e0})) cannot be tested due to different saturation magnetisation and $T_C$ in both samples. Both Refs.~[\onlinecite{Liu:2003_a,Liu:2005_d}] report the coefficient $K_{c1}$ in the 300~nm and 200~nm thick samples an order of magnitude larger than the calculated one which can\cite{Wang:2005_h} be attributed to sample inhomogeneities in these thick epilayers. Ref.~[\onlinecite{Liu:2003_a}] studies also 120~nm thick, annealed and as-grown epilayers with $x=8\%$ deposited on GaAs. The coefficient $K_{[001]}$ doubles its value at low temperature on annealing. Both $K_{[001]}$ and $K_{c1}$ in the thinner samples have values of the order predicted by theory for material with Mn doping $x=8\%$.
%, however, this doping in a fully annealed sample would correspond to higher critical temperature than the measured $T_C=110$~K.

\begin{figure}[h]
\includegraphics[scale=0.38]{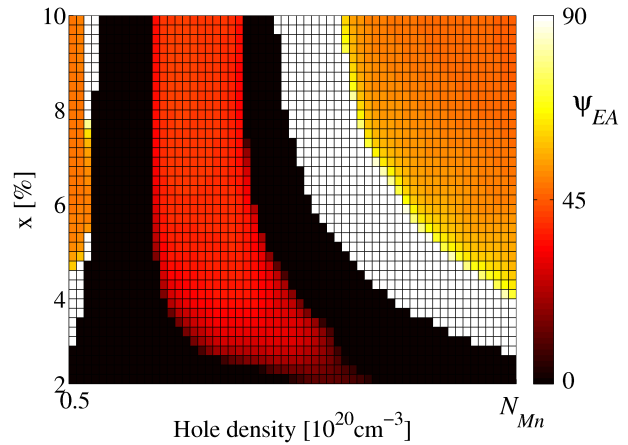} 
%matlab/KCKU/PMA/ABOLF_PHASEDIAGRAM a matlab/KCKU/IMA/TZERO/EA_DIRECTION
\caption{(Color online) Angle $\psi$ of the easy axis with respect to the $[1\overline{1}0]$ axis as function of hole density $p$ (up to zero compensation $p=N_{Mn}$) and Mn local moment concentration $x$ at zero temperature, $e_0 = -0.2\%$, and $\kappa = 0.01\%$;}
\label{gen_phi}
\end{figure}

%%%%%%%%%%%%%experiment in-plane
Now we analyse experiments focusing on the in-plane anisotropy where the relevant anisotropy coefficients are $K_{c1}$ and $K_{[110]}$. Note that the experimental papers discussed below mostly\cite{Wang:2005_e,olejnik:unpub_b,Sugawara:2008_a, Owen:2008_a,Rushforth:2008_a} use the notation with the in-plane magnetisation angle $\psi$ measured from the $[1\overline{1}0]$ axis. To avoid any confusion we write the in-plane form of Eq.~(\ref{free_en_us}) using the original anisotropy coefficients and the angle $\psi=\phi+\pi/4$:
\begin{equation}
F(\hat{M})=-\frac{K_{c1}}{4} \sin^2 2\psi+K_{[110]} \sin^2\psi. 
\label{free_en_inplane}
\end{equation}
%We caution that two experimental studies\cite{Bihler:2008_a,Overby:2008_a} in Sec.~\ref{se_lithopiezo} use the opposite sign convention for the coefficient $K_{[110]}$.
To facilitate the comparison with experiment we use the notation of Eq.~(\ref{free_en_inplane}) consistently in the remaining parts of this paper.

The magnetic easy axes lie closer to the $[100]$ or $[010]$ direction than to any diagonal when $K_{c1} > 0$ and $\sqrt{2}K_{[110]} < K_{c1}$. Negative $K_{c1}$ always leads to diagonal easy axes. We include Fig.~\ref{gen_phi} to elucidate the combined effect of $K_{c1}$ and $K_{[110]}$ on the in-plane direction of the easy axes. The angle $\psi_{EA}(x,p)$, plotted as a function of Mn concentration and hole density at zero temperature minimises the free energy $F(\hat{M})$. The local minima at $\psi = 0^{\circ}$ (black) and $\psi = 90^{\circ}$ (white) are formed for negative $K_{c1}$. When $K_{[110]}$ is positive (negative), the global minimum is at $\psi = 0^{\circ}$ ($\psi = 90^{\circ}$). The higher energy local minimum disappears for $|K_{c1}|=|K_{[110]}|$. Only one energy minimum forms for $|K_{c1}|<|K_{[110]}|$ and for positive (negative) $K_{[110]}$ the easy axis is at $\psi = 0^{\circ}$ ($\psi = 90^{\circ}$). The interface of black and white regions is an evidence of a discontinuity of the function $\psi_{EA}(x,p)$ due to switching of the sign of $K_{[110]}$ when $K_{c1}<0$. The grey (coloured online) regions in Fig.~\ref{gen_phi} correspond to competition of cubic and uniaxial anisotropy when $K_{c1}>0$ and $|K_{c1}|>|K_{[110]}|$. Then there are two easy axes at $\psi_{EA}$ and $180^{\circ}-\psi_{EA}$ forming ``scissors'' closing at the $[1\overline{1}0]$ axis. (The darker the colour, the more closed the scissors.)

\begin{figure}[h]
\begin{center}
%\SetFigLayout{1}{2}
\begin{tabular}{c}
	\includegraphics[scale=0.34]{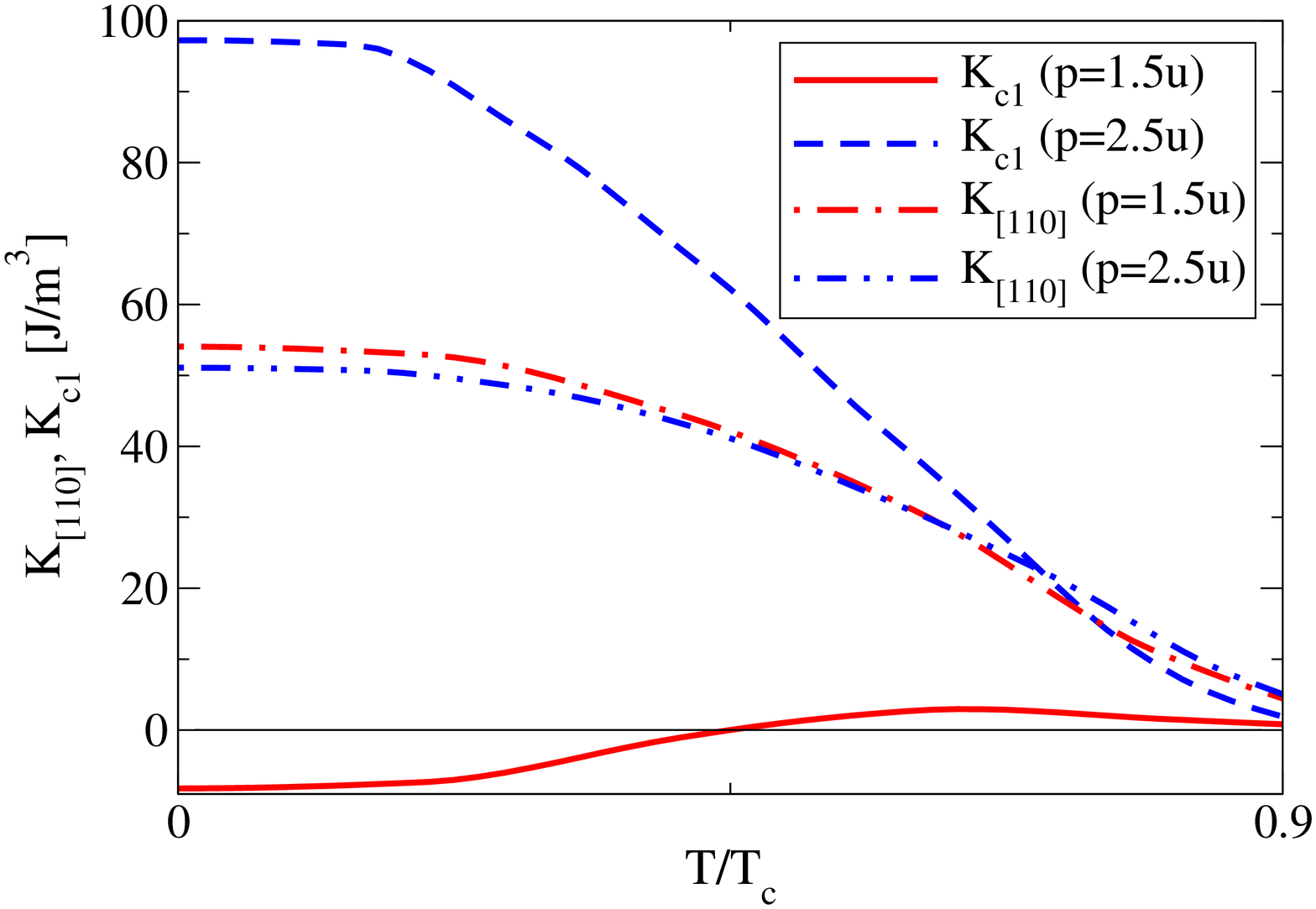} \\
        \includegraphics[scale=0.34]{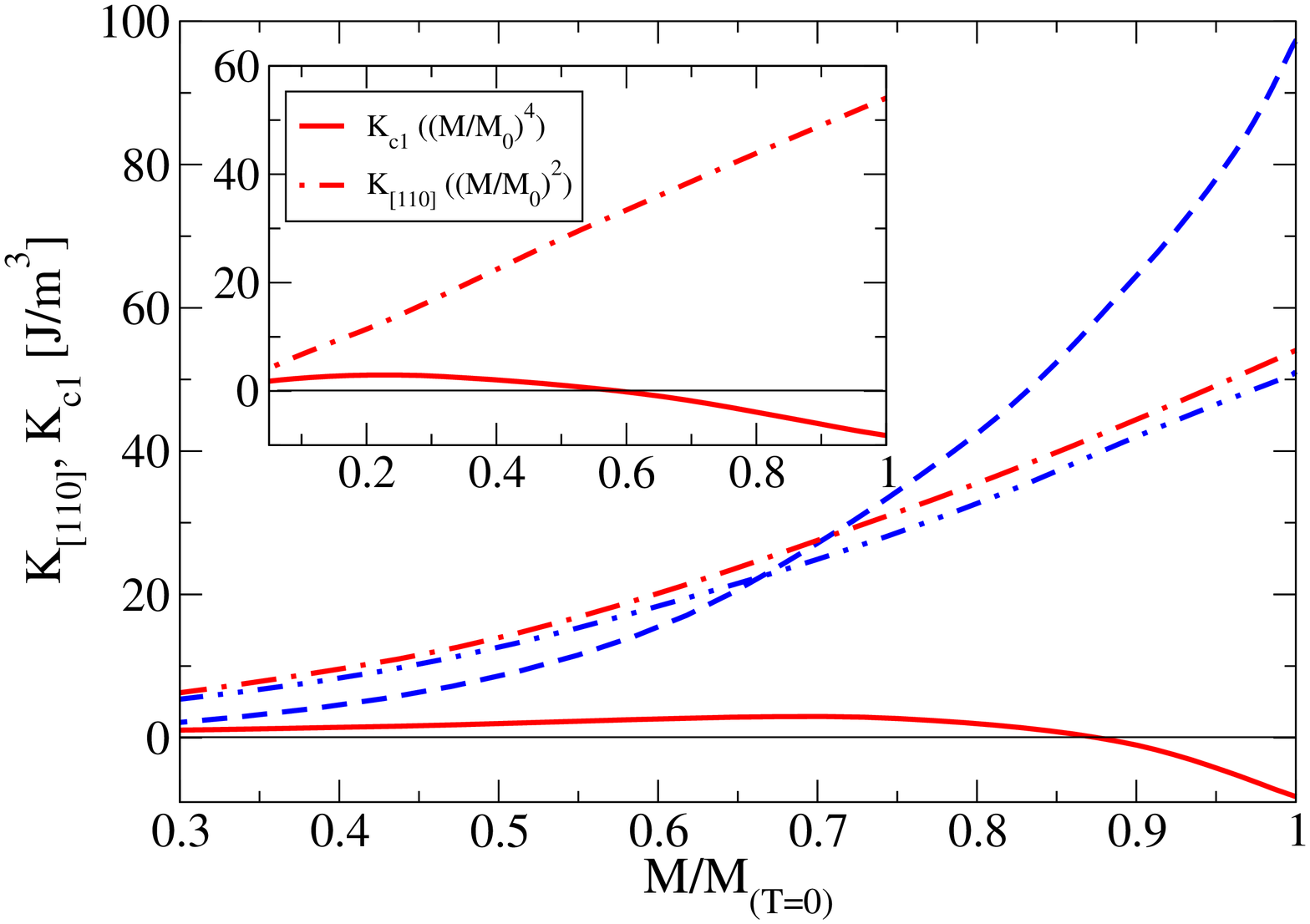} \\
%sources in /home/zemen/WORK/FERMIP/FERMIPSEQ/PRB07/TDEP/EXY01/MAGANISONEWXXYY/KCKU/IMA
\end{tabular}
\end{center}
\caption{Calculated anisotropy fields $K_{c1}$ and $K_{[110]}$ as function of temperature and magnetisation at two hole densities (given in units u~$\equiv 10^{20}$~cm$^{-3}$), Mn concentration $x=2\%$, strains $e_0 = -0.2\%$, $\kappa = 0.005\%$. Irregular behaviour is observed for the lower hole density.}
\label{a_fields_x2}
\end{figure}

To demonstrate the typical scaling of in-plane anisotropy components with temperature, we discuss the 50~nm thick as-grown (Ga,Mn)As epilayer with Mn concentration $x=2.2\%$ determined by x-ray diffraction and secondary ion mass spectrometry, presented in Ref.~[\onlinecite{Wang:2005_e}]. The anisotropy coefficients $K_{[110]}$ and $K_{c1}$ are obtained by fitting to the $M(H)$ loop with magnetic field along the hard direction. They can be compared to Fig.~\ref{a_fields_x2} which shows the calculated anisotropy fields as functions of temperature for two values from the interval of hole densities corresponding to the as-grown sample. For $p = 2.5\times 10^{20}$~cm$^{-3}$ both the calculated and measured $K_{c1}$ is greater than $K_{[110]}$ at low temperatures but becomes smaller than $K_{[110]}$ at $T \approx T_C$. The calculated $K_{c1}$ is an order of magnitude smaller than the experimental one, however, there is agreement on the level of the temperature dependent ratio of $K_{c1}$ and $K_{[110]}$. On the contrary, Fig.~\ref{a_fields_x2} shows a non-monotonous dependence of $K_{c1}$ on temperature for $p=1.5 \times 10^{20}$ cm$^{-3}$. This singular behaviour is not measured in Ref.~[\onlinecite{Wang:2005_e}] but it is reported in a more systematic study in Ref.~[\onlinecite{Thevenard:2007_a}].

The temperature dependence of anisotropy coefficients $K_{[110]}$ and $K_{c1}$ is studied by planar Hall effect in Ref.~[\onlinecite{Shin:2007_a}]. The mutual behaviour of the two coefficients observed in the as-grown (Ga,Mn)As epilayer with nominal Mn concentration $x\approx 4\%$ and $T_C=62$~K is qualitatively the same as in Ref.~[\onlinecite{Wang:2005_e}]. $K_{c1}$ becomes smaller than $K_{[110]}$ at $T=26$~K which is in agreement with our modelling. No sign change of $K_{c1}$ is reported in this experimental work. Again, the calculated $K_{c1}$ is an order of magnitude smaller than the experimental one.

Ref.~[\onlinecite{Thevenard:2007_a}] resolves the in-plane coefficients $K_{c1}$ and $K_{[110]}$ in four samples with nominal Mn doping $x=7\%$ and different hole densities. In samples with lower hole densities the dependence of $K_{c1}$ and $K_{[110]}$ is qualitatively consistent with Ref.~[\onlinecite{Wang:2005_e}], however, both coefficients change sign when temperature is increased in samples with higher hole densities ($p \sim 10^{21}$~cm$^{-3}$, $T_C=130$~K). Our model predicts such sign change for a short interval of high hole compensations and a larger interval of low hole compensations as shown in Fig.~\ref{g_pdep_kcku}(a).

Another type of temperature scaling of $K_{c1}$ and $K_{[110]}$ is observed in a 50~nm thick, annealed sample with nominal Mn doping $x=7\%$ and $T_C=165$~K.\cite{olejnik:unpub_b} $K_{[110]}$ is larger than $K_{c1}$ on the whole temperature interval ($T=4-165$~K). Both coefficients are positive, decrease on increasing temperature, and their magnitudes are of the same order of magnitude as the calculated anisotropies. The stability of sign of $K_{[110]}$ is observed theoretically for higher ``intrinsic'' shear strain as discussed in Fig.~\ref{tdep_strains_b} in Sec.~\ref{su_ima}.

The temperature dependence of domain wall properties of a 500~nm, as-grown (Ga,Mn)As film with Mn doping $x=4\%$ is studied by means of the electron holography in Ref.~[\onlinecite{Sugawara:2008_a}]. The width and angle of the domain walls were determined directly from the high-resolution images. The ratio of the anisotropy coefficients $K_{[110]}/K_{c1}$ was extracted from these observations combined with Landau-Lifshitz-Gilbert simulations. The N\`{e}el type domain walls evolve from near-90$^{\circ}$-walls at low temperatures ($T=10$~K) to large angle $[1\overline{1}0]$-oriented walls and small angle $[110]$-oriented walls at higher temperatures ($T=30$~K). The angles of domain walls aligned with particular crystallographic directions reveal positions of the magnetic easy axes. The ``scissors'' of the easy axes (described in discussion of Fig.~\ref{gen_phi}) are closing around the $[1\overline{1}0]$ axis on increasing temperature consistent with our modelling.
%Calculations were carried out for a range of relevant hole densities ($p=3-4\times 10^{20}$cm$^3$) and Mn local moment concentrations ($x=3-4.5\%$) giving the cubic anisotropy coefficient $K_{c1}$ at $T=0$~K consistent with the $K_{c1}$ determined at $T=10$~K from SQUID measurement. The in-plane shear strain was set to recover the experimental ration $K_{[110]}/K_{c1}$ at low temperature and this value ($\kappa \sim 0.001\%-0.01\%$ for the range of $x$ and $p$) was kept when modelling the uniaxial anisotropy at all higher temperatures.

The domain-wall width is inversely proportional to the effective anisotropy energy barrier between the bistable states on respective sides of the domain wall: $K^{eff}_{[110]}\equiv K_{c1}/4-K_{[110]}/2$ ($[110]$-oriented walls) and $K^{eff}_{[1\overline{1}0]}\equiv K_{c1}/4+K_{[110]}/2$ ($[1\overline{1}0]$-oriented walls).
The width of the $[1\overline{1}0]$-oriented wall in Ref.~[\onlinecite{Sugawara:2008_a}] initially increases with temperature and then saturates at high temperature while the $[110]$-oriented wall width keeps increasing with temperature until it becomes unresolvable. This observation corresponds well to the theoretical prediction and can be qualitatively understood by considering the approximate magnetisation scaling of $K_{c1}\sim M^4$, $K_{[110]}\sim M^2$, and magnetic stiffness~$\sim M^2$. 
%Note that the sample substrate had to be thinned significantly to allow for the electron diffraction, releasing at the same time the original growth strain. The calculations confirm the negligible effect of the typical growth strain in as-grown material on the in-plane anisotropies.

Finally, Refs.~[\onlinecite{Owen:2008_a, Ohno:2008_a}] present (Ga,Mn)As field-effect transistors (FETs), where hole depletion/accumulation is achieved by gating induced changes of the in-plane easy axis alignment. In Ref.~[\onlinecite{Owen:2008_a}] the Mn doped layer is 5~nm thick with Mn doping $x=2.5\%$ and hole density $p \sim 1\times 10^{19}-10^{20}$cm$^{-3}$. The direction of magnetic easy axes was detected by AMR at $T=4$K. The $20\%$ variation of the hole density achieved by applying the gate voltage from $-1$V to $3$V is determined from variation of the channel resistance near $T_C$. This value was a starting point for simulations of the depletion at $T=4$~K giving hole density changes $\Delta p \approx 5\times 10^{19}$cm$^{-3}$. The measured $K_{c1}$ is negative and its magnitude decreases with depletion.
%, whereas the measured $K_{[110]}$ is roughly a factor of 10 smaller than $K_{c1}$ and its magnitude increases with depletion. $K_{c1}$ was calculated for $x=2-3\%$ and $p=2\times 10^{19}-2\times 10^{20}$cm$^{-3}$. Its value is negative on this parameter interval and becomes positive for higher or lower hole densities.
The theoretical magnitude ($\sim 10$~mT) and sign of $K_{c1}$ for the relevant hole density range, as well as the variation of $K_{c1}$ with varying hole density, are consistent with the experiment. Recall that negative $K_{c1}$ corresponds to diagonal easy axes captured by two black/white regions in Fig.~\ref{gen_phi}. Samples reported in Refs.~[\onlinecite{wang:unpub_a,Stanciu:2005_a,Hamaya:2006_a,olejnik:unpub_a}] (see also Sec.~\ref{su_ima}) and in Ref.~[\onlinecite{olejnik:unpub_b}] with diagonal easy axes at low temperatures fall into the right region with lower hole compensations, whereas the sample in Ref.~[\onlinecite{Owen:2008_a}] is a rarely observed example of diagonal easy axes at high compensation and low temperature corresponding to the left black/white region in Fig.~\ref{gen_phi}.

%\section{Bandstructure symmetry considerations}
%\subsection{Universal critical hole density}
%\subsection{Sign-flip of growth strain}

%%%%%%%%%%%%%%%%%%%%%%%%%%%%%%%%%%%%%%%%%%%%
\section{Samples with post-growth controlled strains}
\label{se_lithopiezo}

In the previous section, we discussed three types of lattice strain and calculated corresponding types of uniaxial anisotropy components. In the bare, unpatterned epilayers we could analyse and compare to experiment only anisotropies induced by the growth strain and by the unknown symmetry breaking mechanism modelled by the ``intrinsic'' shear strain. The calculations including the model shear strain allow us also to estimate the magnitude of real in-plane lattice strains, controlled post-growth by patterning or piezo stressing, that can induce sizable changes of anisotropy. In this section we investigate samples where these post-growth techniques are used to apply additional stress along any in-plane direction. We will focus primarily on stresses along the main crystal axes and in-plane diagonals. We will also comment on the procedure for determining the lattice strain from specific geometrical parameters of the experimental setup. Where necessary, we distinguish the externally induced strain and the ``intrinsic'' shear strain, which models the in-plane symmetry breaking mechanism already present in the bare epilayers. Returning to the notation of Sec.~\ref{se_theory} we denote the latter strain by the symbol $e^{int}_{xy}$.
For better physical insight and to relate with discussion in previous section we will map the anisotropies on the phenomenological formulae by decomposing the total strain matrix into the three basis strains (Eqs.~(\ref{strain1}-\ref{strain3})). We will then write the corresponding anisotropy energy terms as in Sec.~\ref{su_afields}, assuming linearity between the respective basis strains and anisotropy energy components (see Eq.~(\ref{axp_e0})). Experiments will be discussed based on microscopic anisotropy calculations with the total strain tensor directly included into the Hamiltonian.

%We estimate the magnitude of the shear strain to be within the interval $0.005\% < e_{xy} < 0.05\%$. Comparison of calculation to experiments which distinguish the two diagonals suggests positive sign of the shear strain (assuming expansion along $[110]$ axis). The magnetic easy direction prefers either diagonal depending mainly on the hole density and Mn concentration as shown by Fig.~\ref{polar}(c,d).

%The diagonal alignment which we use in our model to account for the uniaxial anisotropy of bulk bare samples is common in the fabricated structures. Therefore, calculated data of Fig.~\ref{polar},~\ref{tdep_main} and \ref{tdep_strains} are relevant to this subsection too.

We begin this section by discussion of the in-plane uniaxial strain induced by post-growth lithography treatment of Mn-doped epilayers grown under compressive lattice strain. Narrow bars with their width comparable to the epilayer thickness allow for anisotropic relaxation of the lattice matching strain present in the unpatterned film. An expansion of the crystal lattice along the direction perpendicular to the bar occurs while the epilayer lattice constant along the bar remains unchanged. Parameters sufficient for determination of the induced strain are the initial growth strain $e_0$ and the thickness to width ratio $t/w$ of the bar. In the regime of small deformations the components of the induced strain are linearly proportional to the growth strain. The strain tensor for a bar oriented along the $[100]$ axis reads:
\begin{eqnarray}
\label{strain_relax_comsol} 
{\bf e}^{r}_{[100]} & = & e_0\left(\begin{array}{ccc} 
-\rho + 1 & 0 & 0 \\
0 & 1 & 0 \\ 
0 & 0 & \frac{c_{12}}{c_{11}}\left(\rho - 2\right) \\
\end{array}\right), \;
\end{eqnarray}
where the lattice relaxation is quantified by $\rho$ which is a function of $t/w$ and can vary over the bar cross-section. We calculate the distribution of $\rho$ over the cross-section of the bar using Structural Mechanics Module of Comsol (standard finite element partial differential equation solver, www.comsol.com). Since the macroscopic simulations ignore the microscopic crystal structure, they apply to bars oriented along any crystallographic direction. We therefore introduce a coordinate system fixed to the bar: $x'$-axis lies along the relaxation direction transverse to the bar, $y'$-axis along the bar, and $z'$-axis along the growth direction. We approximate the bar by an infinite rectangular prism with translational symmetry along the $y'$-axis, attached to a thick substrate.

\begin{figure}
\includegraphics[scale=0.75]{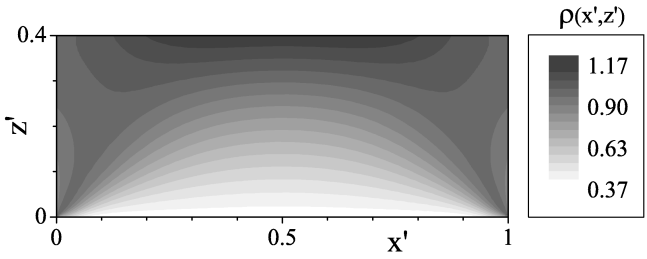}%
\caption{Spatial dependence of the strain coefficient $\rho$ due to lattice relaxation in a narrow bar with $t/w = 0.4$ and compressive growth strain $e_0 < 0$, simulated values of $\rho$ are plotted for the cross-section of the bar.}
\label{e_profile}
\end{figure}

\begin{figure}
\includegraphics[scale=0.34]{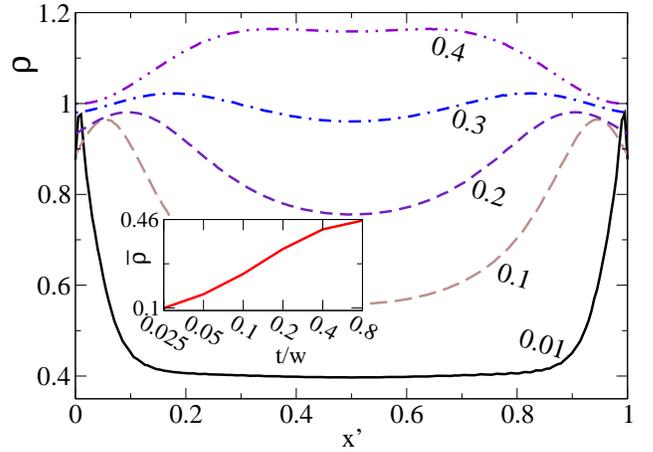}
\caption{(Color online) Sections of $\rho(x',z')$ in Fig.~\ref{e_profile} at fixed values of $z'$ (given next to the curves in relative units) of a thin bar. Inset shows the average strain $\overline{\rho}(t/w)$ as a function of the thickness to width ratio.}
\label{e_cuts}
\end{figure}

Fig.~\ref{e_profile} shows the spatial dependence of the function $\rho(x',z')$ for a given thickness to width ratio and compressive growth strain $e_0 < 0$. Only the area of the bar is plotted, whereas the strain induced in the patterned part of the substrate is not shown. (The substrate relaxation is not directly related to the microscopic simulation of the anisotropy energy). In wide bars ($t/w \ll 1$) the relaxation is very non-uniform, whereas narrow bars ($t/w \gg 1$) are fully relaxed. Fig.~\ref{e_cuts} shows still a fairly non-uniform relaxation for $t/w = 0.4$ with large relaxation at the edges. We point out in this case that the resulting anisotropy can be very sensitive to the details of the etching (vertical under-cut/over-cut profile).

The non-uniform strain distribution in wider bars can in principle force the system to break into magnetically distinct regions. However, experiments show rather that the whole bars behave as one effective magnetic medium. Because of the linearity between the strain and the anisotropy (see Eq.~(\ref{axp_e0})) we can model the mean magnetic anisotropy by considering the spatial average of ${\bf e}^{r}_{[100]}$ over the bar cross-section. The inset of Fig.~\ref{e_cuts} shows the averaged value $\overline{\rho}$ as a function of the width to thickness ratio. It confirms that the effect of relaxation can reach magnitudes necessary to generate significant changes in the magnetic anisotropy. In very narrow bars the induced uniaxial anisotropy can override the intrinsic anisotropies of the unpatterned epilayer and determine the direction of the easy axis.

If the bar is aligned with the $[100]$ or $[010]$ crystal axis, the strain ${\bf e}^{r}_{[100]}$ in Eq.~(\ref{strain_relax_comsol}) with the average relaxation magnitude $\overline{\rho}$ can be used directly as input parameter of the microscopic calculation (see Eq.~(\ref{strain_lutpar}) in the Appendix). Alternatively, the total strain tensor can be decomposed into the growth basis strain from Eq.~(\ref{strain1}) and the uniaxial basis strain introduced in Eq.~(\ref{strain3}):
\begin{eqnarray}
\label{strain_relax_decomp}
{\bf e}^{r}_{[100]}(e_0,\overline{\rho}) & = & {\bf e}^{g}(\tilde{e}_0) + {\bf e}^{u}(\tilde{\lambda}), \\
\tilde{e}_0 & = & e_0\left(1-\frac{\overline{\rho}}{2}\right), \\
\tilde{\lambda} & = & -e_0\frac{\overline{\rho}}{2}. \;
\end{eqnarray}
Their effects on the magnetic anisotropy can be considered separately utilising the results shown in Sec.~\ref{su_afields}.

Now we discuss the introduction of uniaxial in-plane anisotropies by a piezo actuator attached to the sample. In this case, the (Ga,Mn)As film is assumed to follow the deformation of the stressor. (The substrate is usually thinned to achieve better transmission of the piezo-strain to the studied epilayer. Macroscopic Comsol simulations predict transmission of approximately $70\%$ of the piezo-strain in a substrate with thickness to lateral size ratio $t/l \approx 0.1$ and transmission of approximately $90\%$ of the piezo-strain for $t/l \approx 0.02$.) The net effect of the piezo-stressing on normal GaAs epilayers has been investigated experimentally for example in Ref.~[\onlinecite{Shayegan:2003_a}] for a standard PbZrTiO$_3$ (PZT) piezo actuator. The induced strain can reach magnitudes $\sim 10^{-4}$ at low temperatures, which are sufficient to induce observable anisotropies in (Ga,Mn)As, as shown in Sec.~\ref{su_afields}. The deformation is linearly proportional to applied voltage on the transducer and increases with increasing temperature.

The dependence of uniaxial anisotropies due to additional piezo-strains is analogous to the behaviour of relaxed microbars, however, the form of the strain tensor induced by the stressor is typically more complex. Let us first assume a strain tensor with components in the Cartesian coordinate system fixed to the orientation of the piezo stressor: $x'$-axis lies along the principal elongation direction, $z'$-axis is perpendicular to plane of the thin film. We denote the deformation along the $x'$-axis by $\sigma$ and the simultaneous deformation along the $y'$-axis by $\sigma'$. Note that shear strains are typically not considered when describing the action of a piezo-stressor. The third parameter describing the strained (Ga,Mn)As epilayer is the growth strain $e_0$. Our analysis takes into account only structures that can be parametrised by these three values. The strain tensor in the dashed coordinate system reads:
\begin{eqnarray}
\label{strain_piezo_total} 
{\bf e}^{p}_{[100]} & = & \left(\begin{array}{ccc} 
\sigma + e_0 & 0 & 0 \\
0 & \sigma' + e_0 & 0 \\ 
0 & 0 & -\frac{c_{12}}{c_{11}}(2e_0+\sigma+\sigma') \\
\end{array}\right) \;
\end{eqnarray}
Components of this tensor are considered uniform in the studied epilayer. If the principal elongation direction of the piezo stressor is aligned with the $[100]$ crystallographic axis the strain tensor ${\bf e}^{p}_{[100]}$ can be used directly as an input of the microscopic simulation. Similarly to the strain induced by lattice relaxation, ${\bf e}^{p}_{[100]}$ can be decomposed into the growth basis strain and the uniaxial basis strain:
\begin{eqnarray}
\label{strain_piezo_decomp}
{\bf e}^{p}_{[100]}(e_0,\sigma,\sigma') & = & {\bf e}^{g}(\tilde{e}_0) + {\bf e}^{u}(\tilde{\lambda}), \\
\tilde{e}_0 & = & e_0 + \frac{1}{2}(\sigma + \sigma'), \\
\tilde{\lambda} & = & \frac{1}{2}(\sigma - \sigma'). \;
\end{eqnarray}
Again, the results shown in Sec.~\ref{su_afields} can then be used when analysing the resulting magnetocrystalline anisotropies. Recall that ${\bf e}^{g}$ has a minor effect on the in-plane anisotropy and can therefore be omitted when discussing in-plane magnetisation transitions.

So far we have described induced strains aligned with the $[100]$ crystal axis. In case of a lattice relaxation or piezo stressor aligned at an arbitrary angle $\omega$, the following transformation of the total strain tensor ${\bf e}^r_{[100]}$ or ${\bf e}^p_{[100]}$ to the crystallographic coordinate system applies:
\begin{eqnarray}
{\bf e}^{r(p)}_{\omega} & = & R^T_{\omega}{\bf e}^{r(p)}_{[100]}R_{\omega}
\label{strain_ind_rot} \;
\end{eqnarray}
where the rotation matrix reads:
\begin{eqnarray}
\label{rot_matrix} 
R_{\omega} & = & \left(\begin{array}{ccc} 
\cos(\omega -\pi/4) & \sin(\omega -\pi/4) & 0 \\
-\sin(\omega -\pi/4) & \cos(\omega -\pi/4) & 0 \\ 
0 & 0 & 1 \\
\end{array}\right). \;
\end{eqnarray}
The angular shift by $-\pi/4$ is because we measure the angle $\omega$ from the $[1\overline{1}0]$ axis. This convention was introduced in Sec.~\ref{su_afields} before Eq.~(\ref{free_en_inplane})  and is used consistently in this section for all in-plane angles.
The rotated total induced strain can be used directly as the input strain matrix for the microscopic calculation or it can be decomposed into all three elements of the in-plane strain basis. In case of the relaxation-induced strain, we obtain:
\begin{eqnarray}
\label{strain_relax_rot_decomp}
{\bf e}^{r}_{\omega}(e_0,\overline{\rho}) & = & {\bf e}^{g}(\tilde{e}_0) + {\bf e}^{u}(\tilde{\lambda}) + {\bf e}^{s}(\tilde{\kappa}), \\
\tilde{e}_0 & = & e_0\left(1-\frac{\overline{\rho}}{2}\right), \\
\tilde{\lambda} & = & -e_0\frac{\overline{\rho}}{2}\sin 2\omega,
\label{lambda_relax_omega} \\
\tilde{\kappa} & = & e_0\frac{\overline{\rho}}{2}\cos 2\omega.
\label{kappa_relax_omega} \;
\end{eqnarray}
In case of the rotated piezo stressor, the same decomposition follows, however, the effective strain magnitudes $\tilde{\lambda}$ and $\tilde{\kappa}$ depend on different real experimental parameters:
\begin{eqnarray}
\label{strain_piezo_rot_decomp}
{\bf e}^{p}_{\omega}(e_0,\sigma,\sigma') & = & {\bf e}^{g}(\tilde{e}_0) + {\bf e}^{u}(\tilde{\lambda}) + {\bf e}^{s}(\tilde{\kappa}), \\
\tilde{e}_0 & = & e_0 + \frac{(\sigma + \sigma')}{2}, \\
\tilde{\lambda} & = & \frac{(\sigma - \sigma')}{2}\sin 2\omega, 
\label{lambda_piezo_omega} \\
\tilde{\kappa} & = & - \frac{(\sigma - \sigma')}{2}\cos 2\omega. 
\label{kappa_piezo_omega} \;
\end{eqnarray}
Considering the linear dependence of the anisotropy coefficients on the corresponding strain elements (see Eq.~(\ref{axp_e0})), we can write the part due to post-growth induced strains of the phenomenological formula for the free energy as a function of angles $\psi$ and $\omega$:
\begin{eqnarray}
\label{free_piezo}
 F_u(\hat{M}) & = & K_{[110]}(\omega) \sin^2 \psi + \\
 & & + K_{[100]}(\omega) \sin^2(\psi+\pi/4) \nonumber \\
 & \simeq & q_{[110]} \tilde{\kappa}(\omega) \sin^2 \psi + \nonumber \\
 & & + q_{[100]} \tilde{\lambda}(\omega) \sin^2(\psi+\pi/4), \nonumber \;
\end{eqnarray}
where we use the notation analogous to Eq.~(\ref{axp_e0}) in Sec.~\ref{su_afields}. The relation of the effective parameters $\tilde\lambda$ and $\tilde\kappa$ to the experimental parameters of microbars or stressors oriented along arbitrary crystallographic direction is given by Eqs.~(\ref{lambda_relax_omega}-\ref{kappa_relax_omega}) or~(\ref{lambda_piezo_omega}-\ref{kappa_piezo_omega}), respectively. The linearity of the anisotropy constants $K_{[100]}$, $K_{[110]}$, and $K_{[001]}$ on corresponding strain coefficients and the form of the strain tensors in Eqs.~(\ref{strain_relax_rot_decomp}) and~(\ref{strain_piezo_rot_decomp}) allow us to factor out the $\omega$-dependence of $K_u$'s. Figs.~\ref{gen_kc},~\ref{gen_kukxu}, and~\ref{g_pdep_kcku} together with Eqs.~(\ref{strain_relax_rot_decomp}) and~(\ref{strain_piezo_rot_decomp}) can therefore be used for analysing magnetic anisotropies induced by micropatterning or piezo stressors oriented along any crystallographic direction.

\begin{figure}[h]
%\SetFigLayout{1}{2}
\begin{tabular}{c}
      \includegraphics[scale=0.73]{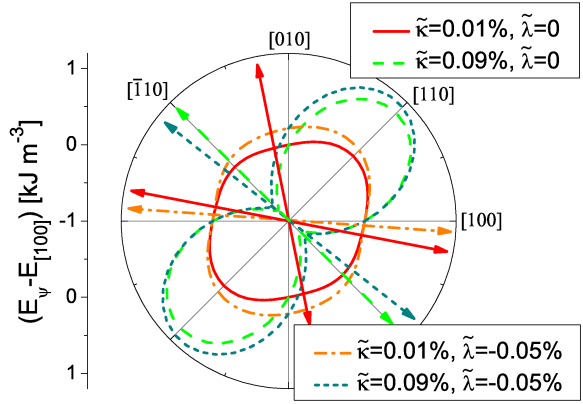} \\
      (a) $x=3\%$, $p=3\times 10^{20}$cm$^{-3}$, $\tilde{e}_0=-0.3\%$ when \\
           $\tilde{\lambda}/2=0$, $\tilde{e}_0=-0.25\%$ when $\tilde{\lambda}/2=0.05\%$ \\
      \includegraphics[scale=0.73]{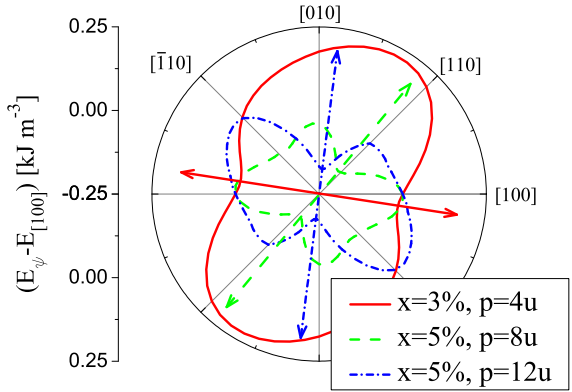} \\
      (b) $\tilde{\lambda}=-0.05\%$, $\tilde{\kappa}=0.01\%$, $\tilde{e}_0=-0.25\%$ \\
\end{tabular}
\caption{(Color online) Magnetic anisotropy energy $\Delta E = E_{\psi}-E_{[100]}$ as a function of the in-plane magnetisation orientation ${\bf M}=|{\bf M}|[\cos\psi,\sin\psi,0]$ and its dependence on material parameters. Effects of the shear strain and the uniaxial strain combine linearly (a). Magnetic easy axes (marked by arrows) change their direction upon change of Mn local moment concentration $x$, and hole density $p$ (in units u~$\equiv 10^{20}$~cm$^{-3}$) for a fixed uniaxial and shear strain (b). Both plots assume zero temperature.}
\label{pattern_exy}
\end{figure}

%%%% polarplot
The full angular dependencies of the anisotropy energy calculated directly from the total strain tensor included into the Kohn-Luttinger kinetic-exchange Hamiltonian for several combinations of $\tilde{\kappa}$ and $\tilde{\lambda}$ are plotted in Fig.~\ref{pattern_exy}. Recall that analogous in-plane angular dependencies of the anisotropy energy were presented in Fig.~\ref{polar_b} - \ref{polar_d}, where only the competition of the growth strain ${\bf e}^{g}$ and shear strain ${\bf e}^{s}$ with the cubic anisotropy of the host lattice was considered.

Fig.~\ref{pattern_exy}(a) shows four angular dependencies of the anisotropy energy for $x=3\%$ and $p=3\times 10^{20}$cm$^{-3}$. The curves are marked by the values of the effective strain components. The solid curve for weak shear strain $\tilde{\kappa}=0.01\%$ and no uniaxial strain $\tilde{\lambda}=0$ has two local minima close to the main crystal axes indicating dominant cubic anisotropy with $K_{c1}>0$ for the considered $x$ and $p$. The easy axes are shifted due to the positive shear strain towards the $[1\overline{1}0]$ axis which is the direction of relative lattice compression, consistently with the discussion in Sec.~\ref{su_afields} for samples with medium hole densities. Additional uniaxial strain $\tilde{\lambda}=-0.05\%$ results in only one global minimum easy axis rotating towards the $[100]$ direction which is again the direction of relative lattice compression.

The dashed curve in Fig.~\ref{pattern_exy}(a) corresponding to strong shear strain $\tilde{\kappa}=0.09\%$ and no uniaxial strain $\tilde{\lambda}=0$ has only one global minimum at the $[1\overline{1}0]$ diagonal, indicating domination of the uniaxial anisotropy over the underlying cubic anisotropy. Addition of the uniaxial strain $\tilde{\lambda}=-0.05\%$ leads to rotation of the easy axis towards the direction of relative compression ($[100]$ for $\tilde{\lambda} < 0$). %We emphasise that the dependence of the in-plane uniaxial anisotropies on the strains ${\bf e}^s$ and ${\bf e}^u$ has the same form (expressed in Eq.~(\ref{axp_axp})) so the easy axis rotation with respect to the direction of lattice compression or expansion has a universal orientation for a given set of material parameters.

Curves plotted in Fig.~\ref{pattern_exy}(b) differ in the material parameters but share the same weak shear strain $\tilde{\kappa}=0.01\%$ and the same uniaxial strain $\tilde{\lambda}=-0.05\%$. The solid curve for $x=3\%$ and $p=4\times 10^{20}$cm$^{-3}$ falls into the range of hole densities where the cubic anisotropy coefficient $K_{c1}$ is positive so the easy axes in the absence of in-plane strains align parallel to the main crystal axes. Adding the uniaxial strain $\tilde{\lambda}$ yields only one global minimum along the $[100]$ direction and the shear strain shifts the easy axis towards the $[1\overline{1}0]$ diagonal. Again, for both strains the easy axes tend to align along the direction of lattice compression for these medium doping parameters.

The dashed curve in Fig.~\ref{pattern_exy}(b) for $x=5\%$ and $p=8\times 10^{20}$cm$^{-3}$ can be described by a negative $K_{c1}$ corresponding to diagonal easy axes in the unstrained bulk epilayer. The additional shear strain $\tilde{\kappa}$ makes the $[110]$ direction the global minimum easy axis. Note that for these values of $x$ and $p$ the easy axis prefers to align with the direction of lattice expansion. Consistently, the uniaxial strain $\tilde{\lambda}$ rotates the easy axis towards the direction of relative lattice expansion, i.e., towards the $[010]$ axis. Finally, the dash-dotted curve for $x=5\%$ and high hole density $p=12\times 10^{20}$cm$^{-3}$ corresponds to positive $K_{c1}$ and again, when the in-plane strains are included the easy axes prefer the direction of relative lattice expansion. To summarise the discussion of Figs.~\ref{pattern_exy}(a) and~(b), the preferred alignment of the in-plane easy axis with either the lattice contraction or expansion direction depends on $x$ and $p$. For a given doping it has always the same sense for both the shear strain $\tilde{\kappa}$ and the uniaxial strain $\tilde{\lambda}$ and is uncorrelated with the sign of the cubic anisotropy component. These conclusions are independent of the growth strain, at least for its typical values $e_0<1\%$.

%%% exp relax
Now we analyse experimental studies that control the in-plane strain by means of post-growth lithography.
Refs.~[\onlinecite{Pappert:2007_a}] and [\onlinecite{Wunderlich:2007_c}] present structures with the shear and uniaxial strain induced locally by anisotropic relaxation of the compressive growth strain. Ref.~[\onlinecite{Wunderlich:2007_c}] studies an L-shaped channel with arms aligned along the $[1\overline{1}0]$ and $[110]$ directions patterned by lithography in a 25~nm thick (Ga,Mn)As epilayer with nominal Mn concentration $x=5\%$. Hole density $p=5 \times 10^{20}$cm$^{-3}$ was estimated from high-field Hall measurements. This patterning allows relaxation of the growth lattice matching strain in direction perpendicular to the channel. Therefore, the generated uniaxial strains in each arm of the L-shaped channel have opposite signs. The induced shear strain is added to (subtracted from) the ``intrinsic'' shear strain in the arm fabricated along the $[1\overline{1}0]$ ($[110]$) axis. The magnitude of the induced strain increases with decreasing width of the channel. A large effect on magnetic easy axes orientation has been observed in a 1~$\mu$m wide channel while only moderate changes have been found in a 4~$\mu$m bar. In both cases the easy axes of the unpatterned epilayer rotated in the direction perpendicular to lattice expansion. The sense and magnitude of the easy-axis reorientations in the relaxed microbars are consistent with theory prediction for the relevant values of $x$, $p$, and microbar geometry.

Refs.~[\onlinecite{Wenisch:2007_a}] and [\onlinecite{Humpfner:2006_a}] show lithographically induced uniaxial anisotropy along the $[100]$ or $[010]$ axis in arrays of narrow bars. Ref.~[\onlinecite{Wenisch:2007_a}] presents 200~nm wide bars fabricated in an as-grown 70~nm thick film with Mn concentration $x=2.5 \%$ determined by x-ray diffraction. Ref.~[\onlinecite{Humpfner:2006_a}] reports lattice relaxation in 200~nm wide, 20~nm thick bars in an as-grown material with nominal Mn concentration $x=4 \%$. In both studies the unpatterned epilayers have two equivalent easy axes close to main crystal axes. After the anisotropic relaxation of the growth strain in the nanobars the easy axis corresponding to the relaxation direction is lost, whereas the other easy axis is maintained. This behaviour is in agreement with our simulations on the relevant interval of dopings and patterning induced strains.

% rotated relax
The anisotropies induced in the relaxed structures in Refs.~[\onlinecite{Pappert:2007_a,Wunderlich:2007_c,Wenisch:2007_a,Humpfner:2006_a}] can be predicted using the results of Sec.~\ref{su_afields} directly. Bearing in mind the negligible effect of the growth strain, the relevant part of the strain tensor describing the relaxation along the main crystal axes has the form of the uniaxial basis strain ${\bf e}^{u}$, as shown in Eq.~(\ref{strain_relax_decomp}), and corresponds to the anisotropy component with the previously calculated coefficient $K_{[100]}$. The relaxation along the diagonals is described by the strain tensor: ${\bf e}^{r}_{[110]}(e_0,\overline{\rho}) = {\bf e}^{s}(\tilde{\kappa})$ with $\tilde{\kappa} = -\frac{1}{2}e_0\overline{\rho}$, where we again neglected the contribution from the growth strain ${\bf e}^g$. It induces uniaxial anisotropy component quantified by the coefficient $K_{[110]}$. Note that the ``intrinsic'' shear strain $e^{int}_{xy}$ in the modelling is independent of the externally introduced lattice distortion and needs to be added to the total strain tensor if the corresponding anisotropy is present in the unpatterned epilayer. As mentioned before, the simulated rotation of easy axis directions in the relaxed microbars is in good agreement with the measured behaviour.

%%% exp piezo
The piezo-strain is also applied in most cases along the main crystal axes or diagonals.
In Ref.~[\onlinecite{Goennenwein:2008_a}] a PZT piezoelectric actuator is attached to %$150 \times 600 \mu$m$^2$ 
a 30~nm thick (Ga,Mn)As epilayer grown on a GaAs substrate thinned to 100~$\mu$m. The principal elongation direction of the actuator is aligned with the $[110]$ crystallographic direction. The nominal Mn concentration of the as-grown epilayer is $4.5\%$. The relative actuator length change is approximately $4 \times 10^{-4}$ at $T=50$~K (measured by a strain gauge) for the full voltage sweep (from -200~V to 200~V). Such piezo-strain induces a rotation of the easy axis by $\Delta \psi_{EA} \approx 65^{\circ}$. Our modelling predicts $\Delta \psi_{EA}$ of the same order for relevant material and strain parameters.
% see /home/zemen/matlab/KCKU/IMA/TZERO/EA_DIRECTION/phase_dphi_dexy.m ``the tange of calculated \Delta \psi_{EA} is quite broad''
The easy axis rotates towards the $[110]$ ($[1\overline{1}0]$) direction upon contraction (elongation) along the $[110]$ axis in agreement with the behaviour observed in the relaxed microbars and with our modelling.

%Quantitative comparison of the shift of magnetisation due to the piezo-induced shear strain is hindered by an unknown strain caused by gluing the actuator to the Hall bar (strain at zero bias).

Ref.~[\onlinecite{Bihler:2008_a}] extends the piezo-stressed (Ga,Mn)As study in Ref.~[\onlinecite{Goennenwein:2008_a}] to low temperatures. Again, PZT piezo actuator is attached to a Hall bar along the $[110]$ crystallographic direction. The 30~nm thick, as-grown (Ga,Mn)As epilayer grown on GaAs substrate has nominal Mn concentration $4.5\%$ and $T_C=85$~K. A strain gauge measurement shows almost linear dependence of the piezo-strain in the Hall bar on temperature (in the range 5~K to 50~K). The anisotropy coefficients $K_{[110]}$ and $K_{c1}$ are extracted from the angle-dependent magnetoresistance measurement as a function of temperature for three voltages (-200~V, 0~V, and 200~V). At high temperatures the relative elongation of the structure is again approximately $4 \times 10^{-4}$ and the corresponding uniaxial anisotropy dominates over the intrinsic uniaxial anisotropy along the $[110]$ axis. Close to 5~K the action of the piezo is negligible so the intrinsic uniaxial anisotropy is stronger than the induced one, however, the total in-plane anisotropy is dominated by the cubic anisotropy.
%The observation of very small uniaxial anisotropy at low temperature reveals absence of a uniaxial strain due to different thermal contraction of the sample and the actuator on cooling. 
The measured and calculated induced anisotropy along the $[110]$ direction are of the same sign and order of magnitude for the considered temperatures. 
%The intrinsic anisotropy measured at low temperature is an order of magnitude smaller than the prediction in this material.

Ref.~[\onlinecite{Overby:2008_a}] presents a 15~nm thick, annealed sample doped to $x=8\%$, subject to piezo stressing along the $[010]$ axis. The anisotropy coefficients are extracted from transverse AMR. The PZT actuator induces relative elongation ranging from $1.1 \times 10^{-3}$ for voltage 200~V to $0.7 \times 10^{-3}$ for -200~V, measured by a strain gauge. The difference of the limits is again  approximately $4 \times 10^{-4}$ but all values are shifted towards tensile strain most likely due to different thermal dilatation in the sample and the actuator. The lattice expansion along the $[010]$ direction leads to alignment of the easy axis along the $[100]$, in agreement with our modelling and with the experimental studies discussed in this section. The extracted cubic anisotropy field is roughly a factor of two lower compared to studies of samples with high hole compensation sharing the value $\approx 1000$~Oe at different nominal Mn concentrations.\cite{Wang:2005_e,Liu:2004_b,Shin:2007_a} The low critical temperature $T_C=80$~K suggests lower effective Mn concentration in Ref.~[\onlinecite{Overby:2008_a}]. Our calculations for lower Mn local moment concentration and high hole compensation predict the anisotropy coefficients $K_{c1}$ and $K_{[100]}$ induced by the piezo strain in correspondence with the measured coefficients.

\begin{figure}
\includegraphics[scale=0.33]{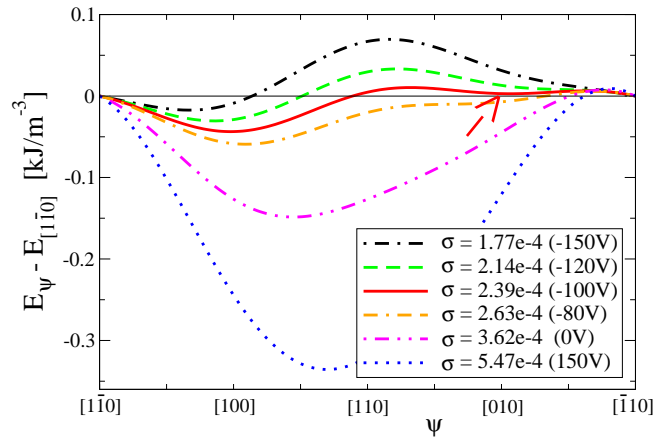}%
\caption{(Color online) Calculated magnetic anisotropy energy $\Delta E = E_{\psi}-E_{[1\overline{1}0]}$ as a function of the in-plane magnetisation angle $\psi$ measured from the $[1\overline{1}0]$ axis at T$=5/8$T$_C$, $e_0=0$, $e_{xy}^{int}=0.017\%$, $x=5\%$, and $p=5\times 10^{20}$cm$^{-3}$. The curves are labelled by $\sigma$, the induced strain along the principal elongation direction of the piezo tilted by angle $\omega=-10^{\circ}$, and by the corresponding voltage. (The relationship of $\sigma$ and voltage is inferred from Ref.~[\onlinecite{Rushforth:2008_a}] to allow for direct comparison with experiment.) The easy axis rotates smoothly upon sweeping the voltage. For -100~V a shallow local energy minimum forms due to the underlying cubic anisotropy (marked by arrow).}
%source in /home/zemen/WORK/FERMIP/FERMIPSEQ/PIEZO_ANDY/MAGANISODIRXXYY/STRES_1_NEW
\label{piezo}
\end{figure}

Finally, we discuss a piezo-strain induced along a general in-plane direction. In Ref.~[\onlinecite{Rushforth:2008_a}] the principal elongation direction of the PZT piezo actuator is tilted by angle $\omega=-10^{\circ}$ (with respect to the $[1\overline{1}0]$ axis). The 25~nm thick, as-grown (Ga,Mn)As epilayer with nominal Mn concentration $x=6\%$ is grown on a GaAs substrate, which was thinned before attaching of the stressor to $\approx 150 \mu$m. The anisotropies are determined from SQUID and AMR measurements at 50~K. The uniaxial strain caused by differential thermal contraction of the sample and the piezo on cooling (at zero applied voltage) is of the order $10^{-4}$. The uniaxial strains generated at the voltage $\pm 150$~V are $\sigma \approx \pm 2 \times 10^{-4}$ and $\sigma' \approx -\sigma/2$ at 50~K. The tilt of the piezo with respect to the crystal diagonal results in a complicated interplay of the intrinsic and induced anisotropy. The easy axis of the bare sample aligns with the $[1\overline{1}0]$ axis due to strong intrinsic uniaxial anisotropy with $K_{[110]}>K_{c1}>0$. This easy axis rotates to an angle $\psi_{EA} = 65^{\circ}$ upon attaching of the piezo and cooling to 50~K. Application of +150~V to the stressor causes the easy axis to rotate further to $\psi_{EA} = 80^{\circ}$ while for -150~V the axis rotates in the opposite direction to $\psi_{EA} = 30^{\circ}$. Note that the negative voltage weakens the total piezo-strain and allows domination of the intrinsic anisotropy with easy axis closer to the $[1\overline{1}0]$ axis.

The hole compensations expected in Ref.~[\onlinecite{Rushforth:2008_a}] are in the range $p/N_{\rm Mn}=0.6-0.4$ and the relevant range of effective Mn concentrations is $x=3-5\%$. $K_{[110]}$ measured in the bare epilayer is modelled by $e_{xy}^{int}=3-2\times10^{-4}$ (slightly weaker than the strain induced in the structure at zero piezo-voltage). The in-plane anisotropy energies calculated on this parameter interval using the total strain tensor (induced and ``intrinsic'' components) are in good quantitative agreement with the easy axis orientations measured at the three piezo voltages. Fig.~\ref{piezo} shows calculated curves for one representative combination of $x$, $p$, and $e_{xy}^{int}$ from the relevant interval, for the fixed tilt of the stressor $\omega=-10^{\circ}$, and for a range of induced strains $\sigma$. The curves are marked also by the voltages as we infer a simple linear relationship between $\sigma$ and the voltage to facilitate comparison with the experimental paper.

The anisotropy behaviour shown in Fig.~\ref{piezo} can be described as a smooth rotation of the global energy minimum upon increase of $\sigma$ rather than the "scissors" effect shown in Fig.~\ref{polar_b} in Sec.~\ref{su_ima}. The total induced strain now contains both components ${\bf e}^{s}$ and ${\bf e}^{u}$ as written in Eq.~(\ref{strain_piezo_decomp}). The uniaxial basis strain ${\bf e}^{u}$ present due to the tilt of the stressor diminishes significantly one of the local minima typically occurring because of interplay of a positive cubic and a small uniaxial anisotropy component along a crystal diagonal. The remainder of the weaker local minimum is observed theoretically for $\sigma$ corresponding to voltages $\approx-100$~V when the ${\bf e}^{s}$ component of the induced strain and the ``intrinsic'' shear strain compensate each other. One would expect domination of cubic anisotropy with two equivalent local minima close to the main crystal axes if the stressor had purely diagonal alignment. The ${\bf e}^{u}$ component of the total strain of the tilted stressor makes the local minimum closer to the $[010]$ axis less pronounced (marked by arrow in Fig.~\ref{piezo}).

For completeness, we discuss the free energy phenomenological formula used in Ref.~[\onlinecite{Rushforth:2008_a}] to describe the in-plane angular dependence of the induced anisotropy. The decomposition of the total induced strain in Eq.~(\ref{strain_piezo_decomp}) into the strain basis introduced in Eqs.~(\ref{strain1}-\ref{strain3}) is not considered in that work. Instead, the induced anisotropy is described by a single uniaxial term $K_{\Omega}\sin^2(\psi-\Omega)$ added to the phenomenological formula rather than terms with coefficients $K_{[110]}$ and $K_{[110]}$ from Eq.~(\ref{free_piezo}). Effectively, this corresponds to a change of variables from $K_{[110]}$ and $K_{[110]}$ to $K_{\Omega}$ and $\Omega$. The angle $\Omega$ is measured from the $[1\overline{1}0]$ axis and it rotates the additional uniaxial anisotropy term so that it describes the effect due to the tilted stressor. One may assume collinearity of the resulting anisotropy component with the principal elongation direction of the piezo.
%, i.e., $\omega = \Omega + \pi/4$, for $|\omega|<\pi/2$. 
However, this simple situation is observed both theoretically and experimentally only when the stressor is aligned with the main crystal axes or diagonals. The missalignment for arbitrary orientation of the induced strain is due to the underlying cubic symmetry of the system incorporated into our microscopic band structure calculation in the form of the band parameters $\gamma_2$, $\gamma_3$, $a_2$, and $a_3$. It has been explained in Sec.~\ref{su_afields} that the collinearity of the in-plane strain and corresponding anisotropy occurs only for the strains ${\bf e}^{s}$ or ${\bf e}^{u}$ (see Eqs.~(\ref{strain2}) and~(\ref{strain3})). For any other stressor orientation, $\Omega\neq\omega$, which is reflected on the level of the anisotropy functions by the inequality, $q_{[100]}(x,p) \neq q_{[110]}(x,p)$. It expresses the difference in the effect on magnetic anisotropy between straining the lattice along the main crystal axis and along the diagonals (see Eq.~(\ref{axp_axp}) in Sec.~\ref{su_afields}).

The transformation from variables $K_{[110]}(x,p,\omega)$ and $K_{[110]}(x,p,\omega)$ to $K_{\Omega}(x,p,\omega)$ and $\Omega(x,p,\omega)$ in the phenomenological formula in Eq.~(\ref{free_piezo}) for $-\pi/2 < \omega < \pi/2$ reads:
\begin{eqnarray}
\label{free_piezo_andy}
 F_u(\hat{M}) & = & K_{[110]}(\omega) \sin^2 \psi + K_{[100]}(\omega) \sin^2(\psi+\pi/4) \nonumber \\
 & = & K_{\Omega} \sin^2(\psi-\Omega), \;
\end{eqnarray}
where:
\begin{eqnarray}
\label{piezo_andy_const}
\Omega(x,p,\omega) & = & \frac{1}{2}\arctan \left( -\frac{K_{[100]}}{K_{[110]}} \right),  \\
K_{\Omega}(x,p,\omega) & = & - K_{[110]}\cos 2\Omega + K_{[100]} \sin 2\Omega. \nonumber \;
\end{eqnarray}
Considering the approximate relation $q_{[100]}=0.43q_{[110]}$ the formulae in Eq.~(\ref{piezo_andy_const}) simplify to:
\begin{eqnarray}
\label{piezo_andy_const_simple}
\Omega(x,p,\omega) & = & \frac{1}{2}\arctan \left(\frac{q_{[100]}(x,p)\sin 2\omega}{q_{[110]}(x,p)\cos 2\omega} \right)  \\
                   & = & \frac{1}{2}\arctan \left(0.43\tan 2\omega \right), \nonumber \\
q_{\Omega}(x,p,\omega) & \equiv & q_{[110]}(x,p) \cos 2\omega \cos 2\Omega + \nonumber \\
                        &     & + 0.43q_{[110]}(x,p) \sin 2\omega \sin 2\Omega, \nonumber \;
% calculation in: /home/zemen/WORK/FERMIP/FERMIPSEQ/PIEZO_ANDY/MAGANISODIRXXYY/INTRINSIC/intrinsic.dat (7.3.2009)
% xmgrace check in: /home/zemen/WORK/FERMIP/FERMIPSEQ/PIEZO_ANDY/MAGANISODIRXXYY/STRES_1_NEW/pom5.agr
\end{eqnarray}
where $K_{\Omega} = q_{\Omega}(\sigma - \sigma')/2$. (The same transformation of variables can be used in case of strains induced along arbitrary in-plane direction by relaxation in a narrow bar (see Eqs.~(\ref{lambda_relax_omega}-\ref{kappa_relax_omega})). Then we obtain $K_{\Omega} = -q_{\Omega}e_0\overline{\rho}/2$.)

Note that in the representation of $F_u(\hat{M})$ via $K_{[110]}$ and $K_{[110]}$ the dependence on $\omega$ can be simply factored out and the dependence on $x$ and $p$ is contained only in the functions functions $q_{[110]}$ and $q_{[100]}$. For our general discussion presented in this paper it is therefore the more convenient form than $F_u(\hat{M})$ expressed via $K_{\Omega}$ and $\Omega$. %When assuming spherical symmetry of the host crystal, discussed in Sec.~\ref{su_afields}, the anisotropy functions are equal $q_{[110]}(x,p) = q_{[100]}(x,p)$ and for the angles we obtain: $\omega = \Omega + \pi/4$. % omega se meri od [1-10]

We conclude that the in-plane alignment of the easy axis in patterned or piezo-stressed samples can be described on a semi-quantitative level by our modelling similarly to the bare (Ga,Mn)As epilayers.
%%%%%%%%%%%%%%%%%%%%%%%%%%%%%%%%%%%%%%%%%%%%%%%%%%

\section{Summary}
\label{se_conclusion}
The objective of this work was to critically and thoroughly inspect the efficiency of a widely used effective Hamiltonian model in predicting the magneto-crystalline anisotropies in (Ga,Mn)As.
We have provided overview of the calculated anisotropies which show a rich phenomenology as a function of Mn concentration, hole density, temperature and lattice strains, and compared it to a wide range of experimental works on the level of the magnetic easy axis direction and on the level of anisotropy fields. The large amount of analysed results compensates for the common uncertainty in sample parameters assumed in experiment and allowed us to make systematic comparisons between theory and experiment on the level of trends as a function of various tunable parameters. Generically, we find this type of comparison between theory and experiment in diluted magnetic semiconductors much more meaningful than addressing isolated samples, given the complexity of these systems and inability of any theoretical approach applied to date to fully quantitatively describe magnetism in these random-moment semiconducting ferromagnets.

In Sec.~\ref{se_theory} we introduced the mean-field model used throughout the study, estimated the relative strength of the shape anisotropy, and discussed the correspondence of the shear strain, modelling the broken in-plane symmetry measured in most (Ga,Mn)As epilayers, with a microscopic symmetry breaking mechanism.

In Sec.~\ref{se_unpatterned} we focused on modelling and experiments in bare unpatterned epilayers. The in-plane and out-of-plane magnetisation alignment was studied. For compressively strained samples the generally assumed in-plane anisotropy is found to be complemented by regions of out-of-plane anisotropy at low hole densities and low temperatures. This observation is corroborated by available experimental data showing in-plane anisotropy in most of the studied epilayers but also the occurrence of the out-of-plane easy axis in materials with high hole compensation. At the same time, the model predicts out-of-plane easy axis for high hole densities at all Mn concentrations which has yet not been observed experimentally.

Next, the competition of cubic and uniaxial in-plane anisotropy components was investigated. Wealth of experimentally observed easy axis transitions driven by change of temperature or hole density finds corresponding simulated behaviour. The following general trend is observed in most samples: at low temperatures the easy axes are aligned close to the main crystal axes, while at high temperatures there is always diagonal alignment. This trend is in good agreement with our calculation, however, at low hole densities the calculated and measured easy axis transitions are more consistent than at higher hole densities where the measured phenomena match the predictions assuming hole densities typically a factor of two lower than in the experiment.
%The analysis of 25 samples allows us to estimate the magnitude and sign of the shear strain assumed by our model: It is positive and scales linearly with Mn doping $e_{xy} \simeq 0.005 \times x [\%]$. Sec.~\ref{se_lithopiezo} confronts this result with in-plane uniaxial strain induced in epilayers by post-growth treatment. 

We next introduced anisotropy fields corresponding to the crystal symmetry and to three distinct uniaxial strains. We extracted these anisotropy fields from the calculated data and found their dependence on material parameters. We observed linear dependence of the uniaxial anisotropy fields on the corresponding strains. Analysing experiments which determine the anisotropy fields from FMR, AMR or SQUID measurements allowed for detailed comparison of the cubic anisotropy component and two uniaxial anisotropy components (due to growth and the $[110/[1\overline{1}0]$ symmetry breaking). The measured and calculated anisotropy fields are of the same order of magnitude ($\sim 10^2-10^3$~Oe) in most samples.

Finally, in Sec.~\ref{se_lithopiezo} we investigated structures where the post-growth patterning or piezo stressing was used to induce additional strains along any in-plane direction. The interplay of the intrinsic and induced anisotropies was studied. We discussed the procedure for obtaining the strain Hamiltonian from the parameters describing the experimental setup and a finite element solver was employed to find the inhomogeneous lattice relaxation in the patterned epilayers. Induced anisotropies were calculated directly using the total strain tensor. Alternatively, we also introduced a decomposition of the total strain matrix for any of the studied materials and device configurations into three basis strains and their additive effect on the total anisotropy. We found an overall semi-quantitative agreement of theory and experiment on the level of easy axis reorientations due to induced strains.

The limitations of the theory approach employed in this paper have been thoroughly discussed in Ref.~[\onlinecite{Abolfath:2001_a}]. The model, which treats disorder in the virtual crystal approximation and magnetic interactions on the mean-field level is expected to be most reliable at lower temperatures and in the (Ga,Mn)As materials with metallic conductivity. We have shown that despite the limitations, the model captures on a semi-quantitative level most of the rich phenomenology of the magnetocrystalline anisotropies observed in (Ga,Mn)As epilayers and microdevices  over a wide parameter  range. We hope that our work will provide a useful guidance for future studies of magnetic and magnetotransport phenomena in (Ga,Mn)As based systems in which magnetocrystalline anisotropies play an important role.

\acknowledgements
We thank K. Y. Wang for providing us with previously unpublished experimental data. We acknowledge fruitful discussions with Richard Campion, Tomasz Dietl, Kevin Edmonds, Tom Foxon, Bryan Gallagher, V\'{\i}t Nov\'{a}k, Elisa de Ranieri, Andrew Rushforth, Mike Sawicki, Jairo Sinova, Laura Thevenard, Jorg Wunderlich. The work was funded through Pr\ae mium Academi\ae{} and contracts number
AV0Z10100521, LC510, KAN400100652, FON/06/E002 of \hbox{GA \v CR},
of the Czech republic, and by
the NAMASTE (FP7 grant No.~214499) and SemiSpinNet projects
(FP7 grant No.~215368).

%We acknowledge financial support from the EU (NANOSPIN, FP6-2002-IST-015728).

\begin{appendix}

\section{Symmetries of the Kohn-Luttinger Hamiltonian}
\label{app_KLHam}
Different representations of the six-band Kohn-Luttinger Hamiltonians are used in literature.
Here, the notation of Ref.~[\onlinecite{Abolfath:2001_a}] is used and extended.

The states at the top of the valence band have $p$-like character and can be represented by the $l$=1 orbital momentum eigenstates $|l,m_l\rangle$. In the basis of combinations of orbital angular momentum eigenstates: 
\begin{eqnarray}
|X\rangle&=&\frac{1}{\sqrt{2}}\big(|1 -1\rangle + |1 1\rangle\big), \nonumber \\ 
|Y\rangle&=&\frac{i}{\sqrt{2}}\big(|1 -1\rangle - |1 1\rangle\big), \nonumber \\ 
|Z\rangle&=&|1 0\rangle\
\label{xyz}
\end{eqnarray}
the Kohn-Luttinger Hamiltonian for systems with no spin-orbit coupling can be written as:
\begin{widetext}
\begin{equation}
{\cal H}_{kp} = \left(\begin{array}{ccc} 
\epsilon_v + Ak_x^2+B(k_y^2+k_z^2) & Ck_xk_y & Ck_xk_z \\ 
Ck_yk_x & \epsilon_v + Ak_y^2+B(k_x^2+k_z^2) & Ck_yk_z \\ 
Ck_zk_x & Ck_zk_y & \epsilon_v + Ak_z^2+B(k_x^2+k_y^2) \\
\end{array}\right),
\label{HKL_no_so}
\end{equation}
where
\begin{eqnarray} 
A &=& \frac{\hbar^2}{2m_0} + \frac{\hbar^2}{m^2_0}\sum_{i\notin \left\lbrace X,Y,Z\right\rbrace}\frac{|\langle X|p_x|i\rangle|^2}{\epsilon_1-\epsilon_i}, 
\label{A} \\
B &=& \frac{\hbar^2}{2m_0} + \frac{\hbar^2}{m^2_0}\sum_{i\notin \left\lbrace X,Y,Z\right\rbrace}\frac{|\langle X|p_y|i\rangle|^2}{\epsilon_1-\epsilon_i}, 
\label{B} \\
C &=& \frac{\hbar^2}{m^2_0}\sum_{i\notin \left\lbrace X,Y,Z\right\rbrace}\frac{\langle X|p_x|i\rangle \langle i|p_y|Y\rangle + \langle X|p_y|i\rangle \langle i|p_x|Y\rangle}{\epsilon_1-\epsilon_i},
\label{C}
\end{eqnarray}
\end{widetext}
and $\epsilon_v$ is the energy of the valence band $p$-orbitals.

The simple form is due to the symmetry of the zinc-blende crystal structure. The summation in elements $A$, $B$, $C$ runs only through the $\Gamma_1$ and $\Gamma_4$ states of the conduction band as other levels are excluded by the matrix element theorem combined with the tetrahedron symmetry.\cite{Yu:2005_a} The only nonzero momentum operator expectation values with neighbouring states are:
\begin{eqnarray} 
\langle X |p_y| \Gamma_4 (z) \rangle = \langle Y |p_z| \Gamma_4 (x) \rangle = \langle Z |p_x| \Gamma_4 (y) \rangle \nonumber \\
\langle X |p_x| \Gamma_1 \rangle = \langle Y |p_y| \Gamma_1 \rangle = \langle Z |p_z| \Gamma_1 \rangle. 
\label{met}
\end{eqnarray}
Due to the reflection symmetry with respect to the $(110)$ planes it holds also:\cite{Yu:2005_a}
\begin{eqnarray} 
\langle X |p_y| \Gamma_4 (z) \rangle = \langle Y |p_x| \Gamma_4 (z) \rangle \rangle
\label{refl}
\end{eqnarray}

If the tetrahedral symmetry of the GaAs lattice is broken by potential $V = xy\xi$ as described in Sec.~\ref{su_beyondcub} the states $\Gamma_1$ and $\Gamma_4 (z)$ of the conduction band considered in the summation in Eq.~(\ref{perturbkp}) are mixed, whereas states $\Gamma_4 (x)$ and $\Gamma_4 (y)$ are left unchanged.
In the perturbed basis $\alpha \Gamma_1 + \beta \Gamma_4(z)$, $-\beta \Gamma_1 + \alpha \Gamma_4(z)$, $\Gamma_4(x)$, $\Gamma_4(y)$ we obtain terms containing the parameter $D$ in the Hamiltonian $\tilde{\cal H}_{kp}$ (See Eq.~(\ref{HKL_new}) in Sec.~\ref{su_beyondcub}.) A weak local potential $V$, $\alpha>>\beta$ was assumed so terms of quadratic and higher order dependence on $V$ could be neglected. Therefore the expression for parameters $A$, $B$, and $C$ does not change. Using Eqs.~(\ref{met}) and (\ref{refl}) allows also for a compact expression of the parameter $D$:
\begin{eqnarray}
\label{D}
D &=& \zeta \langle X |p_y| \Gamma_4 (z) \rangle \langle \Gamma1 |p_x| X \rangle, \\
\zeta &=& \frac{\hbar^2}{m^2_0} \alpha \beta \left[ \frac{1}{\epsilon_v-(\epsilon_{c1}+\Delta)} - \frac{1}{\epsilon_v-(\epsilon_{c4}-\Delta)} \right], \nonumber \;
\end{eqnarray}
where $\epsilon_{c1}$ and $\epsilon_{c4}$ are the energies of the conduction band $\Gamma_1$ and $\Gamma_4$ states, respectively. The small energy $\Delta$ is quadratically dependent on the size of the potential $V$ but we include it to express the shift of the perturbed energy levels.

To include spin-orbit coupling we use the basis formed by combinations of orbital angular momentum:
\begin{eqnarray}
\label{123456}
|1\rangle&\equiv&|j=3/2,m_j=3/2\rangle \nonumber \\
|2\rangle&\equiv&|j=3/2,m_j=-1/2\rangle \nonumber \\
|3\rangle&\equiv&|j=3/2,m_j=1/2\rangle \nonumber \\
|4\rangle&\equiv&|j=3/2,m_j=-3/2\rangle \nonumber \\
|5\rangle&\equiv&|j=1/2,m_j=1/2\rangle \nonumber \\
|6\rangle&\equiv&|j=1/2,m_j=-1/2\rangle \;
\end{eqnarray}
The spin-orbit correction to the 6-band Hamiltonian is diagonal in this basis and can be parametrised only by a single parameter $\Delta_{so}$.\cite{Abolfath:2001_a}
The 6-band Kohn-Luttinger Hamiltonian in the representation of
vectors (\ref{123456}) reads:
\begin{equation}
\hspace*{0cm} {\cal H}_{KL} = \left(\begin{array}{cccccc} {\cal H}_{hh} & -c & -b &
\multicolumn{1}{c|}{0} & \frac{b}{\sqrt{2}} & c\sqrt{2}\\ -c^* & {\cal H}_{lh}
& 0 & \multicolumn{1}{c|}{b} & -\frac{b^*\sqrt{3}}{\sqrt{2}} & -d\\ -b^* & 0 &
{\cal H}_{lh} & \multicolumn{1}{c|}{-c} &   d & -\frac{b\sqrt{3}}{\sqrt{2}} \\
0 & b^* & -c^* & \multicolumn{1}{c|}{{\cal H}_{hh}} &  -c^*\sqrt{2} &
\frac{b^*}{\sqrt{2}}\\ \cline{1-4} \frac{b^*}{\sqrt{2}} &
-\frac{b\sqrt{3}}{\sqrt{2}} & d^* & -c\sqrt{2} & {\cal H}_{so} & 0\\
c^*\sqrt{2} & -d^* & -\frac{b^*\sqrt{3}}{\sqrt{2}} & \frac{b}{\sqrt{2}} & 0 &
{\cal H}_{so}\\
\end{array}\right)
\label{hl}
\end{equation}
The 4-band Hamiltonian is highlighted. The Kohn-Luttinger eigen-energies are hole energies (measured down from the top of the valence band). The matrix elements of ${\cal H}_{KL}$ are listed in Ref.~[\onlinecite{Abolfath:2001_a}]. Here we focus on the modification of these elements due to incorporating the microscopic potential $V = xy\xi$:
%\begin{widetext}
\begin{eqnarray}
\tilde{\cal H}_{hh} &=& \frac{\hbar^2}{2m}\big[(\gamma_1 + \gamma_2)(k_x^2+k_y^2) +
(\gamma_1 - 2\gamma_2)k_z^2 +6\gamma_4k_xk_y \big] \nonumber \\ 
\tilde{\cal H}_{lh} &=& \frac{\hbar^2}{2m}\big[(\gamma_1 - \gamma_2)(k_x^2+k_y^2) + (\gamma_1 +
2\gamma_2)k_z^2 +2\gamma_4k_xk_y \big] \nonumber \\ 
\tilde{\cal H}_{so} &=& \frac{\hbar^2}{2m} \big[ \gamma_1(k_x^2+k_y^2+k_z^2)+4\gamma_4k_xk_y \big]+\Delta_{so} \nonumber \\ 
\tilde{b} &=& \frac{\sqrt{3}\hbar^2}{m} \gamma_3 k_z (k_x - i k_y) \nonumber \\ 
\tilde{c} &=& \frac{\sqrt{3}\hbar^2}{2m}\big[\gamma_2(k_x^2 - k_y^2) - 2i(\gamma_3 k_x k_y+\frac{\gamma_4}{2}(k_x^2+k_y^2))\big] \nonumber \\ 
\tilde{d} &=&-\frac{\sqrt{2}\hbar^2}{2m}\big[\gamma_2(2k_z^2-k_x^2-k_y^2) -2\gamma_4k_xk_y\big]\; 
\label{sym_lutpar}
\end{eqnarray}
%\end{widetext}
where we neglect the higher order effect of broken symmetry on standard Luttinger parameters:
\begin{eqnarray}
\gamma_1 &=& -\frac{2m_0}{3\hbar^2}\left(A+2B\right) \nonumber \\
\gamma_2 &=& -\frac{m_0}{3\hbar^2}\left(A-B\right) \nonumber \\
\gamma_3 &=& -\frac{m_0}{3\hbar^2}C \nonumber \;
\end{eqnarray}
and add a new parameter:
\begin{eqnarray}
\gamma_4 &=& -\frac{2m_0}{3\hbar^2}D.
\label{lutpar_def}
\end{eqnarray}

\section{Lattice strains and microscopic potential}
We incorporate the lattice strain into the $\textbf{k} \cdot \textbf{p}$ theory following Ref.~[\onlinecite{Chow:1999_a}], which shows that $ {\cal H}_{KL}$ and the 6-band strain Hamiltonian have the same structure given in Eq.~(\ref{hl}). Components of the strain tensor $e_{\alpha \beta}$ introduced in Eq.~(\ref{r_exp}) play role of the $k$-vector components. The replacements in of matrix elements of Eqs.~(\ref{sym_lutpar}) (or rather of Eq.~(A9) in Ref.~[\onlinecite{Abolfath:2001_a}]) read:
\begin{eqnarray} 
 & k_{\alpha}k_{\beta} \rightarrow e_{\alpha\beta} & \\
-\frac{\hbar^2}{2m_0}\gamma_1 \rightarrow a_1, & 
-\frac{\hbar^2}{2m_0}\gamma_2 \rightarrow \frac{a_2}{2}, & 
-\frac{\hbar^2}{2m_0}\gamma_3 \rightarrow \frac{a_3}{2\sqrt{3}}, \nonumber \; 
\label{replace}
\end{eqnarray}
where  $a_1$, $a_2$, and $a_3$ are the elastic constants. Their values are different to Luttinger parameters $\gamma_1$, $\gamma_2$, and $\gamma_3$ as they originate from the first order momentum operator perturbation due to strain and second order perturbation treatment of the $\textbf{k} \cdot \textbf{p}$ term, respectively. The strain Hamiltonian has the following elements (in the hole picture):
\begin{eqnarray}
{\cal H}_{hh}^s &=& -\left(a_1 + \frac{a_2}{2}\right)(e_{xx}+e_{yy}) - (a_1 - a_2)e_{zz} \nonumber \\ 
{\cal H}_{lh}^s &=& -\left(a_1 - \frac{a_2}{2}\right)(e_{xx}+e_{yy}) - (a_1 + a_2)e_{zz} \nonumber \\ 
{\cal H}_{so}^s &=& -a_1(e_{xx} + e_{yy} + e_{zz}) \nonumber \\
& & \nonumber \\ 
b^s &=& -a_3(e_{zx} - i e_{zy}) \nonumber \\ 
c^s &=& \frac{a_2}{2}\sqrt{3}(e_{yy} - e_{xx}) + ia_3 e_{xy} \nonumber \\ 
d^s &=& \frac{\sqrt{2}}{2}a_2\left(2e_{zz} - (e_{xx} + e_{yy})\right).\; 
\label{strain_lutpar}
\end{eqnarray}

Now we compare the effect of microscopic symmetry breaking described by including the $\gamma_4$ dependent terms into the Hamiltonian ${\cal H}_{KL}$ to the effect of a uniform lattice strain incorporated as ${\cal H}_{str}$ with matrix elements given in Eqs.~(\ref{strain_lutpar}).  First, we write the strain Hamiltonian ${\cal H}_{str}$ as a sum of a contribution corresponding to the in-plane shear strain along the $[110]$ axis and the growth strain introduced in Sec.~\ref{su_afields} by Eq.~(\ref{strain1}) and Eq.~(\ref{strain2}), respectively. Their magnitudes are denoted by $e_{xy}$ and $e_{xx}=e_{yy}\equiv e_0$. Then we write the correction ${\cal H}_V = \tilde{\cal H}_{KL}-{\cal H}_{KL}$ to the 6-band Kohn-Luttinger Hamiltonian due to the microscopic potential $V = xy\xi$ breaking the tetrahedral symmetry of the crystal as a sum of terms with different dependence on the in-plane direction of the $k$-vector:
\begin{widetext}
\begin{eqnarray}
\label{exy_HKL}
{\cal H}_{str} & = & a_3e_{xy} \left(
\begin{array}{cccccc}
0 & -i & 0 & 0 & 0 & i\sqrt{2} \\
i & 0 & 0 & 0 & 0 & 0 \\
0 & 0 & 0 & -i & 0 & 0 \\
0 & 0 & i & 0 &  i\sqrt{2} & 0 \\
0 & 0 & 0 & -i\sqrt{2} & 0 & 0 \\
-i\sqrt{2} & 0 & 0 & 0 & 0 & 0 \\
\end{array}\right) + a_2e_0\frac{c_{11} + 2c_{12}}{c_{11}} \left(\begin{array}{cccccc} 0 & 0 & 0 & 0 & 0 & 0 \\ 
0 & 2 & 0 & 0 & 0 & \sqrt{2} \\ 
0 & 0 & 2 & 0 & -\sqrt{2} & 0 \\
0 & 0 & 0 & 0 & 0 & 0 \\ 
0 & 0 & -\sqrt{2} & 0 & 1 & 0\\
0 & \sqrt{2} & 0 & 0 & 0 & 1\\
\end{array}\right), \\
{\cal H}_V & = & \frac{\sqrt{3} \hbar^2}{2m_0}\gamma_4(k_x^2+k_y^2) \left(
\begin{array}{cccccc} 
0 & -i &0 & 0 & 0 & i\sqrt{2} \\ 
i & 0 & 0 & 0 & 0 & 0 \\
0 & 0 & 0 & -i & 0 & 0 \\
0 & 0 & i & 0 &  i\sqrt{2} & 0 \\ 
0 & 0 & 0 & -i\sqrt{2} & 0 & 0 \\
-i\sqrt{2} & 0 & 0 & 0 & 0 & 0 \\
\end{array}
\right) + \frac{\hbar^2}{m_0}\gamma_4k_xk_y \left(
\begin{array}{cccccc} 
3 & 0 & 0 & 0 & 0 & 0 \\ 
0 & 1 & 0 & 0 & 0 & -\sqrt{2} \\
0 & 0 & 1 & 0 & \sqrt{2} & 0 \\
0 & 0 & 0 & 3 & 0 & 0 \\ 
0 & 0 & \sqrt{2} & 0 & 2 & 0 \\
0 & -\sqrt{2} & 0 & 0 & 0 & 2 \\
\end{array} 
\right),
\label{Vxy_HKL}
\end{eqnarray}%
\end{widetext}
where $c_{11}$, $c_{12}$ are the elastic moduli.\cite{Chow:1999_a,Vurgaftman:2001_a}

Note that by resetting the reference energy in Eq.~(\ref{Vxy_HKL}) by subtracting $3 \frac{\hbar^2}{m_0}\gamma_4k_xk_y$ from the Hamiltonian ${\cal H}_{V}$ and the following substitutions:
\begin{eqnarray}
a_3e_{xy} & \rightarrow & \frac{\sqrt{3} \hbar^2}{2m_0}\gamma_4(k_x^2+k_y^2) \nonumber \\
-a_2e_0\frac{c_{11} + 2c_{12}}{c_{11}} & \rightarrow & \frac{\hbar^2}{m_0}\gamma_4k_xk_y
\end{eqnarray}
we can identify the two components of the strain Hamiltonian in Eq.~(\ref{exy_HKL}) with the two components of the Hamiltonian ${\cal H}_{V}$  in Eq.~(\ref{Vxy_HKL}). The important difference, however, is the dependence on $k$-vector in case of ${\cal H}_{V}$. The first term of ${\cal H}_{V}$ depends on the magnitude of the $k$-vector, not on its in-plane orientation. The second term of ${\cal H}_{V}$ has the same structure as the second term of ${\cal H}_{str}$ (which incorporates the effect of the growth strain), however, it does depend on the in-plane direction of the $k$-vector so it generates a uniaxial in-plane anisotropy component that contributes to the energy profile (shown in Fig.~\ref{micro}) similarly to the first term of Eq.~(\ref{Vxy_HKL}) (contrary to the negligible uniaxial in-plane anisotropies corresponding to the growth strain).

\section{Cubic anisotropy terms}
\label{cubic_const}
The angular dependence of the magnetocrystalline anisotropy energy can be approximated by a series of terms of distinct symmetry. In Sec.~\ref{su_afields} we introduced a simple phenomenological formula consisting of the low order terms of the cubic and uniaxial symmetry. Here we explain the choice of the independent cubic terms.

We write the terms using the components of the magnetisation unit vector $\hat{M}$: $n_x=\cos\phi\sin\theta$, $n_y=\sin\phi\sin\theta$, $n_z=\cos\theta$, where our angles $\theta$ and $\phi$ are measured from the $[001]$ and $[100]$ axis, respectively. The cubic symmetry requires invariance under permutation of the coordinate indices $x$, $y$, and $z$. The simplest term satisfying the condition is equal to unity: $ n_x^2 + n_y^2 + n_z^2 = 1$. The first order cubic term can be derived from its second power:
\begin{eqnarray}
\label{cubic1}
&   &\left (n_x^2 + n_y^2 + n_z^2 \right)^2 = \\
& = & 2 \left(n_x^2 n_y^2 + n_x^2 n_z^2 + n_y^2 n_z^2\right) \nonumber \\
& + & n_x^4 + n_y^4 + n_z^4. \nonumber
\end{eqnarray}
We obtained two lowest order cubic terms which are mutually dependent. Therefore it is enough to choose only one of them. In case of Eq.~(\ref{free_en_us}) the lowest order cubic anisotropy term reads: $K_{c1}\left(n_x^2 n_y^2 + n_x^2 n_z^2 + n_z^2 n_y^2 \right)$, where $K_{c1}$ is an energy coefficient.

The second order term can be derived from the first order term:
\begin{eqnarray}
&   & \left(n_x^2 n_y^2 + n_x^2 n_z^2 + n_y^2 n_z^2 \right) \left (n_x^2 + n_y^2 + n_z^2 \right) = \\
& = & n_x^4 n_y^2 + n_x^4 n_z^2 + n_x^2 n_y^4 + n_y^4 n_z^2 + n_x^2 n_z^4 + n_y^2 n_z^4 + \nonumber \\
& + & n_x^2 n_y^2 n_z^2. \nonumber
\end{eqnarray}
The two second order terms and the first order term are mutually dependent. Again, only one term describes fully the second order component of the cubic anisotropy. We choose $K_{c2}\left(n_x^2 n_y^2 n_z^2 \right)$ to be included into our approximate formula in Eq.~(\ref{free_en_us}).

The independent third order term is derived as follows:
\begin{eqnarray}
&   & \left(n_x^2 n_y^2 + n_x^2 n_z^2 + n_y^2 n_z^2 \right) \left (n_x^2 + n_y^2 + n_z^2 \right)^2 = \\
& = & 3(n_x^4 n_y^2 n_z^2 + n_x^2 n_y^4 n_z^2 + n_x^2 n_y^2 n_z^4) \nonumber \\
& + & 2(n_x^4 n_y^4 + n_x^4 n_z^4 + n_y^4 n_z^4) \nonumber \\
& + & n_x^6 n_y^2 + n_x^6 n_z^2 + n_x^2 n_y^6 + n_y^6 n_z^2 + n_x^2 n_z^6 + n_y^2 n_z^6. \nonumber
\end{eqnarray}
Note that the first part of the product is proportional to the second order cubic term. Again, we can choose one of the two dependent third order terms. This derivation procedure can be continued but fitting our microscopic data to the phenomenological formula yields a negligible magnitude even for the third order term coefficients.

\section{Used constants}
Let us list all the material parameters used in our codes for (Ga,Mn)As:
\begin{table} [h]
\begin{tabular}{|c|c|c|} \hline
$\gamma_1$ & $\gamma_2$ & $\gamma_3$ \\ \hline
6.85 & 2.1  & 2.9 \\ \hline \hline
$a_1$[eV] & $a_2$[eV] & $a_3$[eV] \\ \hline
-1.16   & -2.0  & -4.8 \\ \hline \hline
$c_{11}$[GPa] & $c_{12}$[GPa] & $a_{lc}$[nm] \\ \hline 
12.21  & 5.66  & 0.565 \\ \hline \hline  
$\Delta_{so}$[eV] & $J_{pd}$ [eVnm$^3$] & \\ \hline
0.341 & 0.055 & \\ \hline
\end{tabular}
\end{table}
\end{appendix}
\bibliography{MSWEBpublications,add_pub_JZ}
\end{document}